\documentclass{aa}  
\usepackage{graphicx}
\usepackage{txfonts}
\usepackage{amsmath}
\usepackage{booktabs} 
\usepackage{dcolumn}  
\usepackage{multirow} 
\usepackage{fixltx2e} 
\usepackage[UKenglish]{isodate}
\usepackage{datetime}
\usepackage{color}
\cleanlookdateon 
\newcolumntype{d}[1]{D{.}{.}{#1}} 
\def\ltsim{\ifmmode\stackrel{<}{_{\sim}}\else$\stackrel{<}{_{\sim}}$\fi}

\providecommand{\Unit}[1]{\ensuremath{\mathrm{~#1}}} 

\providecommand{\Lsun}{\Unit{{\cal L}_{\odot}}}

\providecommand{\Rsun}{\Unit{{\cal R}_{\odot}}}

\providecommand{\mas}{\Unit{mas}}

\providecommand{\parallax}{\ensuremath{\varpi}}
\providecommand{\sigparallax}{\ensuremath{\sigma_{\varpi}}}
\providecommand{\dist}{\ensuremath{r}} 
\providecommand{\gmag}{\ensuremath{G}}
\providecommand{\vmag}{\ensuremath{V}}
\providecommand{\gbp}{\ensuremath{G_{\rm BP}}}
\providecommand{\grp}{\ensuremath{G_{\rm RP}}}

\providecommand{\teff}{\ensuremath{T_{\rm eff}}}
\providecommand{\Tsun}{\ensuremath{T_{{\rm eff}{\odot}}}}
\providecommand{\logg}{\ensuremath{\log g}}
\providecommand{\ag}{\ensuremath{A_{\rm G}}}
\providecommand{\av}{\ensuremath{A_{\rm V}}} 
\providecommand{\relext}{\ensuremath{R_{\rm 0}}}
\providecommand{\mg}{\ensuremath{M_{\rm G}}}
\providecommand{\feh}{\ensuremath{\mathrm{[Fe/H]}}}
\providecommand{\mh}{\ensuremath{\mathrm{[M/H]}}}
\providecommand{\aabun}{\ensuremath{[\alpha/\mathrm{Fe}]}}
\providecommand{\lum}{\ensuremath{{\cal L}}}

\providecommand{\radius}{\ensuremath{{\cal R}}}
\providecommand{\Zabun}{\ensuremath{{\cal Z}}}
\providecommand{\bpminrp}{\ensuremath{G_{\rm BP} - G_{\rm RP}}}
\providecommand{\ebpminrp}{\ensuremath{E({\rm BP}\!-\!{\rm RP})}}

\providecommand{\bcg}{\ensuremath{BC_{\rm G}}}
\providecommand{\mbol}{\ensuremath{M_{\rm bol}}}
\providecommand{\bcvsun}{\ensuremath{BC_{{\rm V}{\odot}}}}
\providecommand{\bcgsun}{\ensuremath{BC_{{\rm G}{\odot}}}}
\providecommand{\mbolsun}{\ensuremath{M_{{\rm bol}{\odot}}}}
\newcommand{\red}{\textcolor{red}}

\def\a0{\ensuremath{A_{\rm 0}}}
\def\deg{\ensuremath{^\circ}}

\providecommand{\gdr}[1]{Gaia~DR{#1}}
\def\gspphot{GSP-Phot\xspace}
\def\priam{Priam\xspace}
\def\flame{FLAME\xspace}
\def\apsis{Apsis\xspace}
\def\extratrees{\textsc{ExtraTrees}\xspace}

\begin{document}


\title{Gaia Data Release 2: first stellar parameters from Apsis}
\titlerunning{\gdr{2}: first stellar parameters from Apsis}
\authorrunning{Andrae, Fouesneau, Creevey et al.}
\author{
Ren\'e Andrae\inst{1}, 
Morgan Fouesneau\inst{1}, 
Orlagh Creevey\inst{2}, 
Christophe Ordenovic\inst{2},
Nicolas Mary\inst{3}, 
Alexandru Burlacu\inst{4},
Laurence Chaoul\inst{5},
Anne Jean-Antoine-Piccolo\inst{5}, 
Georges Kordopatis\inst{2}, 
Andreas Korn \inst{6},
Yveline Lebreton\inst{7,8},
Chantal Panem\inst{5}, 
Bernard Pichon\inst{2},
Frederic Th\'evenin\inst{2},
Gavin Walmsley\inst{5},
Coryn A.L.\ Bailer-Jones\inst{1}\thanks{corresponding author, calj@mpia.de}
}
\institute{
Max Planck Institute for Astronomy, K\"onigstuhl 17, 69117 Heidelberg, Germany \and
Universit\'e C\^ote d'Azur, Observatoire de la C\^ote d'Azur, CNRS, Laboratoire Lagrange, Bd de l'Observatoire, CS 34229, 06304 Nice cedex 4, France \and
Thales Services, 290 All\'ee du Lac, 31670 Lab\`ege, France \and
Telespazio France, 26 Avenue Jean-Fran\c{c}ois Champollion, 31100 Toulouse, France \and
Centre National d'Etudes Spatiales, 18 av Edouard Belin, 31401 Toulouse, France \and
Division of Astronomy and Space Physics, Department of Physics and Astronomy, Uppsala University, Box 516, 75120 Uppsala, Sweden \and
LESIA, Observatoire de Paris, PSL Research University, CNRS UMR 8109,
Universit\'{e} Pierre et Marie Curie, Universit\'{e} Paris Diderot, 5 place Jules Janssen, 92190 Meudon
\and Institut de Physique de Rennes, Universit\'{e} de Rennes 1, CNRS
UMR 6251, F-35042 Rennes, France
\\
}
\date{Submitted to A\&A 21 December 2017. Resubmitted 3 March 2018 and
  3 April 2018. Accepted 3 April 2018}
\abstract{
The second Gaia data release (\gdr{2}) contains, beyond the astrometry,
three-band photometry for 1.38~billion sources.
One band is the G band, the other two were obtained by integrating the Gaia prism spectra (BP and RP). We have used these three broad photometric bands to infer stellar effective temperatures, \teff, for all sources brighter than $\gmag=17$\,mag with \teff\ in the range 3\,000--10\,000\,K (some 161 million sources).
Using in addition the parallaxes, we infer the
line-of-sight extinction, \ag, and the reddening, \ebpminrp, for 88 million sources. Together
with a bolometric correction we derive luminosity and
radius for 77 million sources. These quantities as well as their estimated uncertainties are
part of \gdr{2}. 
Here we describe the procedures by which these quantities were
obtained, including the underlying assumptions, comparison with
literature estimates, and the limitations of our
results. Typical accuracies are of order 324\,K (\teff),
0.46\,mag (\ag), 0.23\,mag (\ebpminrp), 15\%
(luminosity), and 10\% (radius). 
Being based on only a small number of observable quantities and limited training data, our results are necessarily subject to some extreme assumptions that can lead to strong systematics in some cases (not included in the aforementioned accuracy estimates). One aspect is the non-negativity contraint of our estimates, in particular extinction, which we discuss.
Yet in several regions of parameter space our results show very good performance, for example for red clump stars and solar analogues. Large uncertainties render the extinctions less useful at the individual star level, but they show good performance for ensemble estimates.
 We identify regimes in which our parameters should and should not
 be used and we define a ``clean'' sample.
Despite the limitations, this is
the largest catalogue of uniformly-inferred stellar parameters to date.
 More precise and detailed
 astrophysical parameters based on the full BP/RP spectrophotometry are
 planned as part of the third Gaia data release.
}
\keywords{methods: data analysis -- methods: statistical -- stars: fundamental parameters -- surveys: Gaia} 
\maketitle

\newcommand{\vinput}[1]{\noindent {\ \\\red{[#1]\dotfill}}\\ \input{#1}}
\section{Introduction}\label{sec:introduction}

The main objective of ESA's Gaia satellite is to understand the structure,
formation, and evolution of our Galaxy from a detailed study of its constituent
stars. Gaia's main technological advance is the accurate determination of
parallaxes and proper motions for over one billion stars. Yet the resulting
three-dimensional maps and velocity distributions which can be derived from
these are of limited value if the physical properties of the stars remain
unknown. For this reason Gaia is equipped with both a low-resolution prism
spectrophotometer (BP/RP) operating over the entire optical range, and a
high-resolution spectrograph (RVS) observing from 845--872\,nm (the payload is
described in \citealt{2016A&A...595A...1G}).

\begin{figure}
\begin{center}
\includegraphics[width=0.5\textwidth]{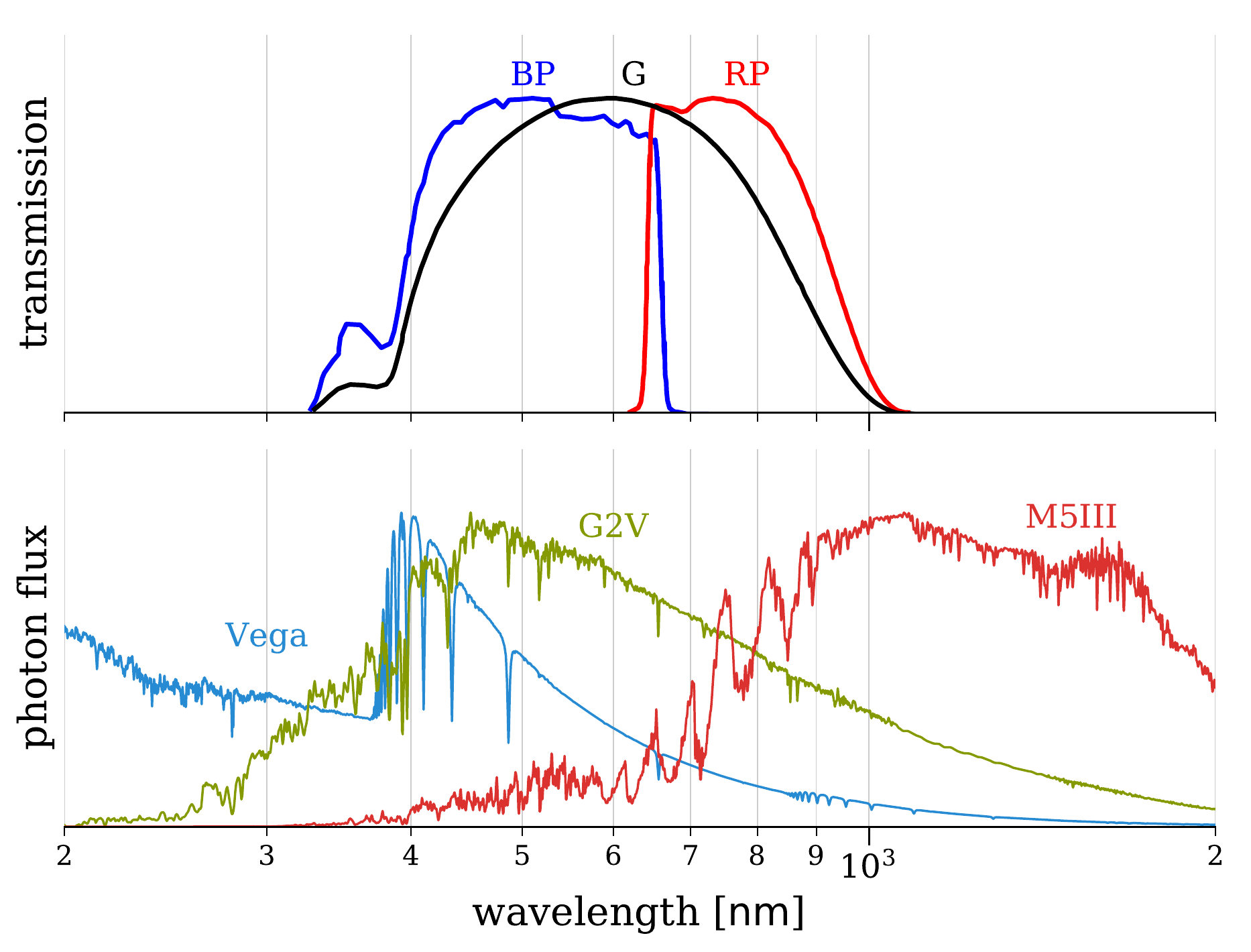}
\caption[]{The nominal transmissions of the three Gaia passbands
\citep{2010A&A...523A..48J, 2012Ap&SS.341...31D} compared with
spectra of typical stars: Vega (A0V), a G2V star (Sun-like star), and an M5III
star. Spectral templates from \citet{1998PASP..110..863P}. All curves
are normalized to have the same maximum.
\label{fig:photbands}}
\end{center}
\end{figure}

The second Gaia data release \cite[\gdr{2},][]{DR2-DPACP-36} contains
a total of 1.69~billion sources with positions and G-band
  photometry based on 22 months of mission observations.
Of these, 1.33~billion sources also have parallaxes and proper motions \citep{DR2-DPACP-51}.
Unlike in the first release, \gdr{2} also includes the integrated fluxes from the BP and
RP spectrophotometers. These prism-based instruments produce low
resolution optical spectrophotometry in the blue and red parts of the
spectra which will be used to estimate astrophysical
parameters for stars, quasars, and unresolved galaxies using the
\apsis\ data processing pipeline (see \citealt{2013A&A...559A..74B}).
They are also used in the chromatic calibration of the astrometry.
The processing and calibration of the full spectra is ongoing, and for
this reason only their integrated fluxes, expressed as the two
magnitudes \gbp\ and \grp, are released as part of \gdr{2} (see
Fig. \ref{fig:photbands}). The production and calibration of these
data are described in \cite{DR2-DPACP-44}.
1.38~billion sources in \gdr{2} have integrated photometry in all three bands, \gmag, \gbp, and \grp\ \citep{DR2-DPACP-40}, and 1.23~billion sources have both five-parameter astrometry and three-band photometry.

In this paper we describe how we use the Gaia three-band photometry and
parallaxes, together with various training data sets, to estimate the effective
temperature \teff, line-of-sight extinction \ag\ and reddening \ebpminrp,
luminosity \lum, and radius \radius, of up to 162 million stars 
brighter than
\gmag=17\,mag (some of these results are subsequently filtered out of the catalogue).
We only process sources for which all three photometric bands are
available. This therefore excludes the so-called bronze sources
\citep{DR2-DPACP-44}. 
Although photometry for fainter sources is available in \gdr{2}, we chose to
limit our analysis to brighter sources on the grounds that, at this stage in the
mission and processing, only these give sufficient photometric and parallax
precision to obtain reliable astrophysical parameters. The choice of
\gmag=17\,mag was somewhat arbitrary, however.\footnote{The
  original selection was $\gmag\leq17$\,mag, but due to a later change
  in the zeropoint, our final selection is actually
  $\gmag\leq17.068766$\,mag.}  The work described here was
carried out under the auspices of the Gaia Data Processing and Analysis
Consortium (DPAC) within Coordination Unit~8 (CU8) (see
\citealt{2016A&A...595A...1G} for an overview of the DPAC).  We realise that more
precise, and possibly more accurate, estimates of the stellar parameters could
be made by cross-matching Gaia with other survey data, such as GALEX
\citep{Morrissey2007}, PanSTARRS \citep{2016arXiv161205560C}, and WISE
\citep{AllWiseCat}. However, the remit of the
Gaia-DPAC is to process the Gaia data. Further exploitation, by including data
from other catalogues, for example, is left to the community at large.
We nonetheless hope
that the provision of these ``Gaia-only'' stellar parameters will assist the
exploitation of \gdr{2} and the validation of such extended analyses. 

We continue this article in section \ref{sec:general} with an overview of our
approach and its underlying assumptions. This is followed by a description of
the algorithm -- called \priam\ -- used to infer \teff, \ag, and \ebpminrp\ in
section \ref{sec:gspphot}, and a description of the derivation of \lum\ and
\radius\ -- with the algorithm \flame\ -- in section \ref{sec:flame}.  The
results and the content of the catalogue are presented in section
\ref{sec:results}. More details on the catalogue itself (data fields etc.) can be
found in the online documentation accompanying the data release. In section
\ref{sec:validation} we validate our results, in particular via comparison with other determinations in the
literature. In section \ref{sec:usage} we discuss the use of the data, focusing
on some selections which can be used to identify certain types of stars,
as well as the limitations of our results. {\em This is mandatory reading for anyone using the catalogue}. 
\priam\ and \flame\ are part of a
larger astrophysical parameter inference system in the Gaia data processing
(\apsis). Most of the algorithms in \apsis\ have not been activated for
\gdr{2}. (\priam\ itself is part of the \gspphot\ software package, which uses
several algorithms to estimate stellar parameters.)  We look ahead in section
\ref{sec:outlook} to the improvements and extensions of our results which can be
expected in \gdr{3}. We summarize our work in section
\ref{sec:summary}. We draw attention to appendix \ref{sec:flags}, where we define a ``clean''
subsample of our \teff\ results. 

In this article we will present both the estimates of a quantity and the
estimates of its uncertainty, and we will also compare the estimated quantity
with values in the literature. The term {\em uncertainty} refers to our computed
estimate of how precise our estimated quantity is. 
This is colloquially (and misleadingly) called an ``error bar''. We provide
asymmetric uncertainties in the form of two percentiles from a distribution
(upper and lower).  We use the term {\em error} to refer to the
difference between an estimated quantity and its literature estimate, whereby
this difference could arise from a mistake in our estimate, in the literature
value, or in both.

\section{Approach and assumptions}\label{sec:general}

\subsection{Overview of procedure}\label{sec:overview}

We estimate stellar parameters source-by-source, using only the three
Gaia photometric bands (for \teff) and additionally the parallax (for
the other four parameters). 
We do not
use any non-Gaia data on the individual sources, 
and we do not make use of any global Galactic
information, such as an extinction map or kinematics.  

The three broad photometric bands -- one of which is near degenerate with the
sum of the other two (see Fig.\ \ref{fig:photbands}) -- provide relatively
little information for deriving the intrinsic properties of the observed Gaia
targets.  They are not sufficient to determine whether the target is really a
star as opposed to a quasar or an unresolved galaxy, for example. According to our
earlier simulations, this will ultimately be possible using the full BP/RP
spectra (using the Discrete Source Classifier in \apsis). As we are only working
with sources down to \gmag=17, it is reasonable to suppose that most of them are
Galactic. Some will, inevitably, be physical binaries in which the secondary is
bright enough to affect the observed signal. We nonetheless proceed as though
all targets were single stars. Some binarity can be identified in the future
using the composite spectrum (e.g.\ with the Multiple Star Classifier in \apsis) or
the astrometry, both of which are planned for \gdr{3}.

\begin{figure}
\begin{center}
\includegraphics[width=0.5\textwidth]{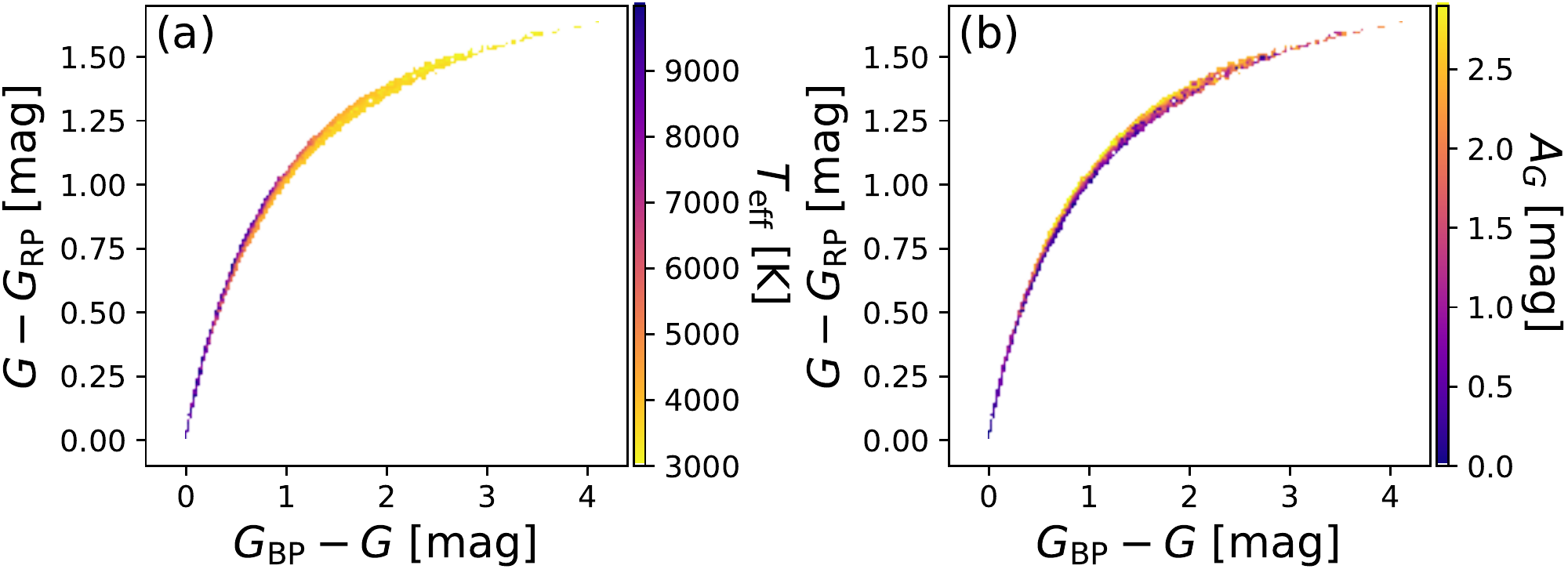}
\caption{Colour--colour diagrams for stars from the PARSEC~1.2S models with an
extinction law from \citet{1989ApJ...345..245C} and $\feh=0$. Both panels use the same data, spanning $\a0=0$--4\,mag. We see that while \teff\ is the
dominant factor (panel a), it is strongly degenerate with extinction (panel b). 
}
\label{fig:teff_ext_degeneracy}
\end{center}
\end{figure}

Unsurprisingly, \teff\ is heavily degenerate with \ag\ in the Gaia colours (see
Fig. \ref{fig:teff_ext_degeneracy}), so it seems near impossible that both
quantities could be estimated from only colours. Our experiments confirm this.
We work around this by estimating \teff\ from the colours on the assumption that
the star has (ideally) zero extinction. 
For this we use an empirically-trained machine learning algorithm 
(nowadays sometimes referred to as ``data driven''). That is, the training data are observed Gaia
photometry of targets which have had their \teff\ estimated from other sources
(generally spectroscopy). This training data set only includes stars which are
believed to have low extinctions.

We separately estimate the interstellar absorption using the three bands
together with the parallax, again using a machine learning algorithm. By using
the magnitudes and the parallax, rather than the colours, the available signal
is primarily the dimming of the sources due to absorption (as opposed to just
the reddening). For this we train on synthetic stellar spectra, because there
are too few stars with reliably estimated extinctions which could be used as an
empirical training set. Note that the absorption we estimate is the extinction
in the G-band, \ag, which is not the same as the (monochromatic) extinction
parameter, \a0. The latter depends only on the amount of absorption in the
interstellar medium, whereas the former depends also on the spectral energy
distribution (SED) of the star (see section 2.2 of
\citealt{2011MNRAS.411..435B}).\footnote{We distinguish between the
  $V$-band extinction \av\ (which depends on the intrinsic source SED)
  and the monochromatic extinction \a0\ at a wavelength of
  $\lambda=547.7$nm (which is a parameter of the extinction law and
  does not depend on the intrinsic source SED).} 
Thus even with fixed \relext\ there is not a
one-to-one relationship between \a0\ and \ag.
For this reason we use a separate model to
estimate the reddening \ebpminrp, even though the available signal is still
primarily the dimming due to absorption. By providing estimates of both
absorption and reddening explicitly, it is possible to produce a de-reddened and
de-extincted colour--magnitude diagram.

\begin{table}
\begin{center}
\caption{
The photometric zeropoints used to convert fluxes to magnitudes via Eq.~\ref{eqn:flux_to_mag} \citep{DR2-DPACP-40}.
\label{table:photometric-zeropoints}
}
\begin{tabular}{ll}
\hline
band & zeropoint ($zp$) [mag] \\
\hline
\gmag & $25.6884 \pm 0.0018$\\
\gbp & $25.3514 \pm 0.0014$\\
\grp & $24.7619 \pm 0.0020$ \\
\hline
\end{tabular}
\end{center}
\end{table}

The inputs for our processing are fluxes, $f$, provided by the upstream processing {\citep{DR2-DPACP-44}}. We convert these to magnitudes, $m$, using  
\begin{equation}
m = -2.5\log_{10} f + zp
\label{eqn:flux_to_mag}
\end{equation}
where $zp$ is the zeropoint listed in Table \ref{table:photometric-zeropoints}. 
All of our results except \teff\ depend on these zeropoints.

We estimate the absolute G-band magnitude via the usual equation
\begin{equation}
\mg \,=\, \gmag - 5\log_{10}\dist + 5 - \ag\ .
\label{eqn:mg}
\end{equation}
This is converted to a stellar luminosity using a bolometric correction (see
section \ref{sec:flame}).  The distance \dist\ to the target is taken simply to
be the inverse of the parallax. Although this generally gives a biased estimate
of the distance \citep{2015PASP..127..994B, DR2-DPACP-38}, the impact of
this is mitigated by the fact that we only report luminosities when the
fractional parallax uncertainty $\sigparallax/\parallax$ is less than 0.2. Thus,
of the 161 million stars 
with \teff\ estimates, only 77 million
have luminosity
estimates included in \gdr{2}. 

Having inferred the luminosity and temperature, the stellar radius is then obtained
by applying the Stefan--Boltzmann law 
\begin{equation}
\lum \,=\, 4\pi\radius^2\sigma\teff^4 \ .
\label{eqn:radius}
\end{equation}

Because our {\em individual} extinction estimates are rather poor for most stars
(discussed later), we chose not to use them in the derivation of luminosities,
i.e.\ we set \ag\ to zero in equation \ref{eqn:mg}.  Consequently, while our
temperature, luminosity, and radius estimates are self-consistent (within the
limits of the adopted assumptions), they are formally inconsistent with our
extinction and reddening estimates.

The final step is to filter out the most unreliable results: these do
not appear in the catalogue (see appendix
\ref{sec:results-filtering}).  We furthermore recommend that for \teff,
only the ``clean'' subsample of our results be used. This is
defined and identified using the flags in appendix
\ref{sec:flags}. When
  using extinctions, users may further want to make a cut
    to only retain stars with lower fractional parallax uncertainties.

\subsection{Data processing}\label{sec:processing}

The software for \apsis\ is produced by teams in Heidelberg, Germany (\priam)
and Nice, France (\flame).  The actual execution of the \apsis\ software on the
Gaia data is done by the DPCC (Data Processing Centre CNES) in
Toulouse, which also
integrates the software. The
processing comprises several operations, including the input and output of data
and generation of logs and execution reports.  The entire process is managed by
a top-level software system called SAGA.
\apsis\ is run in parallel on a multi-core Hadoop cluster system, with data
stored in a distributed file system. The validation results are published on a
web server (GaiaWeb) for download by the scientific software providers. The
final \apsis\ processing for \gdr{2} took place in October 2017. The complete set
of sources (1.69 billion with photometry) covering all Gaia magnitudes was
ingested into the system. From this the 164 million sources
 brighter than
\gmag=17\,mag
were identified and processed. This was done on 1000 cores (with 6
GB RAM per core), and ran in about 5000 hours of CPU time (around five hours wall
clock time). The full \apsis\ system, which involves much more CPU-intensive
processes, higher-dimensional input data (spectra), and of order one billion
sources, will require significantly more resources and time.

\section{Priam}\label{sec:gspphot}

\subsection{General comments}
\label{ssect:priam-general}

Once the dispersed BP/RP spectrophotometry are available, the \gspphot\ software will estimate a number of different stellar parameters for a range of stellar types \citep[see][]{2012MNRAS.426.2463L, 2013A&A...559A..74B}.  For \gdr{2} we use only the \priam\ module within \gspphot\ to infer parameters using integrated photometry and parallax.  All sources are processed even if they have corrupt photometry (see Fig.~\ref{fig:gspphot-quality-cuts-colour-colour}) or if the parallax is missing or non-positive. Some results are flagged and others filtered from the catalogue (see appendix \ref{sec:results-filtering}).

\priam employs extremely randomised trees \citep[][hereafter \extratrees]{Geurts2006}, a machine learning algorithm with a univariate output.  We use an ensemble of 201 trees and take the median of their outputs as our parameter estimate.\footnote{Further \extratrees regression parameters are $k=2$ random trials per split and $n_\textrm{min}=5$ minimal stars per leaf node.} We use the 16th and 84th percentiles of the \extratrees ensemble as two uncertainty estimates; together they form a central 68\% confidence interval.
Note that this is, in general, asymmetric with respect to the parameter estimate.
201 trees is not very many from which to accurately compute such intervals -- a limit imposed by available computer memory -- but our validation shows them to be reasonable.
\extratrees are incapable of extrapolation: they cannot
produce estimates or confidence intervals outside the range of the target variable (e.g.\ \teff) in the training data.  We experimented with other machine learning algorithms, such as support vector machine (SVM) regression \citep[e.g.][]{Deng2012} and Gaussian processes \citep[e.g.][]{Bishop2006}, but we found \extratrees to be much faster (when training is also considered), avoid the high sensitivity of SVM tuning, and yet still provide results which are as good as any other method tried.

\subsection{Effective temperatures}
\label{ssect:gspphot-teff-explanation}

\begin{figure*}
\begin{center}
\includegraphics[width=0.95\textwidth]{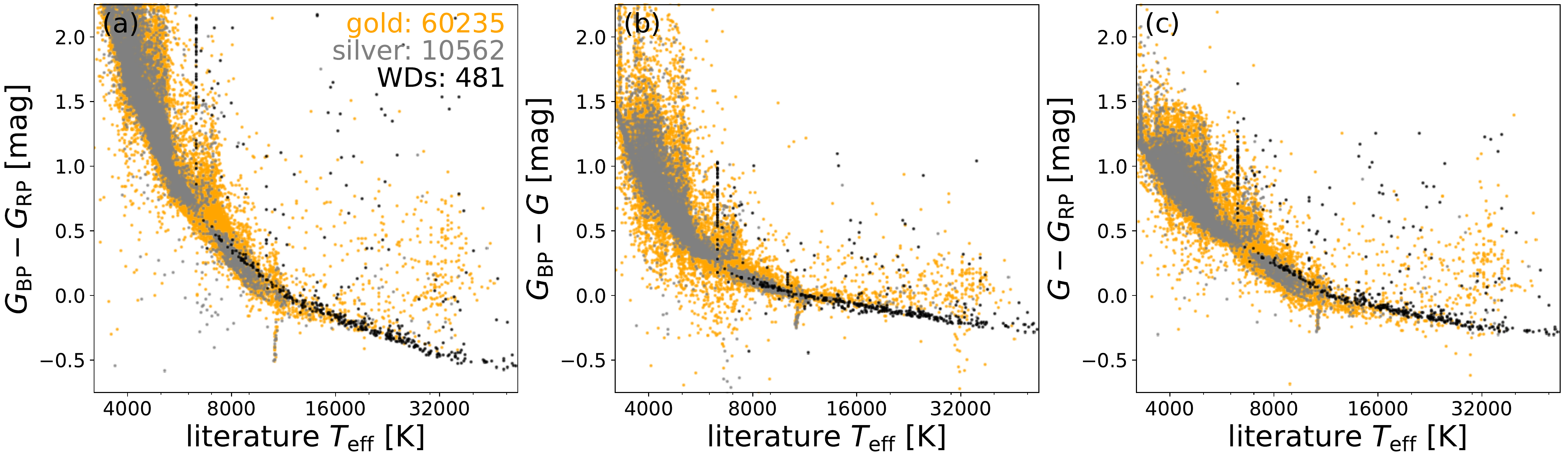}
\caption{Colour--temperature relations for Gaia data (our reference sample
described in sect. \ref{ssect:gspphot-teff-explanation}) with
literature estimates of \teff. Each panel shows a different Gaia colour. Sources
with gold-standard photometry are shown in orange and those with silver-standard
photometry are shown in grey. 
White dwarfs matched to \citet{Kleinman2013} are shown in black.}
\label{fig:gspphot_colours-vs-literature-Teff}
\end{center}
\end{figure*}

Given the observed photometry \gmag, \gbp, and \grp, we use the
distance-independent colours \gbp$-$\gmag~and \gmag$-$\grp\ as the inputs to
\extratrees to estimate the stellar effective temperature \teff. These two
colours exhibit a monotonic trend with \teff\
(Fig.~\ref{fig:gspphot_colours-vs-literature-Teff}). It is possible to form a
third colour, \bpminrp, but this is not independent, plus it is noisier since it does not
contain the higher signal-to-noise ratio \gmag-band. We do not propagate the flux uncertainties through \extratrees. Furthermore, the integrated photometry is calibrated with two different procedures, producing so-called ``gold-standard'' and ``silver-standard'' photometry \citep{DR2-DPACP-44}. As shown in
Fig.~\ref{fig:gspphot_colours-vs-literature-Teff}, gold and silver photometry provide the same colour-temperature relations, thus validating the consistency of the two calibration procedures of \citet{DR2-DPACP-44}.

Since the in-flight instrument differs from its nominal pre-launch prescription \citep{2010A&A...523A..48J, 2012Ap&SS.341...31D}, in particular regarding the passbands (see Fig.~\ref{fig:photbands}), we chose not to train \extratrees on synthetic photometry for \teff. Even though the differences between nominal and real passbands are probably only of the order of $\sim$0.1\,mag or less in the zeropoint magnitudes (and thus even less in colours), we obtained poor \teff~estimates, with differences of around 800\,K compared to literature values when using synthetic colours from the nominal passbands. We instead train \extratrees on Gaia sources with observed photometry and \teff~labels taken from various catalogues in the literature. These catalogues use a range of data and methods to estimate \teff: APOGEE \citep{Alam2015} uses mid-resolution, near-infrared spectroscopy;  the Kepler Input Catalogue\footnote{\url{https://archive.stsci.edu/pub/kepler/catalogs/} file \tt{kic\_ct\_join\_12142009.txt.gz}} \citep{Huber2014} uses photometry; LAMOST \citep{LamostDR1} uses low-resolution optical spectroscopy; RAVE \citep{2013AJ....146..134K} uses mid-resolution spectroscopy in a narrow window around the Ca{\sc ii} triplet. The RVS auxiliary catalogue \citep{SoubiranCS011,DR2-DPACP-47}, which we also use, is itself is a compilation of smaller catalogues, each again using different methods and different data. By combining all these different catalogues we are deliberately ``averaging'' over the systematic differences in their \teff~estimates. The validation results presented in sections~\ref{sec:results} and~\ref{sec:validation} will show that this is not the limiting factor in our performance, however. This data set only includes stars which have low extinctions (although not as low as we would have liked). 95\% of the literature estimates for these stars are below 0.705\,mag for \av~and 0.307\,mag for $E(B-V)$. (50\% are below 0.335\,mag and 0.13\,mag respectively.)
These limits exclude the APOGEE part of the training set, for which no estimates of \av\ or $E(B-V)$ are provided. While APOGEE giants in particular can reach very high extinctions, they are too few to enable \extratrees to learn to disentangle the effects of temperature and extinction in the training process. The training set is mostly near-solar metallicity stars: 95\% of the stars have $\feh> -0.82$ and 99\% have $\feh> -1.89$.

We compute our magnitudes from the fluxes provided by the upstream
processing using equation \ref{eqn:flux_to_mag}. The values of the zeropoints
used here are unimportant, however, because the same zeropoints are used for
both training and application data.

\begin{figure}
\begin{center}
\includegraphics[width=0.48\textwidth]{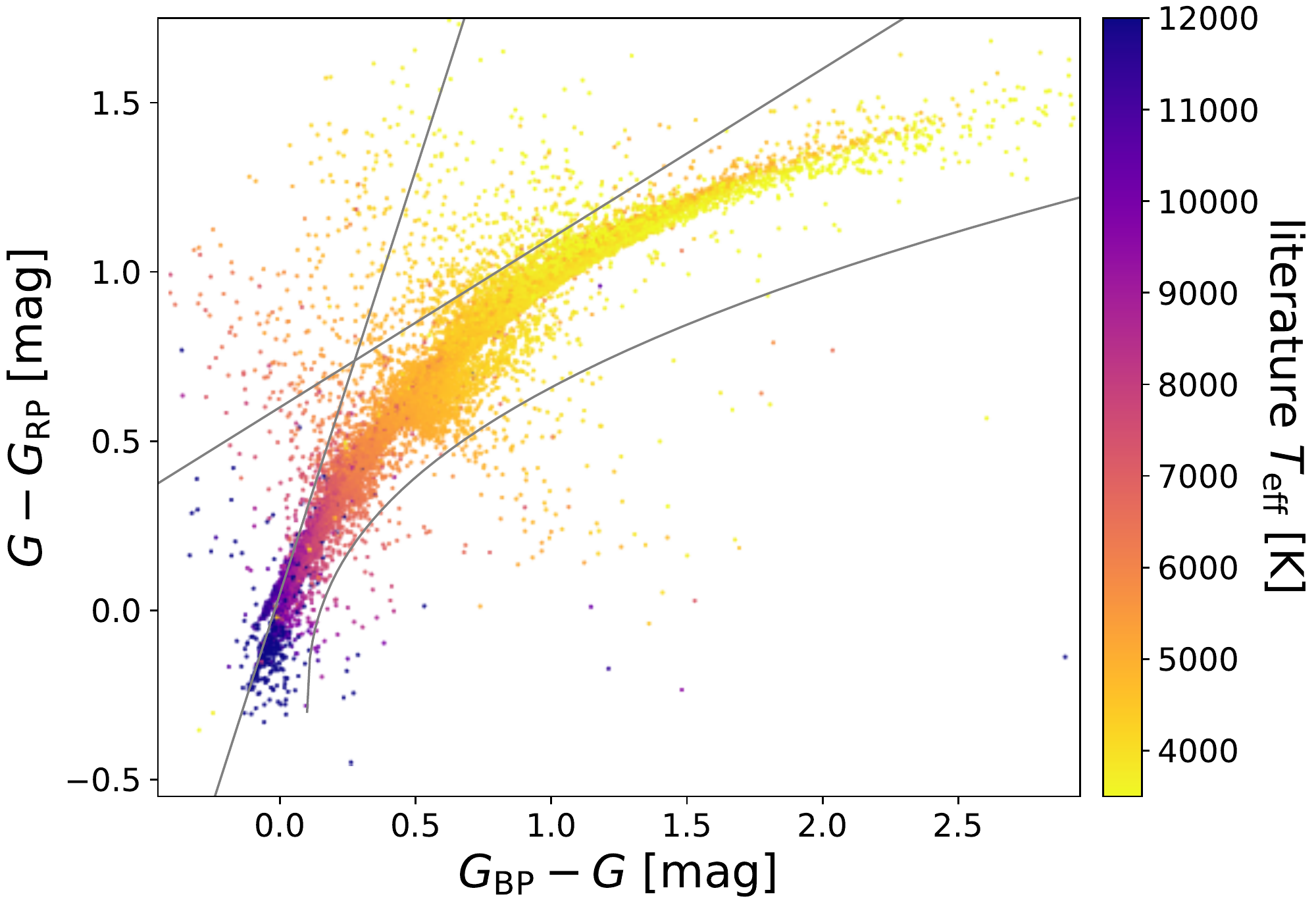}
\caption{Colour--colour diagram for Gaia data (our reference sample described
in sect. \ref{ssect:gspphot-teff-explanation}) with literature estimates of
\teff. Grey lines show quality cuts where bad photometry is flagged (see Table \ref{table:Priam-flags}). Sources with
excess flux larger than 5 have been discarded.}
\label{fig:gspphot-quality-cuts-colour-colour}
\end{center}
\end{figure}

\begin{figure}
\begin{center}
\includegraphics[width=0.48\textwidth]{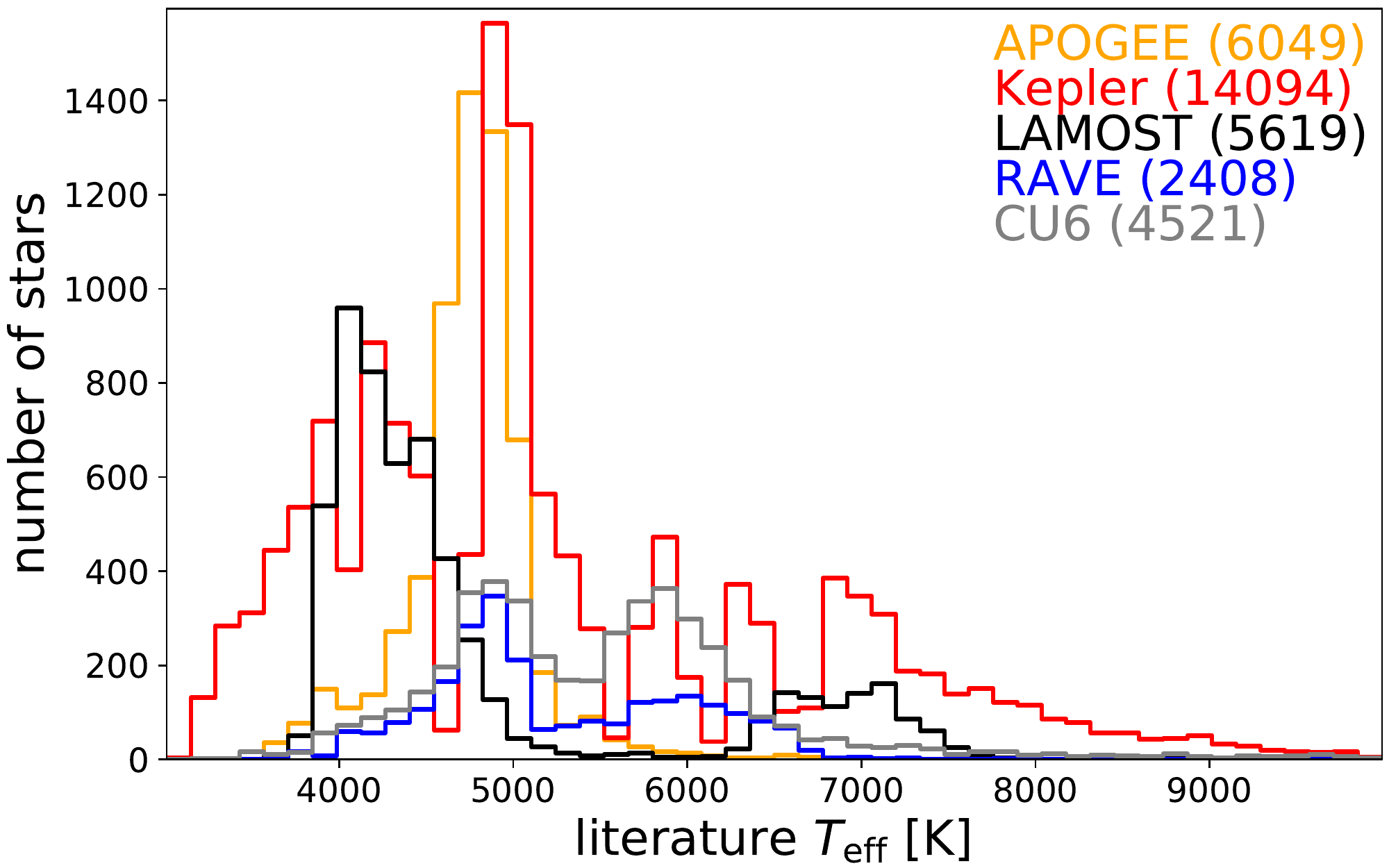}
\caption{Distribution of literature estimates of \teff~for the selected training
sample. The numbers in parenthesis indicate how many stars from each catalogue have been
used. The test sample distribution is almost identical.
}
\label{fig:gspphot-Priam-training-Teff-catalogs}
\end{center}
\end{figure}

\begin{table}
\begin{center}
\caption{Catalogues used for training \extratrees for \teff\ estimation showing
the number of stars in the range from 3\,000K to 10\,000K that we selected and
the mean \teff~uncertainty quoted by the catalogues.
\label{tab:gspphot-teff-catalogs-for-training}
}
\begin{tabular}{lrr}
\hline
                         & number   & mean \teff      \\
 catalogue               & of stars & uncertainty [K] \\
 \hline
 APOGEE                  & 5\,978   & 92              \\
 Kepler Input Catalogue  & 14\,104  & 141             \\
 LAMOST                  & 5\,540   & 55              \\
 RAVE                    & 2\,427   & 61              \\
 RVS Auxiliary Catalogue & 4\,553   & 122             \\
 \hline
 combined                & 32\,602  & 102             \\
 \hline
\end{tabular}
\end{center}
\end{table}

We only retain stars for training if the catalogue specifies a \teff\ uncertainty of less
than 200K, and if the catalogue provides estimates of \logg\ and \feh. The
resulting set of 65\,000 stars, which we refer to as the {\it reference sample},
is shown in Figs.~\ref{fig:gspphot_colours-vs-literature-Teff}
and~\ref{fig:gspphot-quality-cuts-colour-colour}.
We split this sample into near-equal-sized training and test sets.  To make
this split reproducible, we use the digit sum of the Gaia source ID (a long
integer which is always even): sources with even digit sums are used for training, those with odd for testing.
The temperature distribution
of the training set is shown in
Fig.~\ref{fig:gspphot-Priam-training-Teff-catalogs} (that for the test set is
virtually identical). The distribution is very inhomogeneous. The impact of this on the results is discussed in section~\ref{ssec:results-extinction}. Our supervised learning approach implicitly assumes that the adopted training distribution is representative of the actual temperature distribution all over the sky, which is certainly not the case (APOGEE and LAMOST probe quite different stellar populations, for example). However, such an assumption -- that the adopted models are representative of the test data -- can hardly be avoided. We minimise its impact by combining many different literature catalogues covering as much of the expected parameter space as possible.

Table~\ref{tab:gspphot-teff-catalogs-for-training} lists the number of stars (in the training set) from each catalogue, along with their typical \teff~uncertainty estimates as provided by that catalogue (which we will use in section~\ref{ssec:results-teff} to infer the intrinsic temperature error of \priam).\footnote{The subsets in Table~\ref{tab:gspphot-teff-catalogs-for-training} are so small that there are no overlaps between the different catalogues. Also note that the uncertainty estimates provided in the literature are sometimes clearly too small, e.g.\ for LAMOST.} Mixing catalogues which have had \teff\ estimated by different methods is likely to increase the scatter (variance) in our results, but it is a property of \extratrees that this averaging should correspondingly reduce the bias in our results.  Such a mixture is necessary, because no single catalogue covers all physical parameter space with a sufficiently large number of stars for adequate training.  Even with this mix of catalogues we had to restrict the temperature range to 3000K$-$10\,000K, since there are too few literature estimates outside of this range to enable us to get good results.  For instance, there are only a few hundred OB stars with published \teff~estimates \citep{RamirezAgudelo2017,SimonDiaz2017}. We tried to extend the upper temperature limit by training on white dwarfs with \teff~estimates from \citet{Kleinman2013}, but as Fig.~\ref{fig:gspphot_colours-vs-literature-Teff} reveals, the colour-temperature relations of white dwarfs (black points) differ significantly from those of OB stars (orange points with $\teff\gtrsim 15\,000$K). Since \extratrees cannot extrapolate, this implies that stars with true $\teff<3000$K or $\teff>10\,000$K are ``thrown back'' into the interval 3000K--10\,000K (see~section~\ref{ssect:teff-for-OB-and-cool-stars}). This may generate peculiar patterns when, for example, plotting a Hertzsprung--Russell diagram (see section \ref{ssec:results-extinction}).

The colour--colour diagram shown in Fig.~\ref{fig:gspphot-quality-cuts-colour-colour} exhibits substantially larger scatter than expected from the PARSEC models shown in Fig.~\ref{fig:teff_ext_degeneracy}, even inside the selected good-quality region. This is not due to the measurement errors on fluxes, as the formal uncertainties in Fig.~\ref{fig:gspphot-quality-cuts-colour-colour} are smaller than dot size for 99\% of the stars plotted. Instead, this larger scatter reflects a genuine astrophysical diversity that is not accounted for in the models (for example due to metallicity variations, whereas Fig.~\ref{fig:teff_ext_degeneracy} is restricted to $\feh=0$).

\subsection{Line-of-sight extinctions}
\label{ssect:gspphot-extinction-description}

For the first time, \gdr{2} provides a colour--magnitude diagram for hundreds of
millions of stars with good parallaxes. We complement
this with estimates of the \gmag-band extinction \ag~and the \ebpminrp~reddening
such that a dust-corrected colour-magnitude diagram can be produced.

\begin{figure}
\begin{center}
\includegraphics[width=0.5\textwidth]{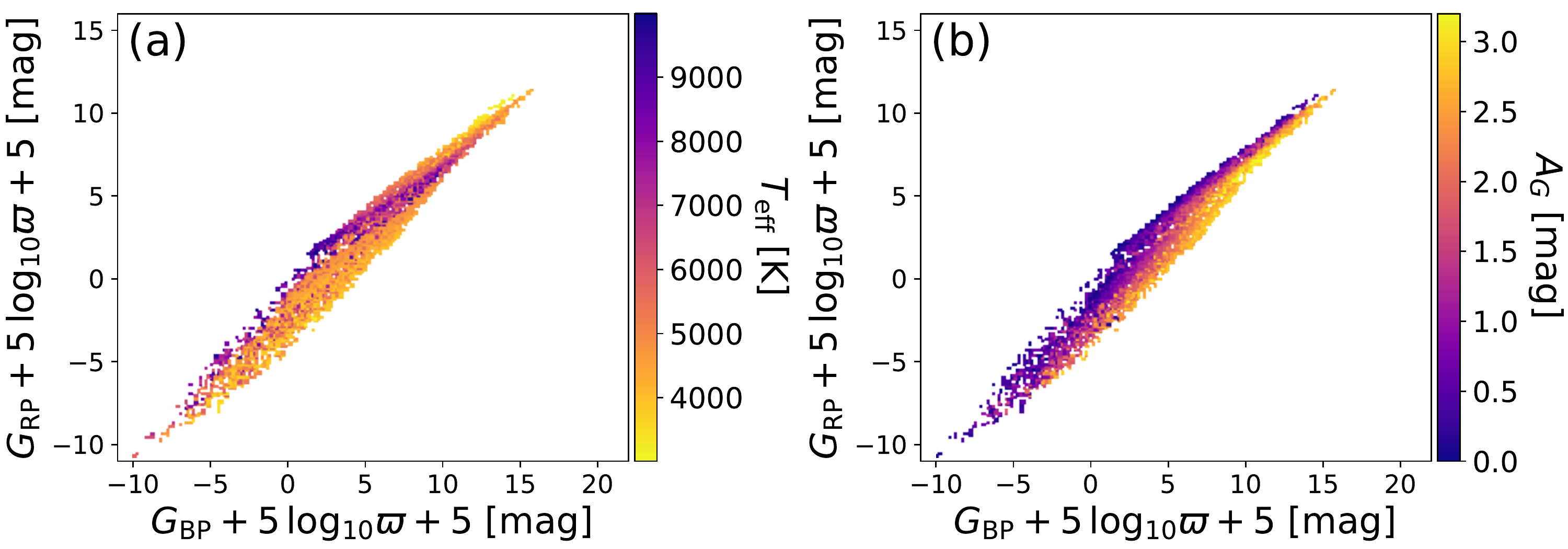}
\caption{Predicted relations between observables $\gbp+5\,\log_{10}\parallax+5=M_\textrm{BP}+A_\textrm{BP}$
and $\grp+5\,\log_{10}\parallax+5=M_\textrm{RP}+A_\textrm{RP}$ using synthetic
photometry including extinction.
\label{fig:gspphot-PARSEC-dimmed-magnitudes-vs-parameters}}
\end{center}
\end{figure}

As expected, we were unable to estimate the line-of-sight extinction from just
the colours, since the colour is strongly influenced by \teff\
(Fig.~\ref{fig:teff_ext_degeneracy}a vs.~b). We therefore use the parallax
$\parallax$ to estimate the distance and then use equation~(\ref{eqn:mg}) to
compute $M_X+A_X$ for all three bands (which isn't directly measured, but for
convenience we refer to it from now on as an observable). We then use the three
observables $\mg+\ag$,
$M_\textrm{BP}+A_\textrm{BP}$, $M_\textrm{RP}+A_\textrm{RP}$ as features for training \extratrees. 
As shown in Fig.~\ref{fig:gspphot-PARSEC-dimmed-magnitudes-vs-parameters}b,
there is a clear extinction trend in this observable space, whereas the
dependence on \teff\ (Fig.~\ref{fig:gspphot-PARSEC-dimmed-magnitudes-vs-parameters}a) is much less
pronounced than in colour-colour space (Fig.~\ref{fig:teff_ext_degeneracy}a). Yet, extinction and temperature are still very degenerate in some parts of the parameter space, and also there is no unique mapping of $M_X+A_X$ to extinction thus leading to further degeneracies (see section \ref{ssec:extinction-in-halo}).
Dependence on the parallax here restricts us to stars with precise parallaxes,
but we want to estimate \ag~and \ebpminrp~in order to correct the
colour-magnitude diagram (CMD), which itself is already limited by parallax
precision. 
We do not propagate the flux and parallax uncertainties through \extratrees.\footnote{We
found that propagating the flux and parallax uncertainties through the \extratrees has no noteworthy impact on our results, i.e.\ our extinction and reddening estimates are not limited by the expected precision of the input data.}

\begin{figure}
\begin{center}
\includegraphics[width=0.5\textwidth]{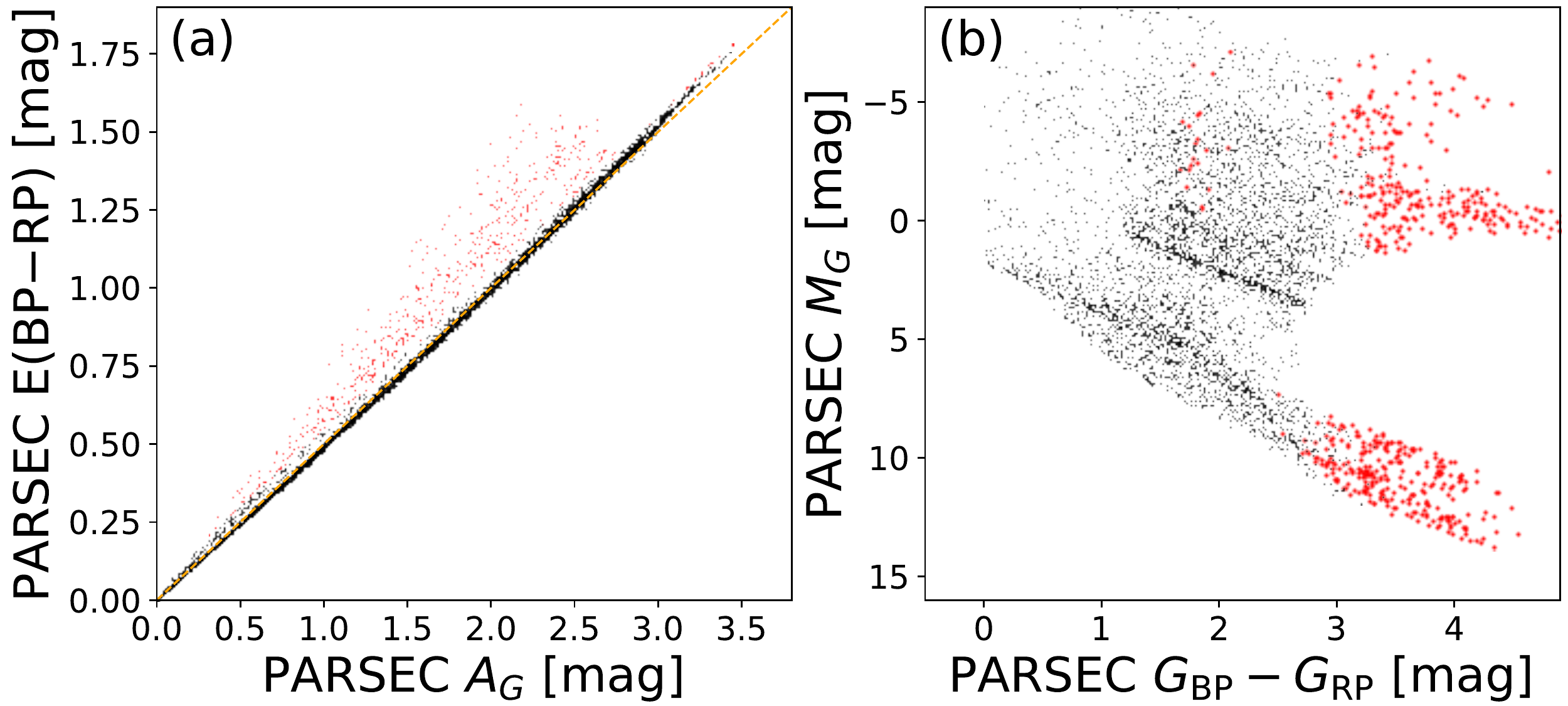}
\caption{Approximate relation between $\ag$ and $\ebpminrp$ (labels of \extratrees training data) for PARSEC~1.2S models \citep{Bressan12} with $0\leq\ag\leq 4$ and 3000\,K$\leq\teff\leq$10\,000\,K. PARSEC models use the extinction law of \citet{1989ApJ...345..245C}. We see in panel (a) that most stars follow the relation $\ag\sim 2\cdot\ebpminrp$ (dashed orange line) while the stars highlighted in red behave differently. Panel (b) shows that these star with different \ag\--\ebpminrp\ relation are very red (i.e.\ cool) sources.}
\label{fig:gspphot-PARSEC-models-AG-vs-EBPminRP}
\end{center}
\end{figure}

In order to estimate extinction we cannot train our models on literature values, for two reasons. First, there are very few reliable literature estimate of the extinction. Second, published estimates are of \av\ and/or E($B-V$) rather than \ag\ and \ebpminrp.  We therefore use
the PARSEC 1.2S models\footnote{\url{http://stev.oapd.inaf.it/cgi-bin/cmd}} to obtain
integrated photometry from the synthetic Atlas~9 spectral libraries
\citep{2004astro.ph..5087C} and the nominal instrument passbands (Fig.~\ref{fig:photbands}). These models use the extinction law from
\citet{1989ApJ...345..245C} and \citet{1994ApJ...422..158O} with a fixed relative
extinction parameter, \relext=3.1.  We constructed a model grid that spans
$\a0=0$--4\,mag, a temperature range of 2\,500--20\,000\,K,
a \logg~range of 1--6.5\,dex, and a fixed solar metallicity ($Z_\odot=0.0152$, $\feh=0$).  
We chose solar metallicity for our models since we could not cover all metallicities and because we expect most stars in our sample to have $\feh\sim 0$.
The extinctions \ag, $A_\textrm{BP}$, $A_\textrm{RP}$~and the reddening
$\ebpminrp=A_\textrm{BP}-A_\textrm{RP}$ are then computed for each star by subtracting from the extincted magnitudes the unextincted magnitudes (which are
obtained for $\a0=0$\,mag). We used the sampling of the PARSEC evolutionary models as is, without further rebalancing or interpolation. Since this sampling is optimised to catch the pace of stellar evolution with time, the underlying distribution of temperatures, masses, ages, and extinctions is not representative of the Gaia sample. Therefore, as for \teff, this will have an impact on our extinction estimates. However, while the Gaia colours are highly sensitive to \teff, the photometry alone hardly allows us to constrain extinction and reddening, such that we expect that artefacts from this mismatch of the distributions in training data and real data will be washed out by random noise.

We use two separate \extratrees models, one for \ag~and
one for \ebpminrp.  The input observables are $\mg+\ag$,
$M_\textrm{BP}+A_\textrm{BP}$, and $M_\textrm{RP}+A_\textrm{RP}$ in both cases. That is, 
we do not infer \ebpminrp~from colour measurements. On account of
the extinction law, \ebpminrp\ and \ag\ are strongly correlated to the relation
$\ag\sim 2\cdot\ebpminrp$ over most of our adopted temperature range, as can be seen in
Fig.~\ref{fig:gspphot-PARSEC-models-AG-vs-EBPminRP}.\footnote{Using different stellar atmosphere models with different underlying synthetic SEDs, \citet{2010A&A...523A..48J} found a slightly different relation between \ag\ and \ebpminrp.} The finite scatter is due
to the different spectral energy distributions of the stars: the largest deviations occur
for very red sources.  Note that because \extratrees
cannot extrapolate beyond the training data range, we avoid negative estimates
of \ag~and \ebpminrp. This non-negativity means the likelihood cannot be Gaussian, and as discussed in 
appendix \ref{appendix:mean-cluster-extinction}, a truncated Gaussian is more appropriate.

Evidently, the mismatch between synthetic and real Gaia photometry, i.e.\
differences between passbands used in the training and the true passbands (and
zeropoints), will have a detrimental impact on our extinction estimates, possibly leading to systematic errors.
Nonetheless, this mismatch is only $\sim$0.1mag in the zeropoints \citep{DR2-DPACP-40} and as shown in \citet{DR2-DPACP-31}, the synthetic photometry (using inflight passbands) of
isochrone models actually agrees quite well with the Gaia data.  Indeed, as will be shown
in sections \ref{ssec:results-extinction} and~\ref{ssec:red-clump-stars},
this mismatch appears not to lead to obvious systematic
errors.\footnote{The situation for \teff\ would be different, where using
synthetic colours results in large errors.}

Although we cannot estimate temperatures from these models with our data,
the adopted \teff~range of 2500--20\,000\,K for the PARSEC models allows us to
obtain reliable extinction and reddening estimates for intrinsically very blue
sources such as OB stars, even though the method described in
section \ref{ssect:gspphot-teff-explanation} cannot provide good \teff~estimates
for them.

\section{FLAME}\label{sec:flame}

The Final Luminosity, Age, and Mass Estimator (\flame) module aims to infer fundamental parameters of stars.
In \gdr{2} we only activate the components for inferring luminosity and radius. Mass and age will follow in the next data release, once \gspphot\ is able to estimate \logg\ and \feh\ 
from the BP/RP spectra and the precision in \teff\ and \ag\ improves.
We calculate luminosity \lum\ with
\begin{equation}
-2.5\log_{10}\lum = \mg + \bcg(\teff) - \mbolsun
\label{eqn:lum}
\end{equation}
where  \lum\ is in units of \Lsun\ (Table~\ref{tab:refparameters}), 
\mg\ is the absolute magnitude of the star in the G-band,
$\bcg(\teff)$ is a temperature dependent bolometric correction (defined below), and
$\mbolsun = 4.74$\,mag is the 
solar bolometric magnitude
as defined in IAU Resolution 2015 B2\footnote{\url{https://www.iau.org/static/resolutions/IAU2015_English.pdf}}. 
The absolute magnitude is computed from the G-band flux and parallax using equations \ref{eqn:flux_to_mag} and \ref{eqn:mg}.
As the estimates of extinction provided by \priam\ were shown not to be 
sufficiently accurate on a star-to-star basis for many of our brighter validation 
targets, we set \ag\ to zero when computing \mg.
The radius \radius\ is then calculated from equation~\ref{eqn:radius} using this luminosity and \teff\ from \priam.
These derivations are somewhat trivial; at this stage \flame simply provides
easy access for the community to these fundamental parameters.

Should a user want to estimate luminosity or radius assuming a non-zero extinction 
$A_{G,\rm{new}}$ and/or a change in the bolometric correction of $\Delta \bcg$, one can use the following expressions
\begin{equation}
\lum_{\rm new} = \lum \, 10^{0.4 (A_{G,\rm{new}} - \Delta \bcg)} 
\label{eqn:newlum}
\end{equation}
\begin{equation}
\radius_{\rm new} = \radius \, 10^{0.2 (A_{G,\rm{new}} - \Delta \bcg)}  \ .
\label{eqn:newradius}
\end{equation}

\begin{table}
\begin{center}
\caption{
Reference solar parameters. 
\label{tab:refparameters}
}
\begin{tabular}{rrd{4.6}}\hline
quantity & unit & \text{value} \\
\hline
\Rsun\ &m & 6.957e+08 \\
\Tsun\ &K & 5.772e+03 \\
\Lsun\ &W & 3.828e+26 \\
\mbolsun\ & mag & 4.74 \\
\bcgsun\ &mag & +0.06\\
$\vmag_{\odot}$ &mag & -26.76 \\
\bcvsun\ &mag & -0.07\\
$M_{V\odot}$ &mag & 4.81 \\
\hline
\end{tabular}
\end{center}
\end{table}

\subsection{Bolometric Correction}\label{sec:bc}

We obtained the bolometric correction \bcg\ on a grid as a function of \teff, \logg, \mh, and \aabun, derived from the MARCS synthetic stellar spectra \citep{2008AA...486..951G}.
The synthetic spectra cover a \teff\ range from 2500K to 8000K,
\logg\ from $-$0.5 to 5.5 dex, 
\feh\ from $-$5.0 to +1.0 dex, and \aabun\  from +0.0 to +0.4 dex.
Magnitudes are computed from the grid spectra using the G filter (Fig.
\ref{fig:photbands}).
These models assume local thermodynamic equilibrium (LTE), with plane-parallel geometry for dwarfs and spherical symmetry for giants.  We extended the \teff\ range using the \bcg\ from \cite{2010A&A...523A..48J}, but with an offset added to achieve continuity with the MARCS models at 8000 K.  However, following the validation of our results (discussed later), we choose to filter out \flame\ results for stars with \teff\ outside the range 3300 -- 8000\,K (see appendix \ref{sec:results-filtering}).

For the present work we had to address two issues. First, \bcg\ is a function of
four stellar parameters, but it was necessary to project this to be a function
of just \teff, since for \gdr{2} we do not yet have estimates of the other three stellar
parameters. 
Second, the bolometric correction needs a reference point to set the
absolute scale, as this is not defined by the models. We will refer to
this as the {\em offset} of the bolometric correction, and it has been defined here so that the solar bolometric correction \bcgsun\ is +0.06 mag.  Further details are provided in appendix~\ref{sec:flame-app}.

To provide a 1-D bolometric correction, we set
\aabun=0
and select the \bcg\ corresponding\footnote{Choosing $|\feh| < 1.0$ or including $\aabun=+0.4$ 
only changed the \bcg\ in the third decimal place, well below its final uncertainty. 
Fixing \feh\ to a single value (e.g.\ zero) had just as little impact relative to the uncertainty.} 
to $|\feh| < 0.5$.
As there is still a dependence on \logg, we adopt for each 
\teff\ bin the mean value of the bolometric correction. We also compute the standard deviation
$\sigma(\bcg)$ as a measure of the uncertainty due to the dispersion in \logg.
We then fit a polynomial to these values to define the function
\begin{equation}
\bcg(\teff) =  \sum_{i=0}^4 a_i (\teff - \Tsun)^i.
\label{eqn:bcgfgkstars}
\end{equation}
The values of the fitted coefficients are given in Table~\ref{tab:bcgcoeff}.
The fit is actually done with the offset parameter $a_0$ fixed to
$\bcgsun = +0.06$\,mag, the reference bolometric value of the Sun (see
appendix \ref{sec:flame-app}).
{We furthermore make two independent fits, one for the \teff\ in the range 4000--8000\,K and another for the range 3300--4000\,K.}

\begin{table}[h]
\begin{center}
\caption{Polynomial coefficients of the model \bcg(\teff) defined in equation \ref{eqn:bcgfgkstars} (column labelled \bcg).
A separate model was fit to the two temperature ranges.
The coefficient $a_0$ was fixed to its value for the 4000--8000 K temperature range.
For the lower temperature range $a_0$ was fixed to ensure continuity at 4000 K.
The column labelled $\sigma(\bcg)$ lists the coefficients for a model of the uncertainty due 
to the scatter of \logg.
\label{tab:bcgcoeff}
}
\begin{tabular}{lrr}
\hline
& \bcg & $\sigma(\bcg)$ \\
\hline
& \multicolumn{2}{c}{4000 -- 8000~K} \\
\hline
$a_0$    & 6.000e$-02$ & 2.634e$-02$ \\
$a_1$   & 6.731e$-05$ &  2.438e$-05$ \\
$a_2$ & $-$6.647e$-08$ & $-$1.129e$-09$ \\
$a_3$  &  2.859e$-11$ &$-$6.722e$-12$ \\
$a_4$ &  $-$7.197e$-15$ & 1.635e$-15$ \\
\hline
& \multicolumn{2}{c}{3300 -- 4000~K} \\
\hline
$a_0$  & 1.749e$+00$ & $-$2.487e$+00$ \\
$a_1$  & 1.977e$-03$ & $-$1.876e$-03$ \\
$a_2$  & 3.737e$-07$ & 2.128e$-07$ \\
$a_3$  & $-$8.966e$-11$ & 3.807e$-10$ \\
$a_4$  & $-$4.183e$-14$ & 6.570e$-14$ \\
\hline
\end{tabular}
\end{center}
\end{table}
 
\begin{figure}
\center{\includegraphics[width=\columnwidth]{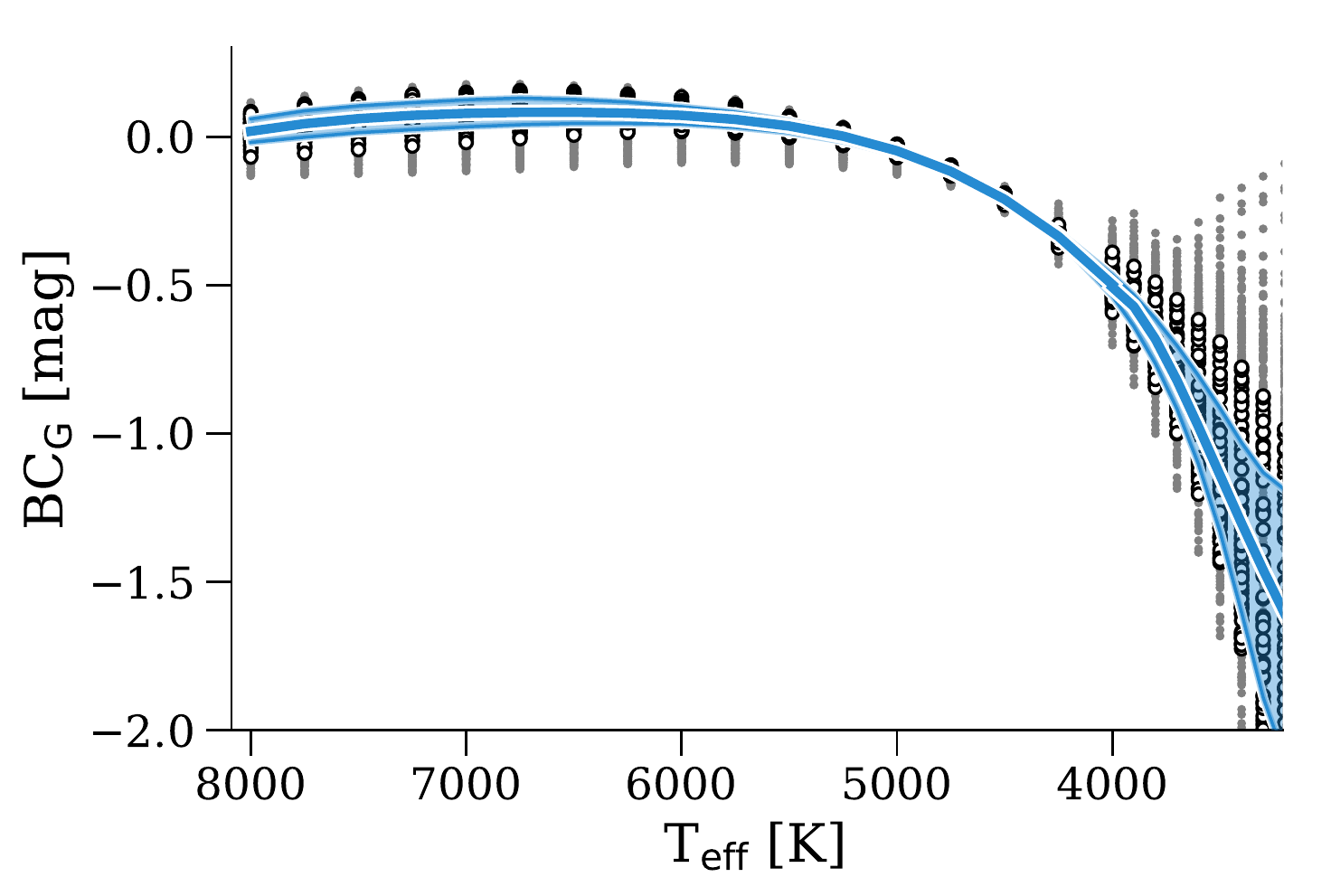}}
\caption{Bolometric corrections from the MARCS models (grey dots) and
  the subset we selected (open circles) to fit the polynomial model
  (equation~\ref{eqn:bcgfgkstars}, with fixed $a_0$), to produce the
  thick blue line and the associated 1-$\sigma$ uncertainty indicated by the blue shaded region.}
\label{fig:bc_summary}
\end{figure}

Fig.~\ref{fig:bc_summary} shows \bcg\ as a function of \teff.
The largest uncertainty is found for \teff\ $ < 4000$\,K where the spread in the 
values can reach up to $\pm$0.3 mag, due to not distinguishing between giants and 
dwarfs\footnote{We could have estimated a mass from luminosities and colours in order  
to estimate \logg, and subsequently iterated to derive new luminosities and radii.  
However, given the uncertainties in our stellar parameters, we decided against doing this.}.
We estimated the uncertainty in the bolometric correction
by modelling the scatter due to \logg\ as a function of \teff, using the same polynomial model as in equation~\ref{eqn:bcgfgkstars}.
The coefficients for this model are  also listed in Table~\ref{tab:bcgcoeff}.

\subsection{Uncertainty estimates on luminosity and radius}

\begin{figure}
  \includegraphics[width=\columnwidth]{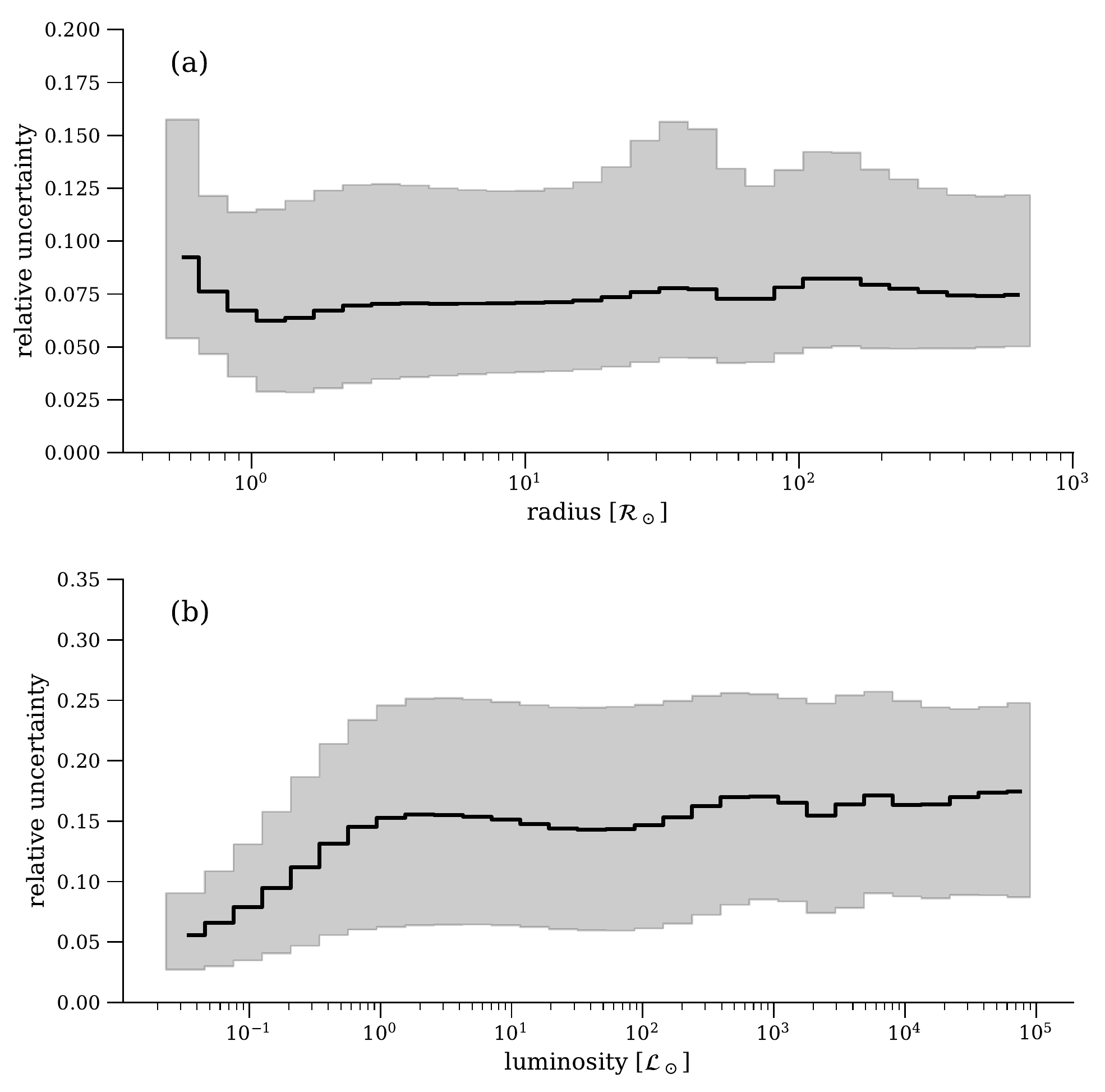}
\caption{Distribution of FLAME relative uncertainties for (a) radius and (b) luminosity,
after applying the GDR2 filtering (Table \ref{tab:results-filtering}). In both panels the black line shows the median value of the uncertainty, and the shaded regions indicate the 16th and 84th percentiles.\label{fig:flameradiuserror} }
\end{figure}

The upper and lower uncertainty levels for \lum\ are defined
symmetrically as $\lum \pm \sigma$, where $\sigma$ has been calculated
using a standard (first order) propagation of the uncertainties in the
G-band magnitude and parallax\footnote{A revision of the
    parallax uncertainties between processing and the data release
    means that our fractional luminosity uncertainties are incorrect
    by factors varying between 0.6 and 2 (for 90\% of the stars), with
    some dependence on magnitude (see appendix A of
    \cite{DR2-DPACP-51}, in particular the upper panel of Figure
    A.2). Although there was no opportunity to rederive the luminosity
    uncertainties, these revised parallax uncertainties (i.e.\ those in
    \gdr{2}) were used when filtering the FLAME results according to
    the criterion in appendix \ref{sec:results-filtering}.}. Note,
however, that we do not include the additional uncertainty arising
from the temperature which would propagate through the bolometric
correction (equation \ref{eqn:lum}).
For \radius, the upper and lower uncertainty levels correspond to the radius computed using the upper and lower uncertainty levels for \teff.  As these \teff\ levels are 16th and 84th percentiles of a distribution, and percentiles are conserved under monotonic transformations of distributions, the resulting radius uncertainty levels are also the 16th and 84th percentiles. This transformation neglects the luminosity uncertainty, but in most cases the \teff\ uncertainty dominates for the stars in the published catalogue (i.e.\ filtered results; see Appendix~\ref{sec:results-filtering}).  The distribution of the uncertainties in \radius\ and \lum\ for different parameter ranges are shown as histograms in Figs.~\ref{fig:flameradiuserror}.  
The radius uncertainty defined here is half the difference between the upper and lower uncertainty levels. 
It can be seen that the median uncertainties in \lum, which considers just the uncertainties in \gmag\ and \parallax, is around 15\%. For radius it's typically less than 10\%.  While our uncertainty estimates are not particularly precise, they provide the user with some estimate of the quality of the parameter.

\section{Results and catalogue content}\label{sec:results}

We now present the \apsis results in \gdr{2} by looking at the performance on various test data sets.
We refer to summaries of the differences between our results and their literature values as ``errors'', as by design our algorithms are trained to achieve minimum differences for the test data. 
This does not mean that the literature estimates are ``true'' in any absolute sense.
We ignore here the inevitable inconsistencies in the literature values, since we do not expect our estimates to be good enough to be substantially limited by these.
 
\subsection{Results for \teff}
\label{ssec:results-teff}

We use the test data set (as defined in section \ref{ssect:gspphot-teff-explanation}) to examine the quality of our \teff\ estimates.
We limit our analyses to those 98\% of sources which have ``clean'' \priam flags for \teff\ (defined in appendix \ref{sec:flags}).
Our estimated values range from 3229\,K to 9803\,K on this test set.
The smallest lower uncertainty level is 3098\,K and the highest upper uncertainty level is 9985\,K.  As the uncertainties are percentiles of the distribution of \extratrees outputs, and this algorithm cannot extrapolate, these are constrained to the range of our training data (which is 3030\,K to 9990\,K).

\begin{figure}
\begin{center}
\includegraphics[width=0.45\textwidth]{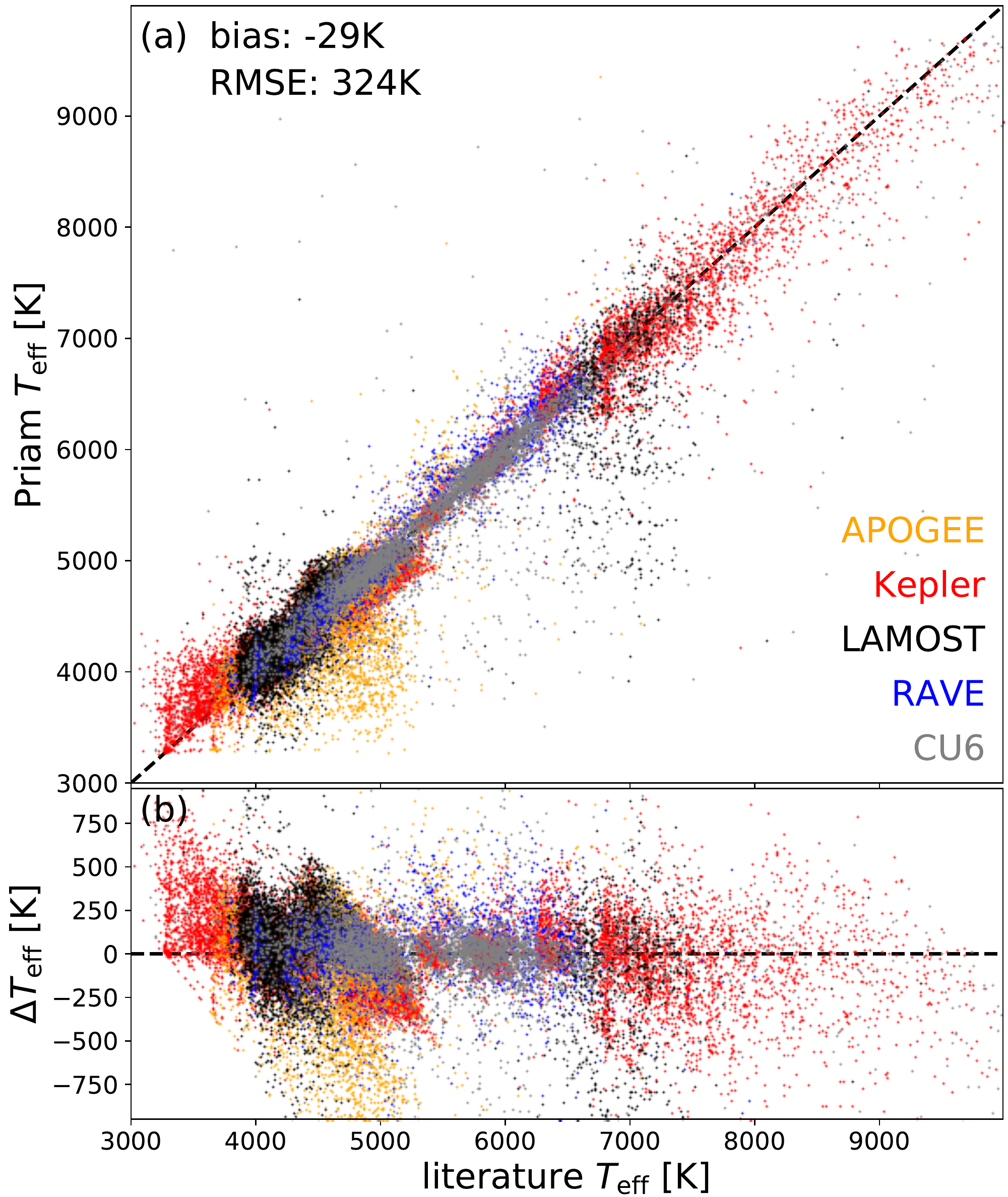}
\caption{Comparison of \priam\ \teff\ estimates with literature values on the test data set for sources with clean flags, colour coded according to catalogue.
The upper panel plots the \priam outputs; the lower panel plots the residuals  $\Delta\teff=\teff^\textrm{Priam}-\teff^\textrm{literature}$.}
\label{fig:gspphot-Teff-test-errors}
\end{center}
\end{figure}

\begin{table}[h]
\begin{center}
\caption{\teff\ error on various sets of test data for sources which were not used in training. We also show test results for 8599 sources with clean flags from the GALAH catalogue \citep{2017MNRAS.465.3203M}, a catalogue not used in training at all. The bias is the mean error.}
\label{tab:gspphot-Teff-test-error-by-reference-catalogs}
\begin{tabular}{lrr}
\hline
reference catalogue & bias [K] &  RMS error [K]\\
\hline
APOGEE 	 & $-$105 	 & 383 \\
Kepler Input Catalogue 	 & $-$6 	 & 232 \\
LAMOST 	 & $-$9 	 & 381 \\
RAVE 	 & 21 	 & 216 \\
RVS Auxiliary Catalogue 	 & $-$50 	 & 425 \\
GALAH & $-$18 & 233  \\
\hline
\end{tabular}
\end{center}
\end{table}

Fig.~\ref{fig:gspphot-Teff-test-errors} compares our \teff\ estimates with the literature estimates for our test data set. The root-mean-squared (RMS) test error is 324\,K, which includes a bias (defined as the mean residual) of $-29$\,K. 
For comparison, the RMS error on the training set is 217\,K, with a bias of $-22$\,K
(better than the test set, as expected). We emphasise that the RMS test error of
324\,K is an average value over the different catalogues, which could have
different physical \teff\ scales. Moreover, since our test sample, just like our training sample, is not representative for the general stellar population in \gdr{2}, the 324\,K uncertainty estimate is likely to be an underestimate. Nevertheless, given this RMS test error of 324\,K, we can subtract (in quadrature) the 102\,K literature uncertainty (Table~\ref{tab:gspphot-teff-catalogs-for-training}) to obtain an internal test error estimate of 309\,K for \priam.

\begin{figure}
\begin{center}
\includegraphics[width=0.49\textwidth]{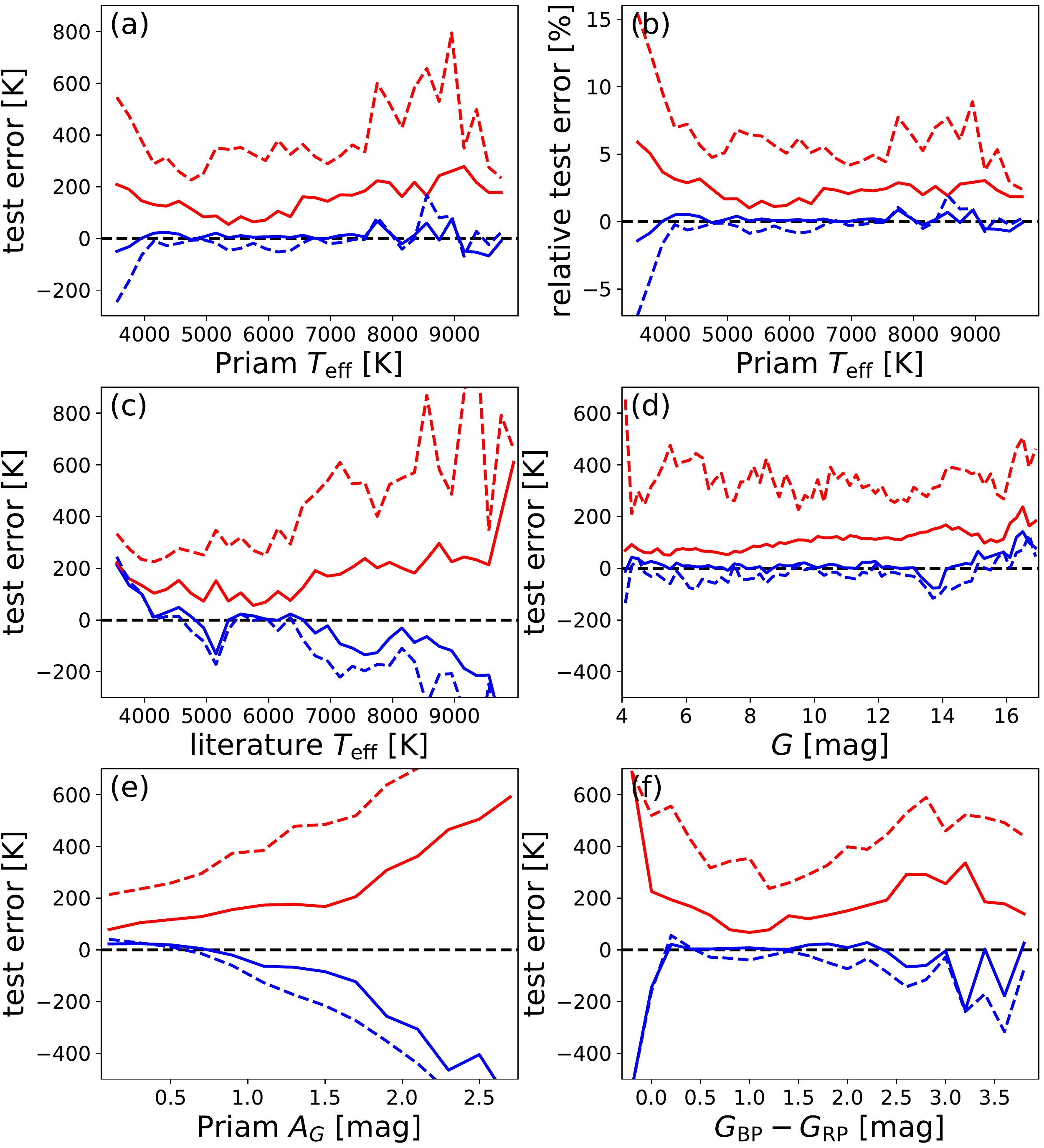}
\caption{Dependence of \teff\ test errors on estimated \teff~(panel a and relative errors in panel b), on literature \teff~(panel c), on \gmag~(panel d), estimated \ag~(panel e) and $\gbp-\grp$~colour (panel f).  Red lines shows root-mean-squared errors (dashed) and root-median-squared errors (solid). Blue lines show mean errors (dashed) and median errors (solid), as measures of bias.}
\label{fig:gspphot-Teff-test-error-vs-other-data}
\end{center}
\end{figure}

Table~\ref{tab:gspphot-Teff-test-error-by-reference-catalogs} shows that the errors vary considerably for the different reference catalogues. Consequently, the temperature errors for a stellar population with a restricted range of \teff\ could differ from our global estimates (see sections~\ref{ssect:teff-for-OB-and-cool-stars} and~\ref{ssec:Solar-analogs}). 
This is illustrated in panels (a) and (b) of Fig.~\ref{fig:gspphot-Teff-test-error-vs-other-data}. If the estimated temperature is below about 4000\,K, we can expect errors of up to 550\,K. Likewise, if the estimated temperature is above 8000\,K, the absolute error increases while the relative error is consistently below 10\% for $\teff\gtrsim 4\,000$\,K. The dependence of test error on literature temperatures (Fig.~\ref{fig:gspphot-Teff-test-error-vs-other-data}c) shows the same behaviour. Note that the errors are dominated by outliers, since when we replace the mean by the median, the errors are much lower (solid vs.\ dashed lines in Fig.~\ref{fig:gspphot-Teff-test-error-vs-other-data}).

\begin{figure}
\begin{center}
\includegraphics[width=0.49\textwidth]{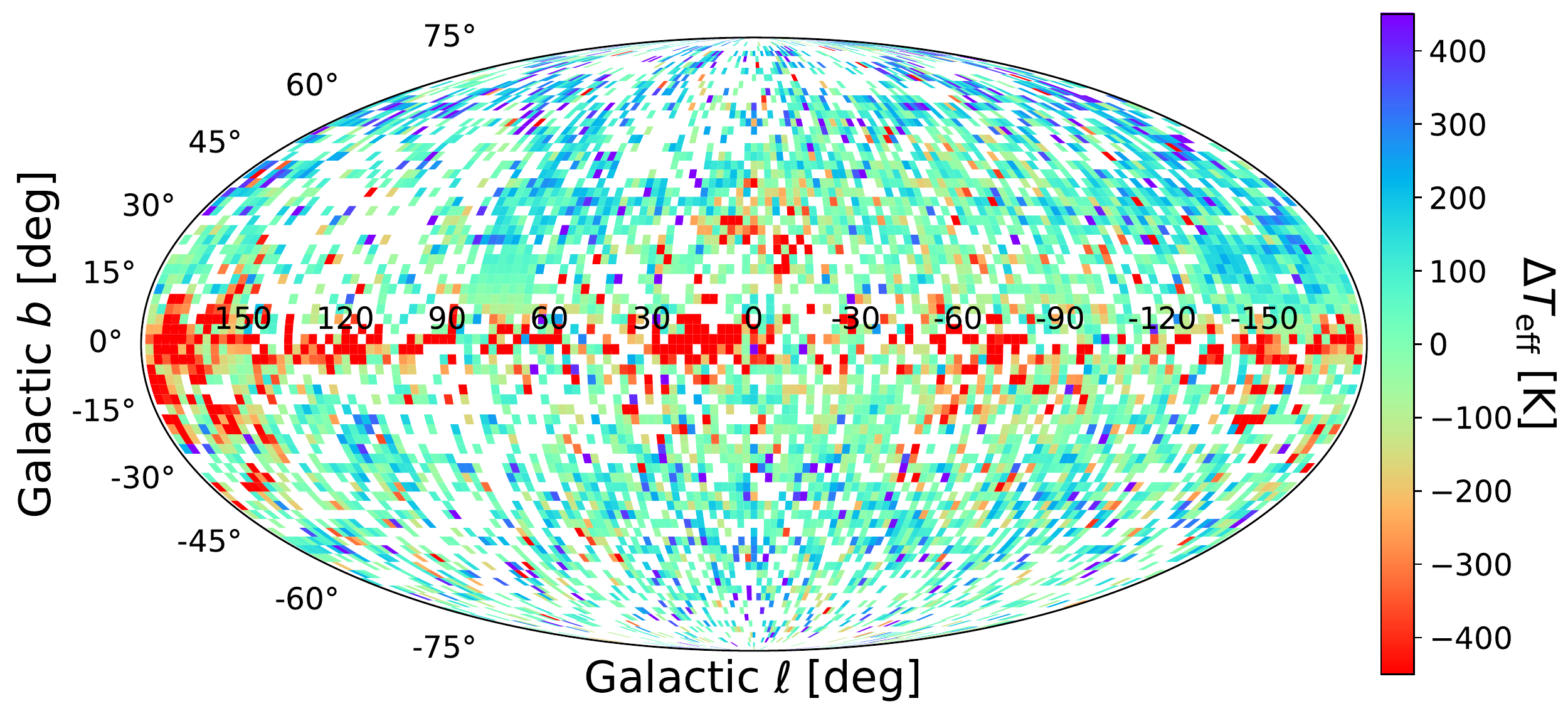}
\caption{Mean difference of \priam\ \teff~from literature values for test data, plotted in Galactic coordinates (Mollweide projection).
}
\label{fig:gspphot-Teff-test-errors-skymap}
\end{center}
\end{figure}

As we can see from Fig.~\ref{fig:gspphot-Teff-test-error-vs-other-data}d, the temperature error increases only very slightly with \gmag~magnitude, which is best seen in the medians since outliers can wash out this trend in the means. Fig.~\ref{fig:gspphot-Teff-test-error-vs-other-data}e shows that the temperature error is weakly correlated with the estimated \ag~extinction, but now more dominant in the mean than the median. This is to be expected since our training data are mostly stars with low extinctions. Stars with high extinctions are under-represented, and due to the extinction--temperature degeneracy they are assigned systematically lower \teff\ estimates. This is particularly apparent when we plot the temperature residuals in the Galactic coordinates (Fig.~\ref{fig:gspphot-Teff-test-errors-skymap}): stars in the Galactic plane, where extinctions are higher, have systematically negative residuals. Finally, Fig.~\ref{fig:gspphot-Teff-test-error-vs-other-data}f shows that the temperature error also depends on the $\gbp-\grp$~colour of the star. Very blue and very red stars were comparatively rare in the training data. For the bluest stars, we see that we systematically underestimate \teff. This is a direct consequence of the upper limit of 10\,000\,K in the training sample (but not in the Galaxy) and the inability of \extratrees to extrapolate.

Fig.~\ref{fig:gspphot-Teff-test-errors-skymap} also suggests a slight tendency to systematically overestimate \teff~at high Galactic latitudes. Halo stars typically have subsolar metallicity, hence tend to be bluer for a given \teff\ than solar metallicity stars.
This may lead \priam, which is trained mostly on solar-metallicity stars, to overestimate \teff\ (see Sect.~\ref{subsec:teff-error-vs-logg-and-FeH}). Alternatively, although the extinction in our empirical training sample is generally low, it is not zero, such that for high latitude stars with almost zero extinction, \priam would overestimate \teff. Most likely, both effects are at work, with the latter presumably dominating.

\begin{figure}
\begin{center}
\includegraphics[width=0.49\textwidth]{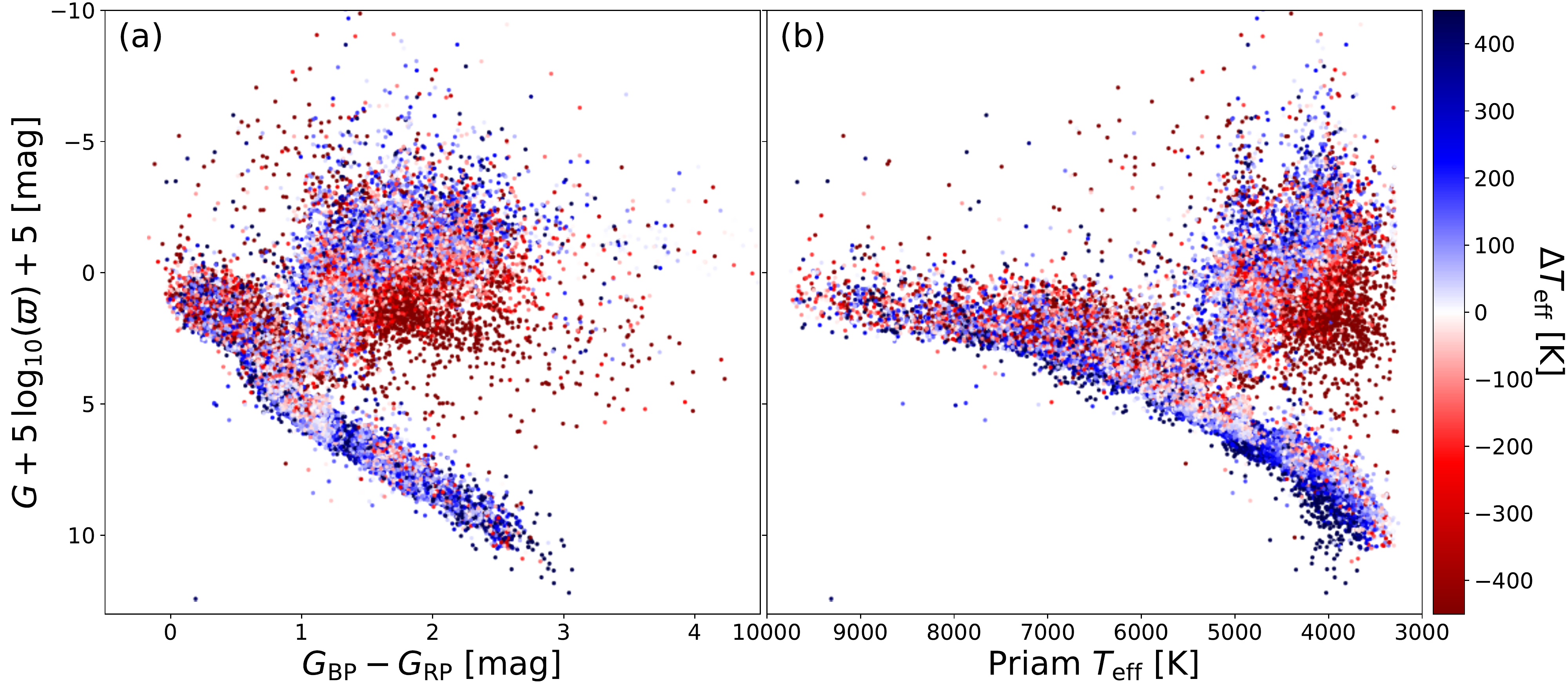}
\caption{Difference between \priam\ \teff\ and literature values for the test data shown in a colour-magnitude diagram (left panel) and Hertzsprung-Russell diagram (right panel).
\label{fig:gspphot-Teff-errors-in-HRD}
}
\end{center}
\end{figure}

The differences between our temperature estimates and the literature values are shown in a CMD in Fig.~\ref{fig:gspphot-Teff-errors-in-HRD}a.  
\priam\ predicts lower \teff\ in those parts of the CMD where we suspect the extinction may be high (e.g.\ the lower part of the giant branch). Conversely, the overestimation of \teff\ in the lower part of the main sequence may be due nearby faint stars having lower extinctions than the low but non-zero extinction in our empirical training sample. These systematics with extinction agree with Fig.~\ref{fig:gspphot-Teff-test-error-vs-other-data}e. This will be discussed further in the next section.

\begin{figure}
\begin{center}
\includegraphics[width=0.5\textwidth]{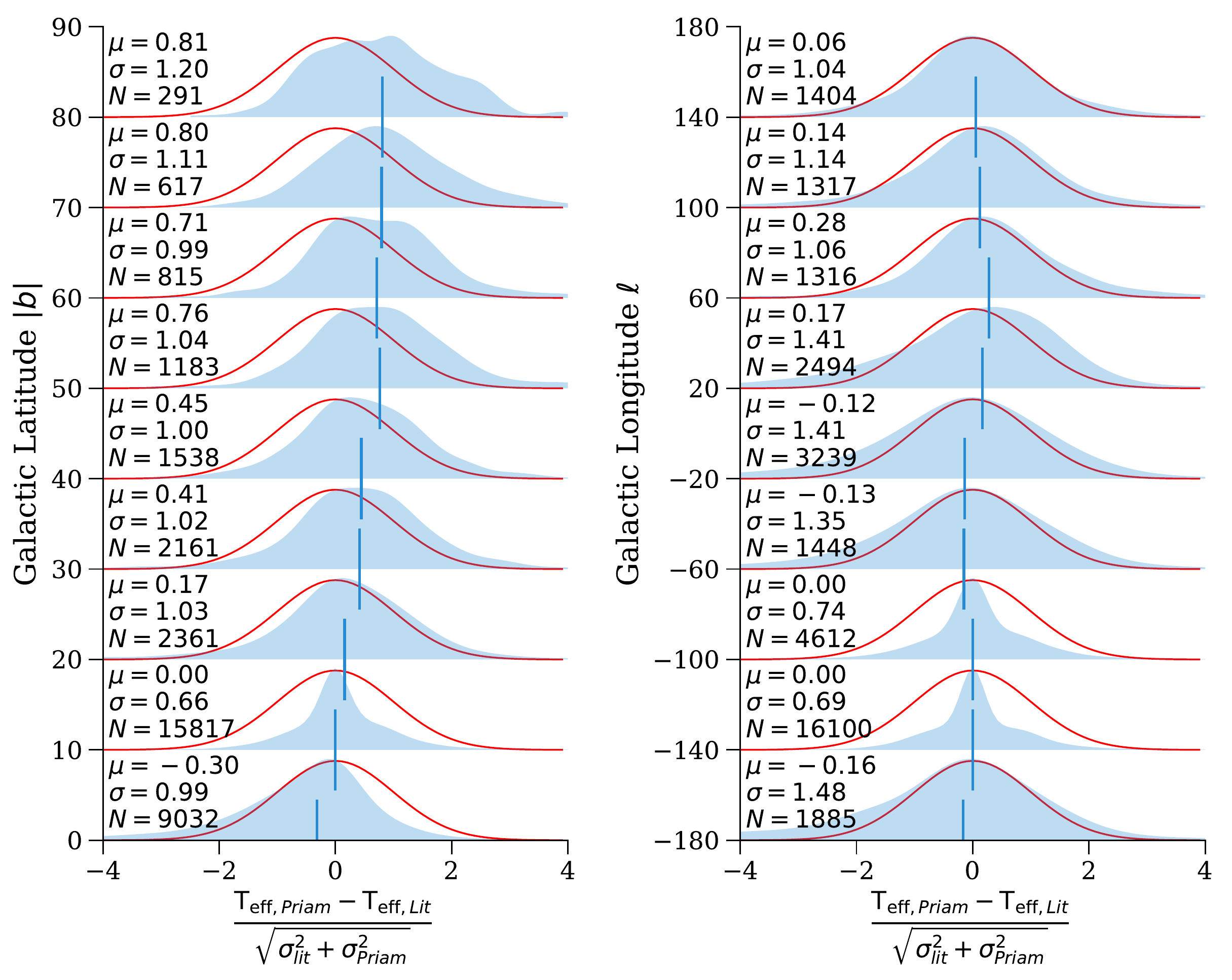}
\caption{Distribution of the \teff\ residuals (\priam minus literature) normalized
by their combined uncertainties for the test data set, for different Galactic latitudes (left column) and longitudes (right column).  The \priam uncertainty
$\sigma_\textrm{\priam}$ used in the computation for each star is formed from the lower uncertainty
level if $T_\textrm{eff,\priam}>T_\textrm{eff,lit}$, and from the upper uncertainty
level otherwise.  The upper left corner of each panel reports the mean
$\mu$ and standard deviation ($\sigma$) of these normalized residuals.  The red
curves are unit Gaussian distributions. The vertical lines indicate
the median of each distribution. 
For unbiased estimates and correct uncertainties in both the literature
and our work, the histograms and the red Gaussians should match.
}
\label{fig:gspphot-normedteff-error-RAVE}
\end{center}
\end{figure}

In order to assess our uncertainty estimates, we again use the test data. Ideally, the distribution of our uncertainty estimates should coincide with the distribution of the errors. We find that for 23\% of test stars, their literature values are below our lower uncertainty levels (which are 16th percentiles), whereas for 22\% of test stars the literature values are above our upper uncertainty levels (84th percentiles). One interpretation of this is that our uncertainty intervals are too narrow, i.e.\ that the supposed 68\% central confidence interval (84th minus 16th) is in fact more like a 55\% confidence interval. However, the literature estimates have finite errors, perhaps of order 100--200\,K, and these will increase the width of the residual distribution (compared to computing residuals using perfect estimates). 
We investigate this more closely by plotting the distribution of residuals normalised by the combined (\priam and literature) uncertainty estimates.
This is shown in 
Fig.~\ref{fig:gspphot-normedteff-error-RAVE} for all our test data and different directions in the Galaxy.
If the combined uncertainties were Gaussian measures of the residuals, then the histograms should be Gaussian with zero mean and unit standard deviation (the red curves). This is generally the case, and
suggests that, although we do not propagate the flux uncertainties, the \priam\ uncertainty estimates may indeed provide 68\% confidence intervals and that the 55\% obtained above arose only from neglecting literature uncertainties.
The left column in Fig.~\ref{fig:gspphot-normedteff-error-RAVE}
shows a systematic trend in the residuals (mean of the
histogram) as a function of Galactic latitude, which is also evident from Fig.~\ref{fig:gspphot-Teff-test-errors-skymap}. This most likely reflects a systematic overestimation of \teff\ for zero-extinction stars at high latitudes.
Also note that the panels for $\ell=60^\circ-100^\circ$ and $|b|=10^\circ-20^\circ$ exhibit narrow peaks. These two panels are dominated by the Kepler field, which makes up 43\% of the training sample (see Table~\ref{tab:gspphot-teff-catalogs-for-training}). The fact that these two peaks are sharper than the unit Gaussian suggests overfitting of stars from the Kepler sample.\footnote{Propagating the flux errors through the \extratrees gives slightly lower test errors (supporting the idea that we may overfit the Kepler sample) and brings the normalised residuals closer to a unit Gaussian.}
Concerning the asymmetry of
the conference intervals, we find that for 57\% of sources, 
upper minus median and median minus lower
differ by less than a factor of two, while for 2.5\% of sources
these two bands can differ by more than a factor of ten.

\subsection{Results for \ag~and \ebpminrp}
\label{ssec:results-extinction}

We look now at our estimates of line-of-sight extinction \ag~and reddening
\ebpminrp. Where appropriate we will select on parallax uncertainty. As
will be discussed in section~\ref{ssec:extinction-in-halo}, some of our
estimates of \ag\ and \ebpminrp\ suffered from strong degeneracies. These (about
one third of the initial set of estimates) were filtered out of the final
catalogue.

\begin{figure}
\begin{center}
  \includegraphics[width=0.45\textwidth]{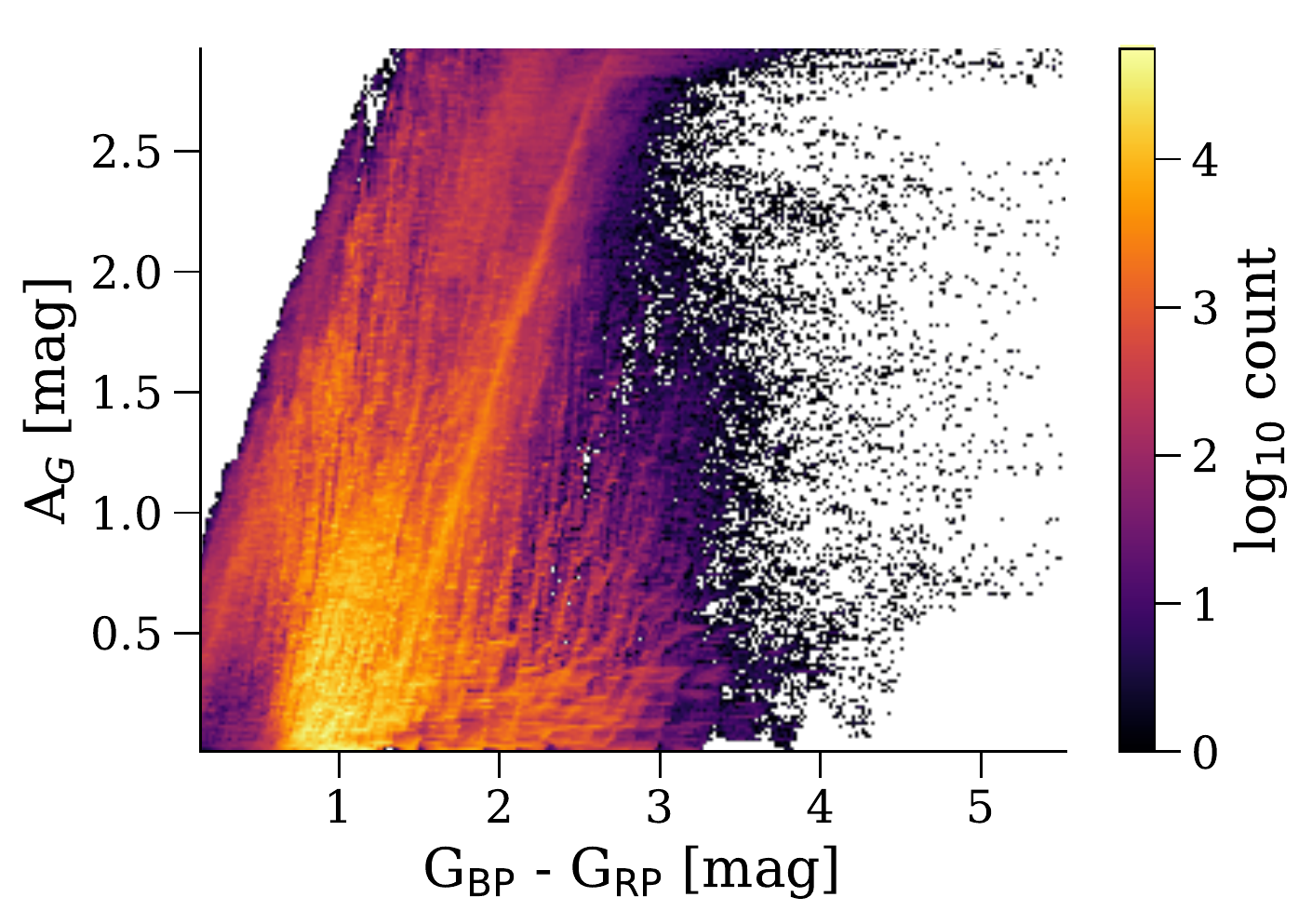}
  \caption{Estimates of \ag\ versus source colours over the
    entire sky.  While there is an expected overall trend of redder
    sources being more extinct, the very broad dispersion shows
    that \bpminrp\ is not a good proxy for
  extinction.  Note that the saturation of the extinction arises from our model
  assumptions.
}
\label{fig:gspphot-AG-vs-color}
\end{center}
\end{figure}

As explained in section \ref{ssect:gspphot-extinction-description}, we
are unable to estimate the line-of-sight extinction from just the
colours. 
Fig.~\ref{fig:gspphot-AG-vs-color} demonstrates that 
neither $\ag$ nor $\ebpminrp$ has a one-to-one relation with the
colour. (Plots against the other two colours are shown in the online
\gdr{2} documentation.)
This complex distribution is the combined result of having both
very broad filters and a wide  range of stellar types.  It may be possible to find an approximative
colour--extinction relation only if one 
can a priori restrict the sample to a narrow part of the HRD, such as
giant stars.

\begin{figure}
\begin{center}
\includegraphics[width=0.5\textwidth]{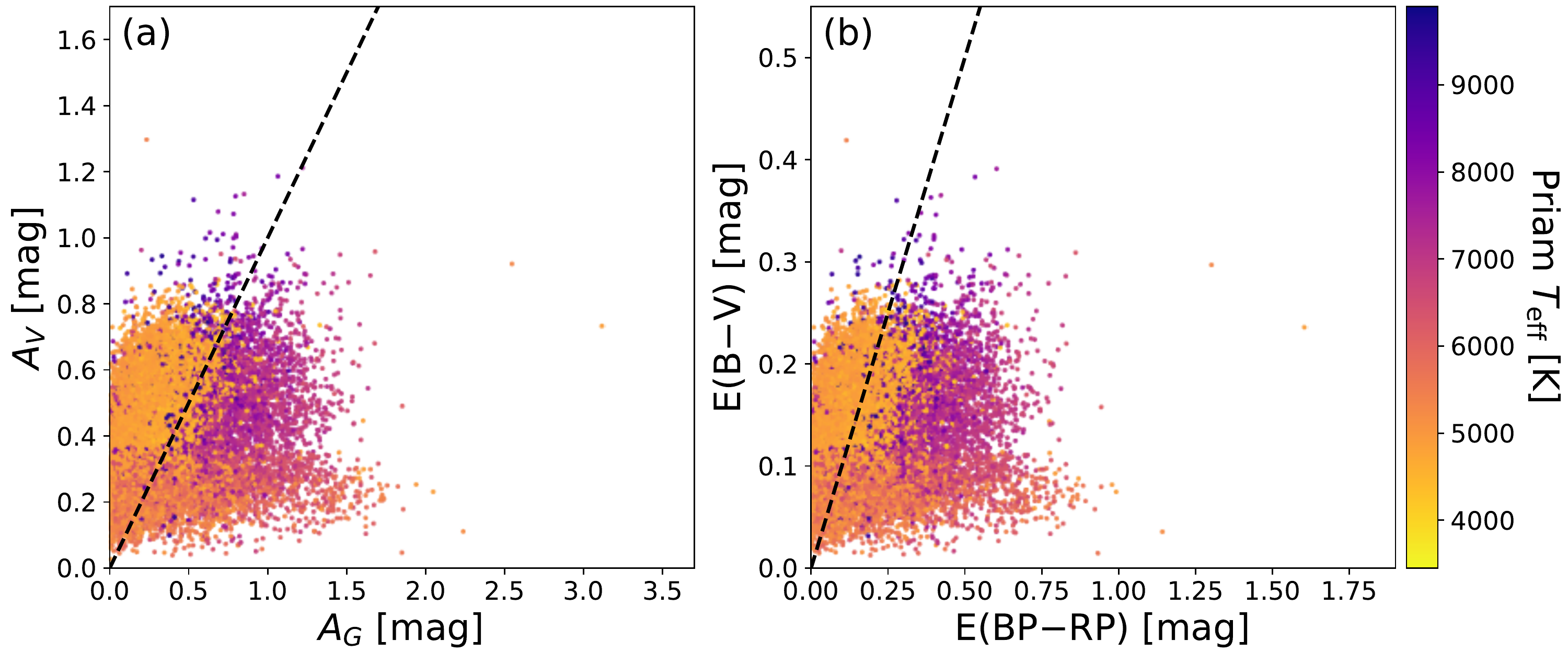}
\caption{\ag~vs.~\av~(panel a) and \ebpminrp~vs.~$E(B-V)$ (panel b) for 22\,894
stars from the Kepler Input Catalog with parallax uncertainty less than 20\%.
The dashed line shows the identity relation. As discussed in the main text, this
comparison is inconclusive since \ag~and \ebpminrp~are subject to large
uncertainties but they cannot scatter into negative values (thus causing seeming
biases). Furthermore, the Gaia passbands are very broad and thus strongly depend
on the intrinsic source SED. Dashed lines indicate one-to-one relations.}
\label{fig:gspphot-extinction-vs-literature-KIC}
\end{center}
\end{figure}

\begin{figure}
\begin{center}
\includegraphics[width=0.45\textwidth]{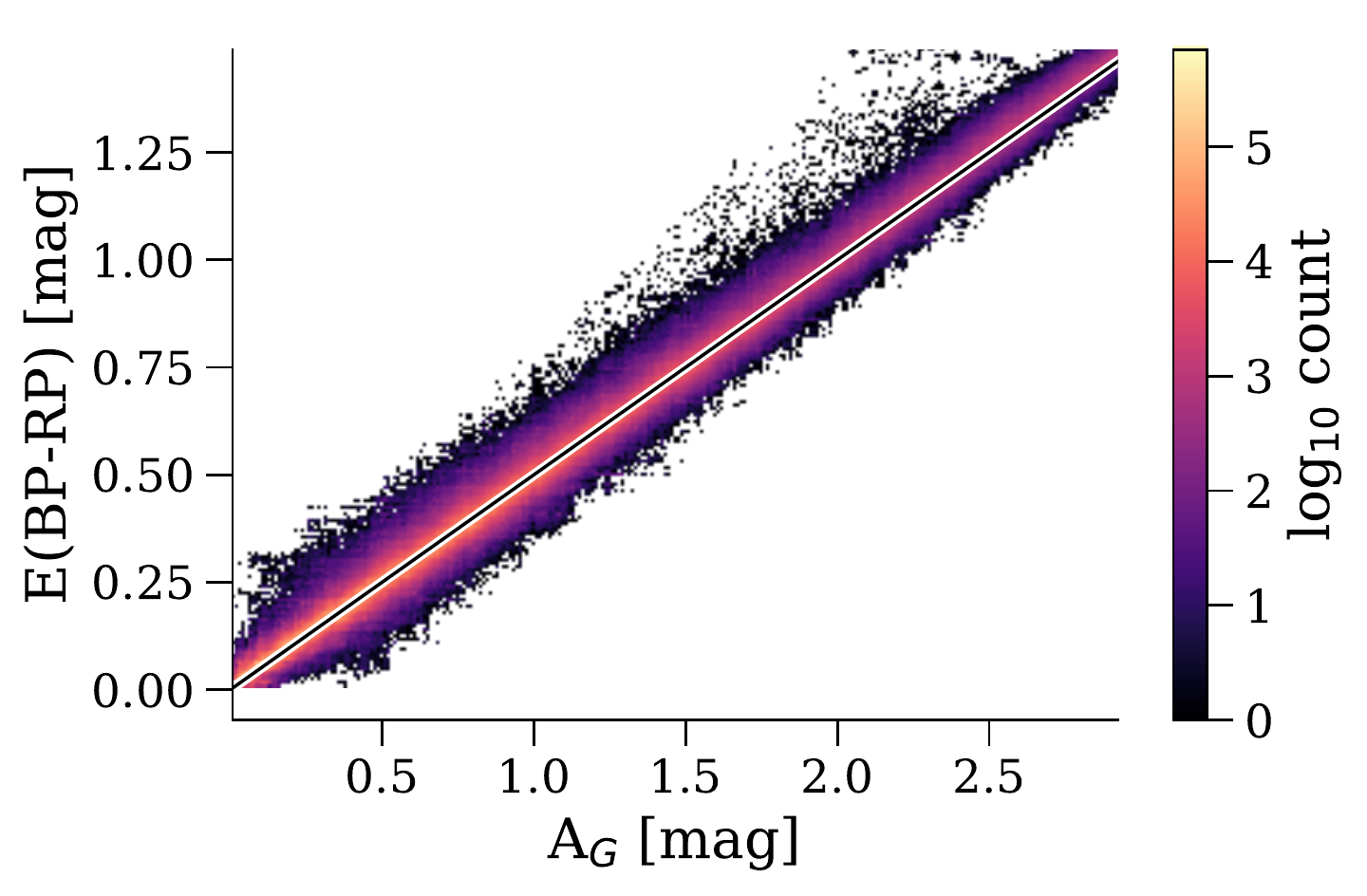}
\caption{\ag~vs \ebpminrp~for sources with $\sigparallax<1$\mas\ (no selection on flags).
The dash black line shows the approximation $\ag\sim 2\cdot\ebpminrp$.}
\label{fig:gspphot-AG-vs-EBPminRP}
\end{center}
\end{figure}

In addition, using only three optical bands (and parallax), we do not
expect very accurate extinction estimates.
A direct comparison to the literature is complicated by the fact that the
literature does not estimate \ag~or \ebpminrp~but rather \a0, \av, or $E(B-V)$.
We compare them nonetheless on a star-by-star basis in
Table~\ref{table:compare-extinctions-to-literature}, and
Fig.~\ref{fig:gspphot-extinction-vs-literature-KIC} shows the results for stars
for the Kepler Input Catalog.  The largest RMS difference for these samples is
0.34\,mag between \ag~and \av\ and 0.24\,mag between \ebpminrp~and
$E(B-V)$. This appears to be dominated primarily by systematically larger values
of \ag~and \ebpminrp. The differences between nominal and real passbands are
only of order 0.1\,mag in the zeropoints \citep{DR2-DPACP-40} and thus are
unlikely to explain this. Instead, these large values arise from the
degeneracies in the extinction estimation (see
section~\ref{ssec:extinction-in-halo}) and the non-trivial transformation
between \av~and \ag\ and between $E(B-V)$ and \ebpminrp. To mitigate this
problem we validate \ag~using red clump stars in
section~\ref{ssec:red-clump-stars}. For now, we conclude from
Table~\ref{table:compare-extinctions-to-literature} that the scatter in \ag\ may
be as high as 0.34\,mag and the scatter in \ebpminrp\ as high as 
0.24\,mag. Given such a large scatter, we can only verify the extinction
estimates at an ensemble level. Let us also emphasise that since \extratrees
cannot produce negative results for \ag~or \ebpminrp, the large random scatter
may give rise to an \textit{apparent} systematic error\footnote{We show
in appendix \ref{appendix:mean-cluster-extinction} that an apparent bias can
arise if one uses the mean as an estimator when the likelihood is skewed.} that
can also be seen in Fig.~\ref{fig:gspphot-extinction-vs-literature-KIC}. We also find an approximate relation
$\ag\sim 2\cdot\ebpminrp$ (Fig.~\ref{fig:gspphot-AG-vs-EBPminRP}). This is
essentially by construction, as we use the same PARSEC models for the
determination of both quantities
(see~Fig.~\ref{fig:gspphot-PARSEC-models-AG-vs-EBPminRP}).

\begin{table}
\begin{center}
\caption{Comparison of our \ag~and \ebpminrp~estimates with literature values of \av~and $E(B-V)$ (for sources with $\sigparallax/\parallax<0.2$, but no selection on flags). In each case we quote the mean difference and the RMS difference.
}
\label{table:compare-extinctions-to-literature}
\begin{tabular}{llcc}
\hline
& &  & \ebpminrp \\
& & $\ag-\av$ & $-E(B-V)$ \\
\hline\hline
Kepler Input Catalog						& RMS 	& 0.34mag 	& 0.18mag \\
(15\,143)								& mean 	& 0.00mag 	& 0.07mag \\
\hline
\citet{2014AA...561A..91L} 			& RMS 	& --					& 0.24mag \\
(1431 stars)									& mean 	& --				 	& 0.16mag \\
\hline
\citet{2014MNRAS.445.2758R} 	& RMS		& 0.21mag		& -- \\
(1315 stars)									& mean 	& 0.08mag 	& -- \\
\hline
\end{tabular}
\end{center}
\end{table}

\begin{figure}
\begin{center}
\includegraphics[width=0.45\textwidth]{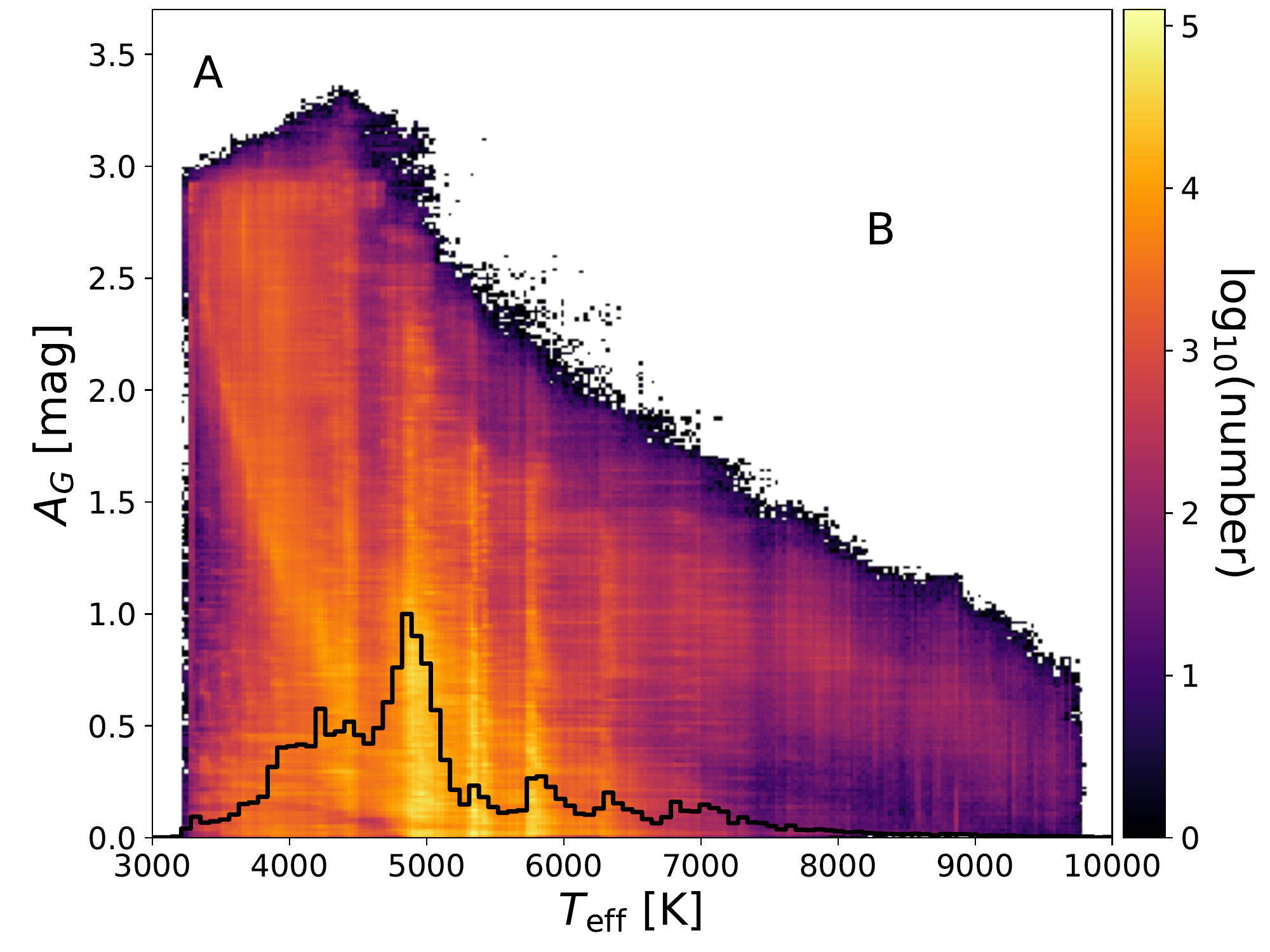}
\caption{Variation of \priam \ag\ with \priam\ \teff~for all stars with clean flags and parallax uncertainty less than 20\%.
Two unoccupied regions in this plane are marked by ``A'' and ``B''.  The black histogram at the bottom shows the total distribution of \teff~in our training sample, i.e., the sum over all catalogues shown in Fig.~\ref{fig:gspphot-Priam-training-Teff-catalogs}. The histogram peaks coincide with the vertical stripes.}
\label{fig:gspphot-Teff-vs-AG-bias}
\end{center}
\end{figure}

Fig.~\ref{fig:gspphot-Teff-vs-AG-bias} shows the distribution of \teff~vs.~\ag~for all stars with clean \priam\ flags.
Since \extratrees cannot extrapolate from the \teff\ training data range of 3000K--10\,000K, the results are restricted to this range. We see unoccupied regions on this plot, labelled ``A'' and ``B''.
The empty region A is due to the apparent magnitude limit of $G\leq 17$, which
removes stars with lower \teff~already at lower \ag, since they are fainter.
This is expected. However, the second empty region labelled B in
Fig.~\ref{fig:gspphot-Teff-vs-AG-bias} is more interesting. There are seemingly
no hot stars with high extinctions. This void is an artefact and is due to our
\extratrees training sample (for extinctions) comprising only low-extinction stars. Therefore, if hot
stars in the overall sample are reddened by dust, they have no counterparts in
the training sample and are thus assigned systematically lower temperatures
which, given the {\teff} training sample, is the only way that \extratrees can match their reddish colours. We also note the vertical stripes in Fig.~\ref{fig:gspphot-Teff-vs-AG-bias}, which are a consequence of the  inhomogeneous temperature distribution in our training sample (see Fig.~\ref{fig:gspphot-Priam-training-Teff-catalogs}). Unfortunately, a desirable rebalancing of our training sample fell victim to the tight processing schedule for \gdr{2}. However, these results are not our final data products and revised training sets will be used for \gdr{3} (section \ref{sec:outlook}).

\begin{figure*}
\begin{center}
\includegraphics[width=0.99\textwidth]{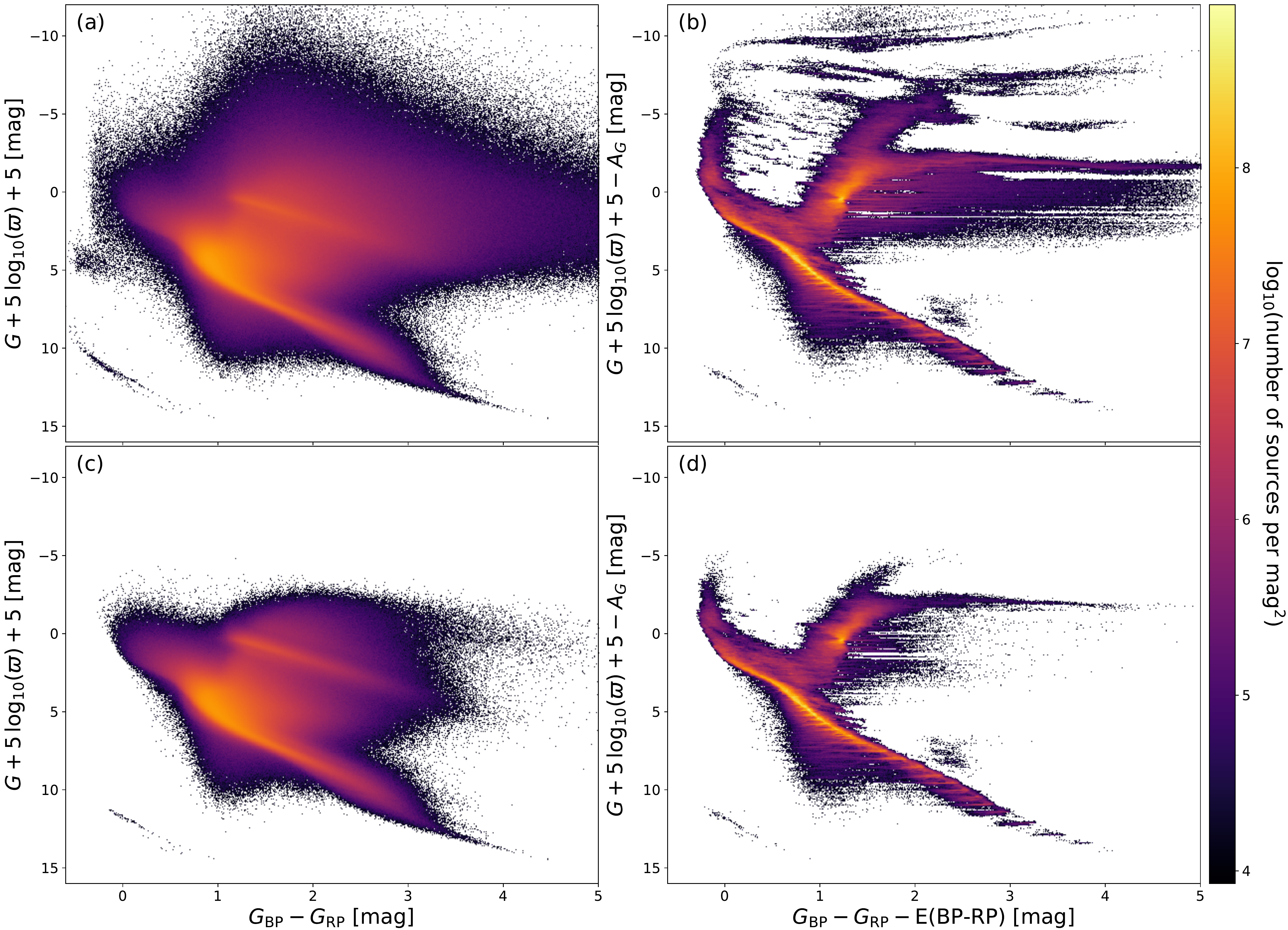}
\caption{Observed colour-magnitude diagrams (panels a and c) and dust-corrected colour-magnitude diagrams (panels b and d). Using our estimate of \ag, we obtain \mg\ from $\mg+\ag$ ($=\gmag+5\log_{10}\!\parallax+5$). The upper panels~a and~b show all sources with $\gmag\leq 17$ and $\parallax>0$. The lower panels~c and~d restrict this further to sources with parallax uncertainties lower than 20\%.
}
\label{fig:gspphot-de-reddened-CMD}
\end{center}
\end{figure*}

\begin{figure}
\begin{center}
\includegraphics[width=0.45\textwidth]{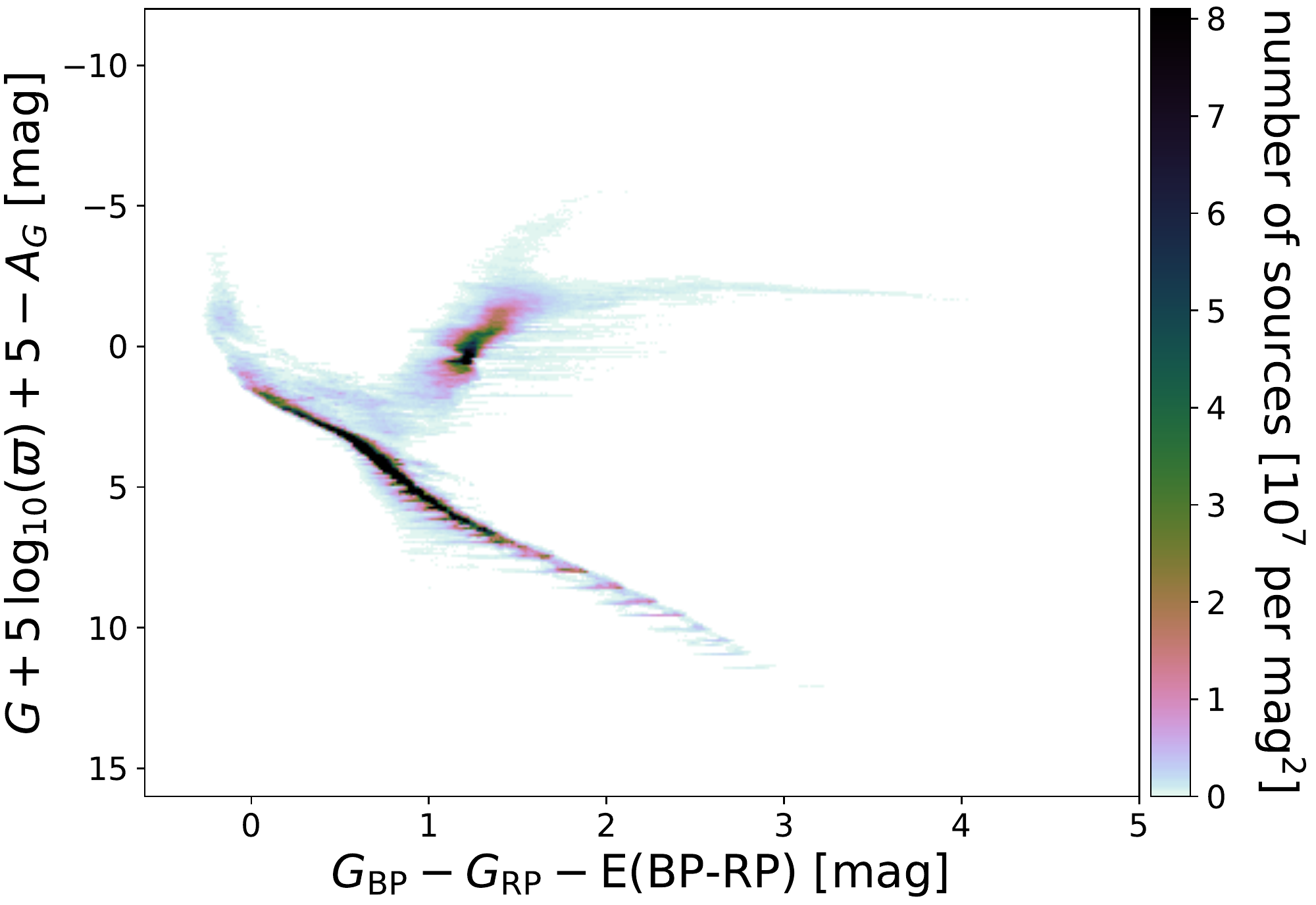}
\caption{Same as Fig.~\ref{fig:gspphot-de-reddened-CMD}b but on linear density scale. Some sources, e.g.\ white dwarfs, have such a low density as to now be invisible on this scale.}
\label{fig:gspphot-de-reddened-CMD-linear}
\end{center}
\end{figure}

Although our extinction estimates are inaccurate on a star-by-star
level, our main goal in estimating \ag~and \ebpminrp~is to enable a
dust correction of the observed CMD. To this end, it is sufficient if
our extinction estimates are mostly unbiased such that they are
applicable at the ensemble level. This is often the case, as
is shown in Fig.~\ref{fig:gspphot-de-reddened-CMD}. The observed CMD
in Fig.~\ref{fig:gspphot-de-reddened-CMD}a exhibits a very diffuse
source distribution. In particular, the giant branch is completely
washed out, while the red clump is visible as a thin line above the
main sequence, which is the result of dust extinction and
reddening. If we use our estimates of \ag~and \ebpminrp~in order to
correct the observed CMD, then we achieve
Fig.~\ref{fig:gspphot-de-reddened-CMD}b. This dust-corrected CMD is
much less diffuse. In particular, the red clump is now an actual
clump, the main sequence is more compact, and we can identify the
giant branch. The horizontal stripes along the main sequence in
Fig.~\ref{fig:gspphot-de-reddened-CMD}b are artefacts that originate
from the sampling of PARSEC evolutionary tracks which we took directly
without further interpolation or smoothing. There are also
  horizontal streaks above the red giant branch, which are sources
  with poor parallaxes leading to a clustering in our
  results. Furthermore, there is a small group of 8587 sources
  ($\sim$0.01\% of all sources that have extinction estimates) just
  above the main sequence, which either are outliers that we failed to
  remove (see section~\ref{ssec:extinction-in-halo}) or which may be
  genuine binaries. If we additionally require the relative parallax
uncertainty to be less than 20\% \citep{2015PASP..127..994B}, then the
observed CMD becomes Fig.~\ref{fig:gspphot-de-reddened-CMD}c, which is
much cleaner than Fig.~\ref{fig:gspphot-de-reddened-CMD}a. The
dust-corrected CMD corresponding to this --
Fig.~\ref{fig:gspphot-de-reddened-CMD}d -- is likewise more
distinct. In particular, the horizontal streaks above the giant
  branch are removed by cutting in relative parallax uncertainty.
Nevertheless, Fig.~\ref{fig:gspphot-de-reddened-CMD} panels~b and~d
also exhibit clear artefacts, some of which are due to bad parallaxes,
although most are introduced by our methods and the training
data. As all our models are only for single stars, binaries
  will receive systematically wrong extinction estimates. As most
  binaries reside above the main sequence, \priam will typically
  misinterpret them as highly reddened single stars from the upper
  part of the main sequence. Finally, we note that the logarithmic
  scale in Fig.~\ref{fig:gspphot-de-reddened-CMD} overemphasises the
  low-density regions, intentionally drawing the reader's attention to
  the various artefacts. Nevertheless, our results produce a very thin
  main sequence, as is obvious from
  Fig.~\ref{fig:gspphot-de-reddened-CMD-linear} which is exactly
  identical to Fig.~\ref{fig:gspphot-de-reddened-CMD}b apart from a
  linear density scale.

\begin{figure*}
\begin{center}
\includegraphics[width=0.90\textwidth, clip]{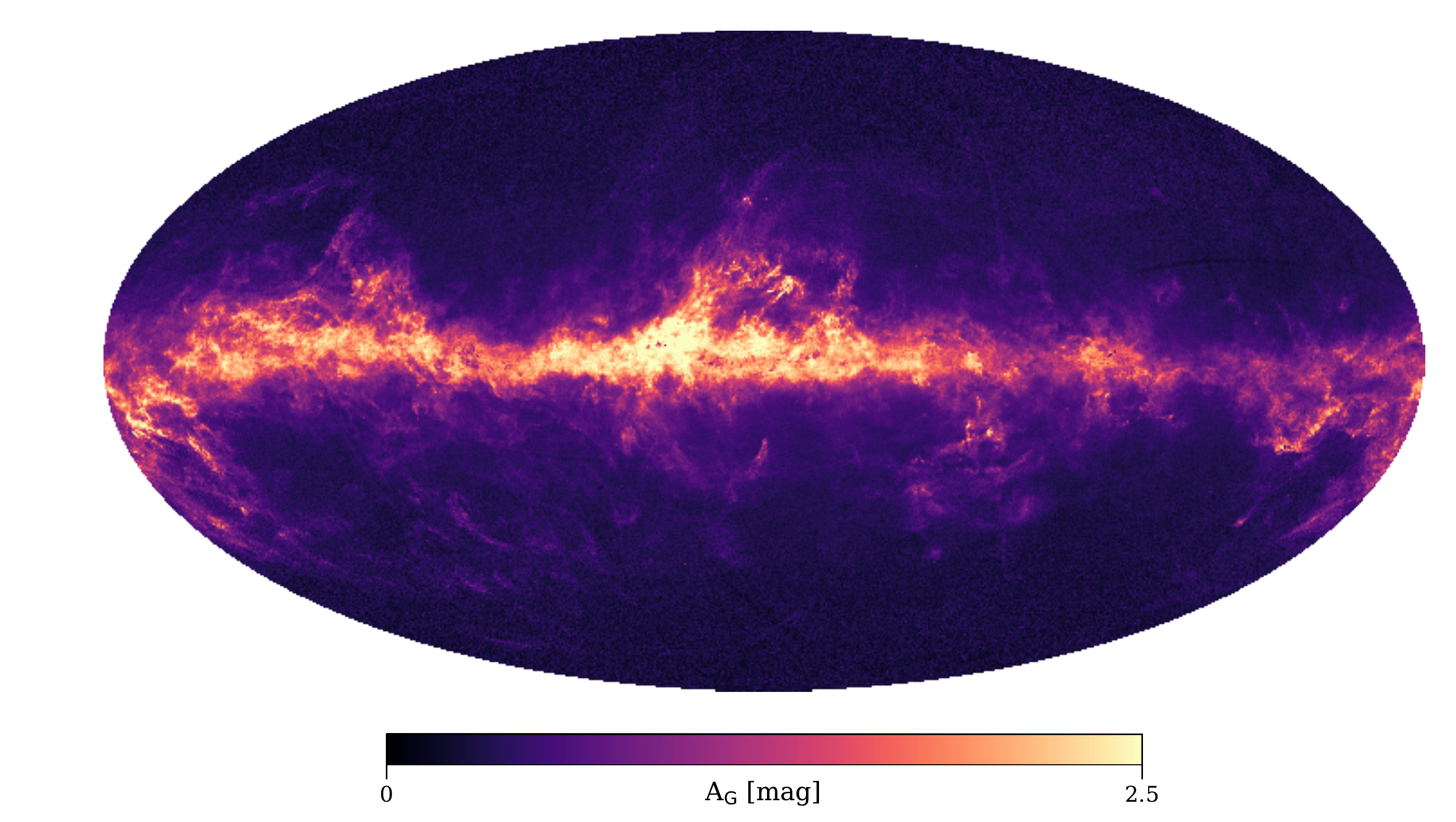}
\caption{Distribution of \ag\ (averaged over all parallaxes) in Galactic
coordinates (Mollweide projection). The map is centered on the Galactic Center,
with longitudes increasing towards the left.}
\label{fig:gspphot-AG-skymap-Galactic}
\end{center}
\end{figure*}

The statistical validity of our \ag~estimates is further attested to
by Fig.~\ref{fig:gspphot-AG-skymap-Galactic}.  Recall that we do not
use any sky position during our inference: each star estimate remains
independent of any other.\footnote{On account of this independence,
  plus the finite variance (guaranteed by the inability of \extratrees
  to extrapolate beyond the training range), the Central Limit Theorem
  applies to any average of our extinction or reddening estimates.}
This plot shows features quite distinct from just plotting the Gaia
colour, as should be obvious from their lack of correlation (shown in
Fig.~\ref{fig:gspphot-AG-vs-color}). 
Plotting our extinction estimates on the sky not only highlights the Milky Way disk, but also numerous detailed substructures.
Apart from the Small and Large Magellanic Clouds, we also recover a wealth of
structure across a wide range of scales, from thin
filaments to large cloud complexes. 
The Perseus, Taurus, and Auriga
cloud complexes dominate the anticentral region (far left and right sides of the
map, respectively), while the Orion molecular cloud complex ($\ell \sim 210, b\sim -15 \deg$)
 and the California nebula ($\ell \sim 160, b\sim -8 \deg$) show
exquisite substructures, as does Ophiuchus just above the Galactic Center. More will be shown in section \ref{sec:dust_clusters}.

\subsection{Results for \lum \label{ssec:results-Lbol}}

In this and the next subsection
we describe the contents and the quality of the catalogue entries for
\lum\ and \radius.
We remind readers that upon validation of \flame\ astrophysical parameters, 
several filters were put in place to remove individual entries, 
e.g. 
stars with \radius\ $\le$ 0.5\Rsun\ have no published radii or luminosities: 
see Appendix \ref{sec:results-filtering} for details. 
Only 48\%
of the entries with \teff\ also have \lum\ and \radius\ (77 million stars).
Unless otherwise specified, we present the results for the published catalogue.

\begin{figure}[h]
  \includegraphics[width=\columnwidth]{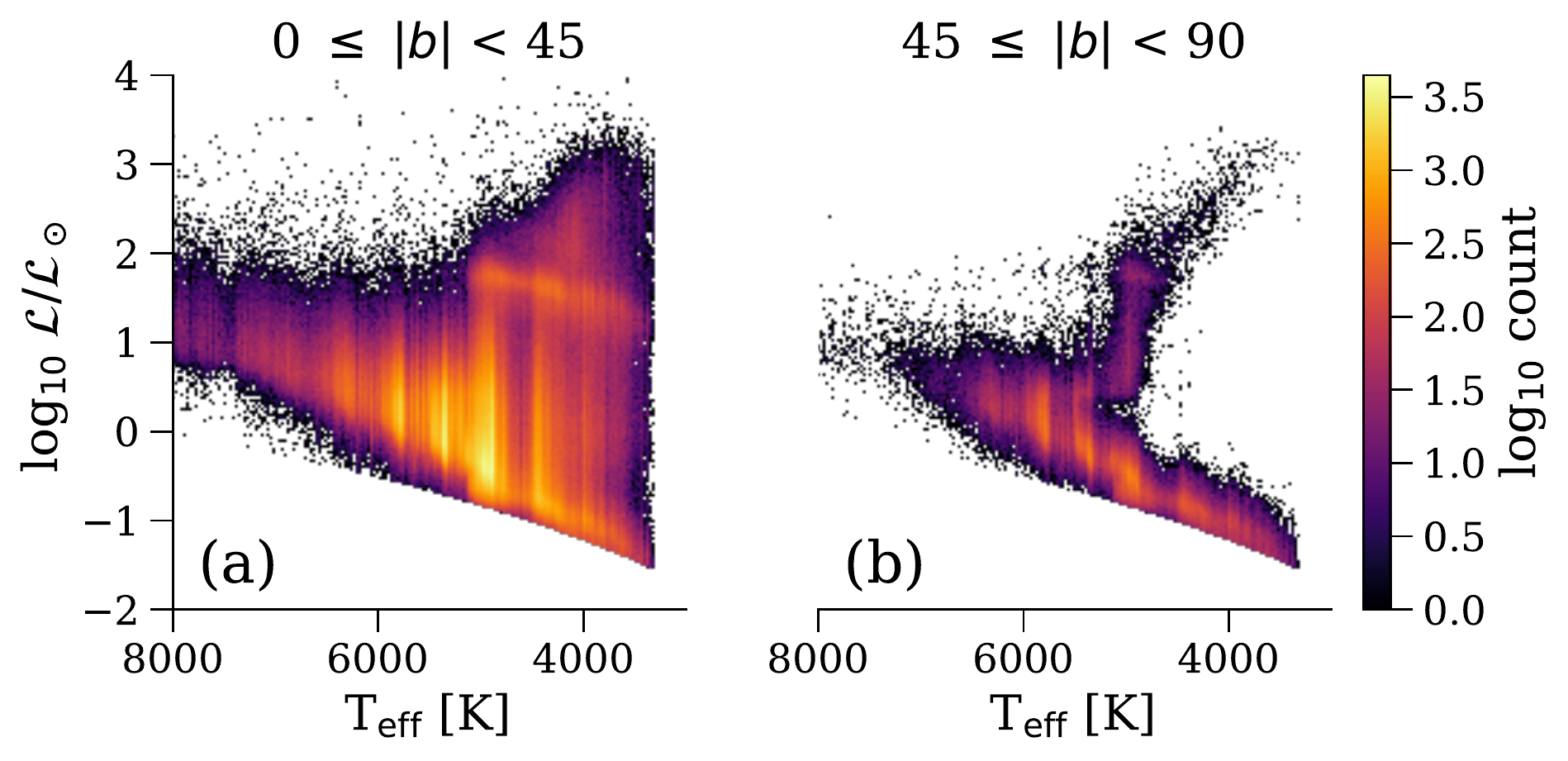}
  \includegraphics[width=\columnwidth]{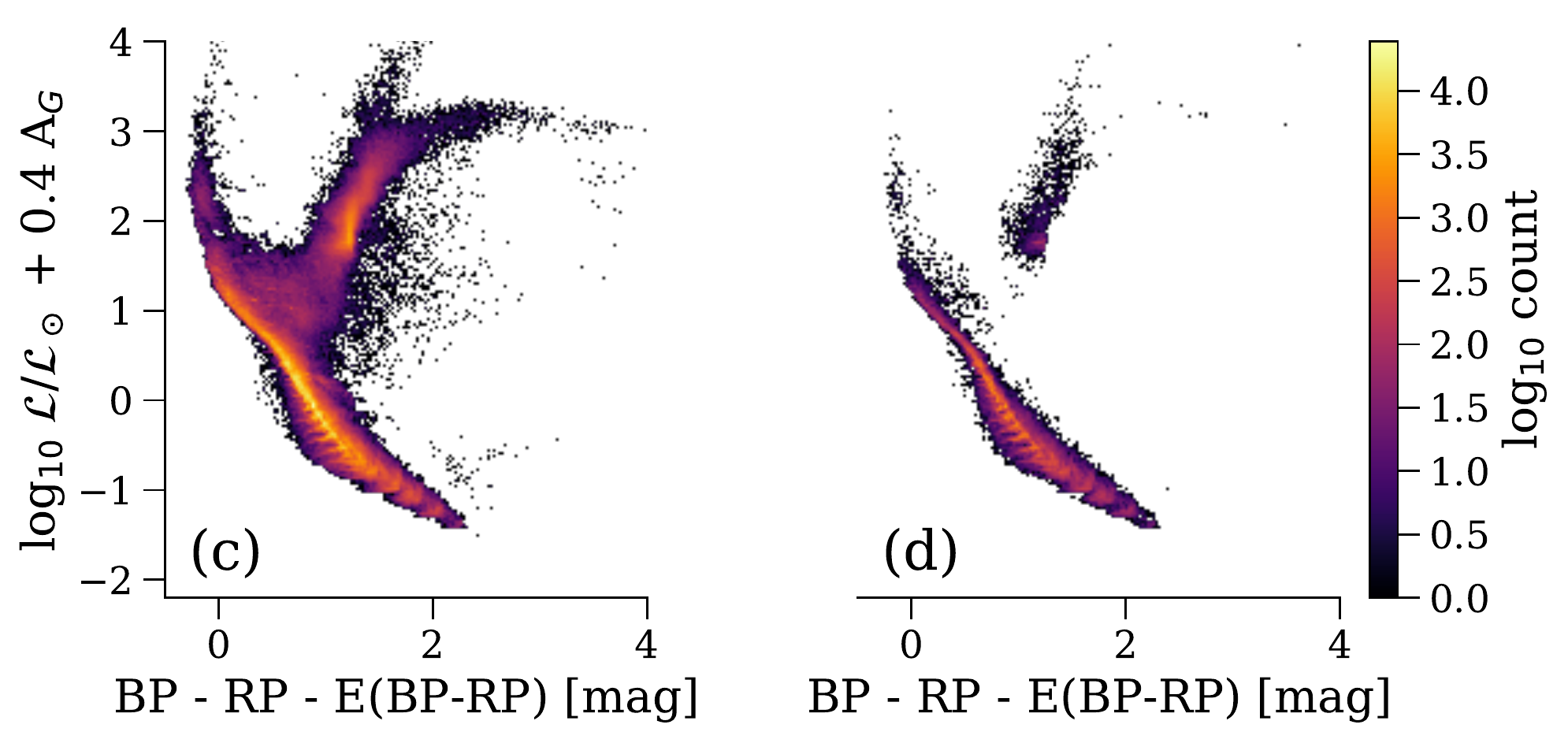}
  \caption{HRD from \gdr{2} separated in galactic latitude (left and right).  The top panels show \lum\ against \teff\ and the lower panels show \lum\ against colour, but corrected for extinction. \label{fig:flame-hrdiag}}
\end{figure}

The quality and distribution of the luminosities in the catalogue 
can be best examined by constructing Hertzsprung--Russell Diagrams (HRD). The 
HRD using \flame\ \lum\ and \priam\ \teff\ is shown in the 
top panels of Figure~\ref{fig:flame-hrdiag} separated by galactic latitude $b$ 
($|b| \le 45$ and $|b| \ge 45$).
For stars at lower galactic latitudes, our neglect of extinction in the
luminosity estimation can lead to misinterpretations for
individual stars or populations of stars.
This can be seen in panel (a) where in particular the red giant branch is extended towards 
lower \teff, and their luminosities appear lower (see also section~\ref{ssec:results-teff}).
The vertical stripes at distinct \teff\ values is a result 
of the inhomogeneous temperature distribution in the 
training sample (discussed in section~\ref{ssec:results-extinction}, 
see Fig~\ref{fig:gspphot-Priam-training-Teff-catalogs}).
The clean diagonal cut on the lower end is a direct result of our
filtering out of
sources with \radius\ < 0.5 \Rsun.
Replacing \teff\ by de-reddened color in the 
abscissa, and including \ag\ as given in equation~\ref{eqn:newlum},
we see (in panel c) that the HRD tightens up nicely with 
a clear structure defining the main expected components.
These results clearly highlight the degeneracy between \ag\ and \teff\ when only three 
photometric bands are available, 
but it also provides a positive validation of the extinction parameters.

For stars at higher galactic latitude we find a very different distribution, where 
extinction no longer plays a dominant role.  Using \lum\ and \teff\ directly 
from the catalogue yields a clean HRD with clearly defined components, as shown in
panel (b).  For reference we show the same sources in panel (d) while including \ag\ and
replacing \teff\ by de-reddened color.

\subsection{Results for \radius \label{sec:results-radius}}

\begin{figure*}
  \center{\includegraphics[width=\textwidth]{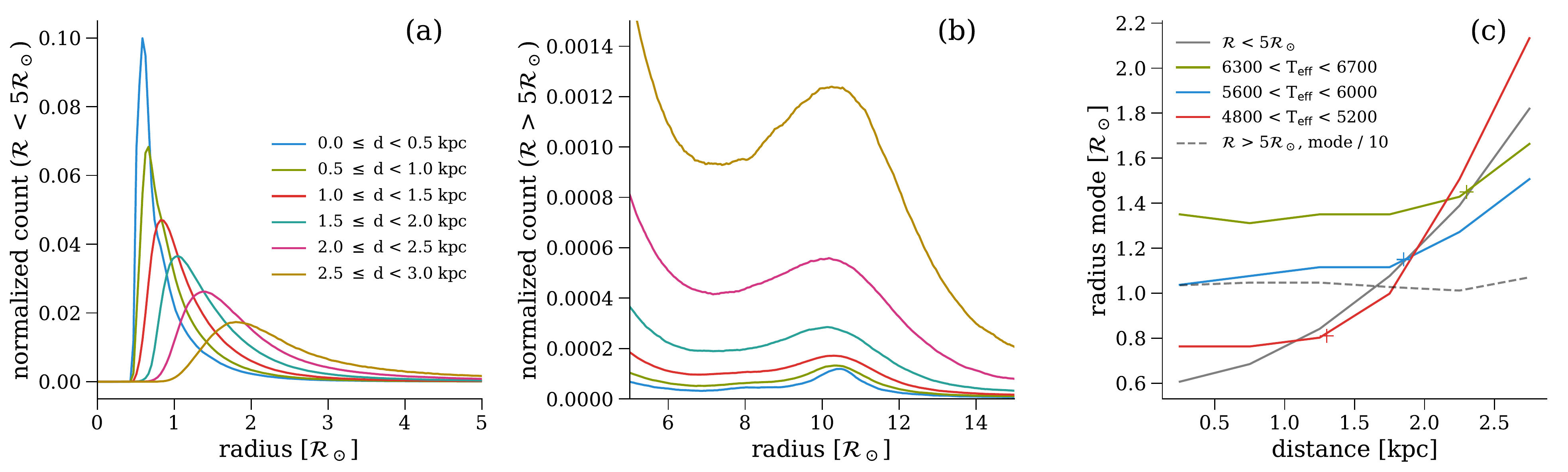}
  \caption{Distribution of radii for different distance bins (different colours) 
  for stars with (a) $\radius < 5$ \Rsun, and (b) $5 < \radius /\Rsun < 15$.
  Panel (c) highlights the variation of the mode of the distributions with distance.
  The crosses indicate to approximately what distance this population of main sequence stars have published
  radii and luminosities.
  The solid grey lines and the dashed line show the results corresponding to panels (a) and (b)
  respectively (the latter scaled by a factor of ten).
  The coloured continuous lines show the results for $\radius < 5\Rsun$ but for 
  narrower \teff\ ranges, in order to isolate 
  the impact of the assumptions on extinction from the  \flame\ filtering criteria.}
      \label{fig:flame-radii-distance}}
\end{figure*}

The distribution of the radii of our sources 
for different distances from the Sun are shown in Fig.~\ref{fig:flame-radii-distance}.
Here we assume the distance is the inverse of the parallax.
Panels (a) and (b) show sources with $\radius < 5 \Rsun$ and 
$0.5 < \radius/\Rsun <15$, corresponding roughly to main-sequence and giant stars
respectively.

In panel (a) we see that the radius distribution changes with distance
and in particular that the mode 
of the distribution is found at larger radii as we move to larger distances.  Such a change in the 
distributions to within a few kpc should not exist (not least because we are not considering
any specific direction).
This is in fact a selection effect due to the filtering imposed on \flame\ parameters 
(see appendix~\ref{sec:results-filtering}).
The broader distributions at larger distances are a direct result of
this filtering, whereby stars with
sufficiently small diameters and luminosities are removed.
For example, due to the magnitude cut at $G=17$ mag for all
astrophysical parameters, no solar-like star will exist in the catalogue for
distances larger than approximately $2.5$\,kpc.
Likewise, the \flame\ filtering will also remove smaller, fainter stars that
have large parallax uncertainties ($\sigma_\varpi/\varpi > 0.2$).

We do not expect to find the same selection effect for more evolved stars, however, and 
this can be seen in panel (b) where we find distributions that peak between 
10 and 11 \Rsun\ at all distances.  
This is shown more clearly in panel (c) where we plot the mode of 
the distributions as a function of distance.

As we have chosen not to include extinction in our calculations of
luminosity, we investigate the impact of this assumption on the
characterisation of the local population.  In general, \lum\ will be
underestimated for most stars and as a consequence \radius\ for a
fixed \teff\ will also be underestimated.  However, \teff\ is also
partially degenerate with extinction, and a hotter extincted star
could appear cooler (this is shown in section~\ref{ssec:flame-validation} for
a sample of giant stars).  For a fixed \lum\ this would imply a larger
\radius.  For a group of similar stars the impact of the zero
extinction assumption should manifest itself as a slow change in the
peak radius as the distance increases.  We performed a similar
analysis as shown in panel (a), but now for less evolved stars in
three different temperature ranges: $4800 < \teff < 5200$,
$5600 < \teff < 6000$, and $6300 < \teff < 6700$\,K.  The modes of
these distributions as a function of distance are shown as the colour
lines in panel (c).  For each of the temperature ranges, we can
identify at what distance the population is filtered out by the
\flame\ criteria from a rapid increase in the mode; these are
denoted by the '+' symbol.  
For values below
these limits, however, it can also be seen that even with the
assumption of zero extinction, the peak increases very slowly and
remains within 5--7\% of the value at the closest distances to us,
a value consistent with our typical uncertainties. 
We therefore
conclude that the published radii can be safely
used (considering their uncertainties) for the less evolved stars,
without correcting for extinction.

For the evolved stars (grey dashed line in panel c) we find that the
mode of the distribution 
remains essentially flat as a function of distance.
Here it is possible that the impact of setting
\ag\ = 0.0 mag on \lum\ and \radius\ is more pronounced than a
possible \teff\ bias from \priam.  For these stars, one should
consider this fact when using the \lum\ values. However, we expect
\radius\ to be affected to a much lesser extent.

\section{Validation and comparison with external data}\label{sec:validation}

We now validate our results, primarily through comparison with results from non-Gaia sources. Recall that we use the term ``error'' to refer to the difference between an estimated quantity and its literature estimate, even though one or both could contain errors.

\subsection{Temperature errors vs.~\logg~and \feh}
\label{subsec:teff-error-vs-logg-and-FeH}

\begin{figure}
\begin{center}
\includegraphics[width=0.49\textwidth]{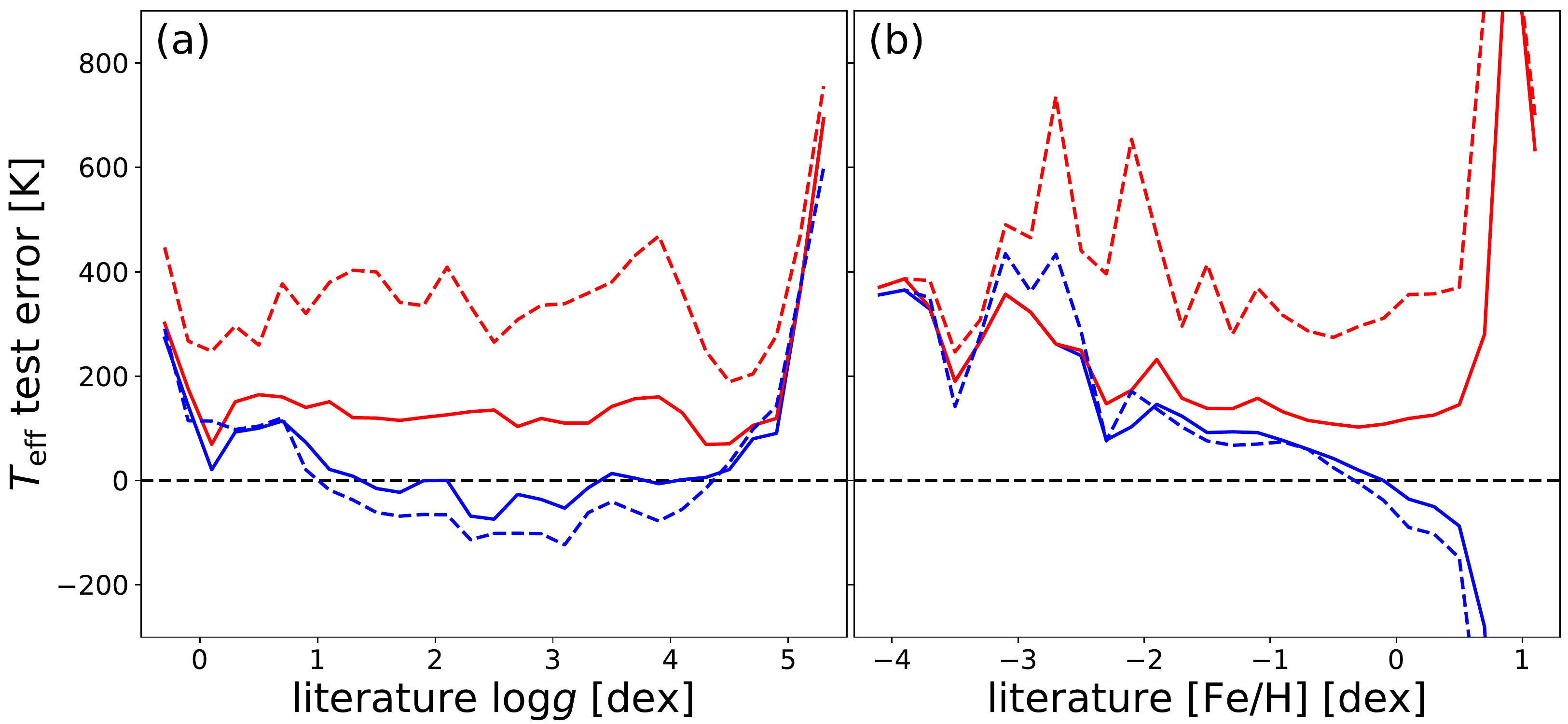}
\caption{Dependence of temperature error on literature values of \logg~(panel a) and \feh~(panel b) for test data (\textit{not} used for training \extratrees) with clean flags.  Red lines are root-mean-squared error (dashed) and root-median-squared error (solid) and blue lines are mean error (dashed) and median error (solid).}
\label{fig:gspphot-Teff-test-error-vs-logg-and-FeH}
\end{center}
\end{figure}

Using the test set with literature values for \teff, \logg, and \feh, Fig.~\ref{fig:gspphot-Teff-test-error-vs-logg-and-FeH} shows how the differences between our temperatures and those in the literature vary with \logg\ and \feh. 
Other than for the extreme \logg\ values, the test errors (RMS and bias) show no significant dependence on \logg. In particular, dwarfs and giants have the same quality of temperature estimates. For $\logg\gtrsim 4.8$, our \teff~estimates are strongly overestimated. 
This might be because our high \logg\ stars are generally cool, with spectra dominated by molecular absorption which may complicate the estimation of \teff\ even when dealing with broad-band integrated photometry. Alternatively it's due to dwarfs being preferentially nearby and thus generally having a low extinction compared to the mean of the \teff\ training sample, resulting in an overestimation of \teff\ (see Fig.~\ref{fig:gspphot-Teff-errors-in-HRD}).

The right panel of Fig.~\ref{fig:gspphot-Teff-test-error-vs-logg-and-FeH} shows that the \teff\ RMS error increases with decreasing metallicity, with the smallest RMS error (and bias) around solar metallicity. In the range $-2\lesssim\feh\lesssim 0.5$ our estimates are good, which simply reflects the metallicity distribution in our training sample.  Outside this interval, \priam\ estimates of \teff~(and thus the derived FLAME parameters) are more biased. Note that for metal-poor stars, \priam\ systematically overestimates \teff, which will play a role for halo stars but is probably secondary to the impact of extinction in Fig.~\ref{fig:gspphot-Teff-test-errors-skymap}.

\subsection{Stellar types in CMD and HRD}

\begin{figure}
\begin{center}
\includegraphics[width=0.49\textwidth]{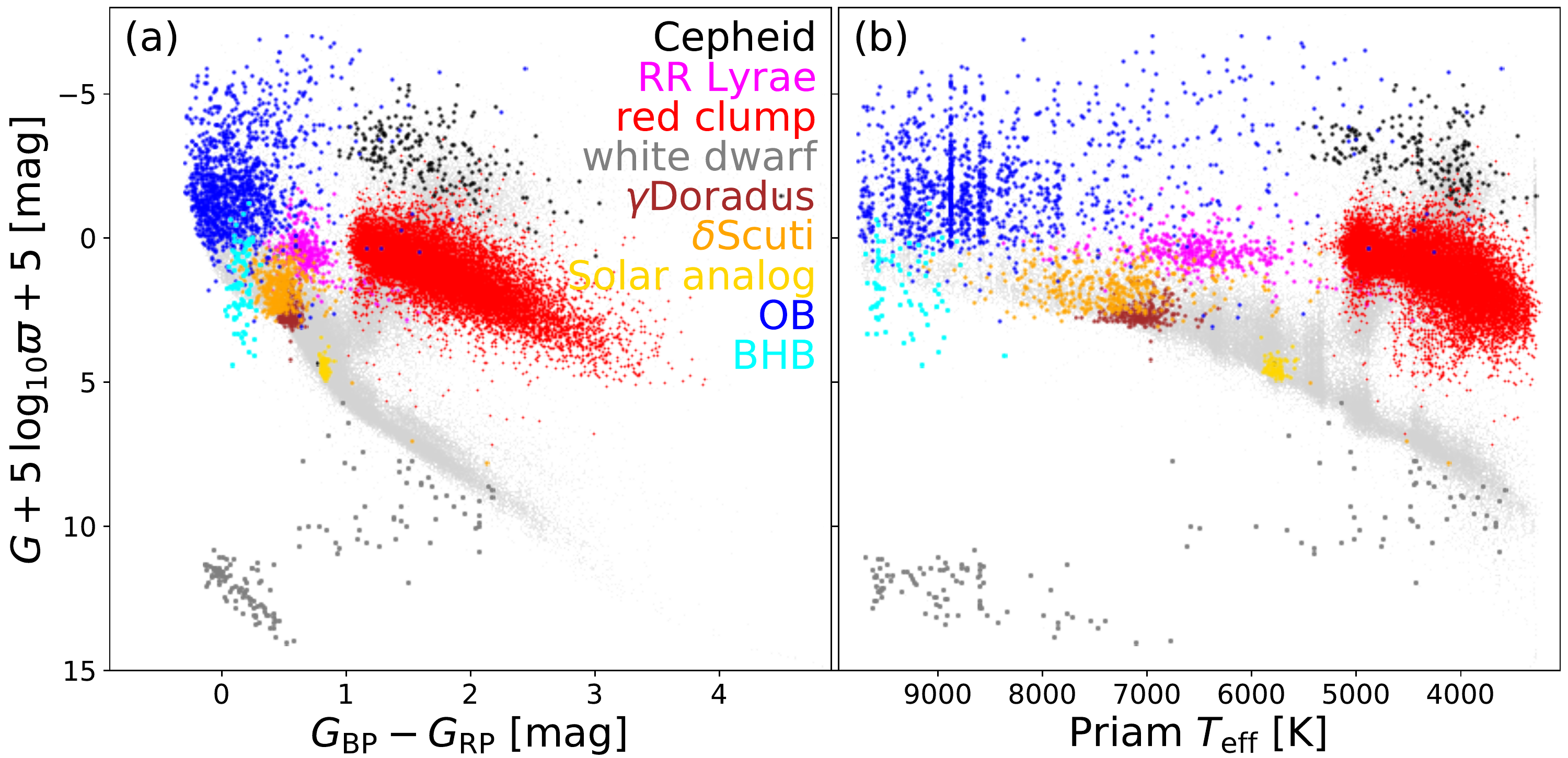}
\caption{Colour-magnitude diagram (panel a) and Hertzsprung-Russell diagram
(panel b) highlighting stars of known classes (from the literature): Cepheids
\citep{2003yCat..34040423T}, RR Lyrae \citep{2006MNRAS.368.1757W}, red clump
\citep{2014ApJ...790..127B}, white dwarfs \citep{Kleinman2013}, $\gamma$~Doradus and $\delta$~Scuti \citep{SarroLSB015}, Solar analogues \citep{TucciMaia2016}, OB \citep{RamirezAgudelo2017,SimonDiaz2011,SimonDiaz2017}, and BHB \citep{2004AJ....127..899S}. We show only stars with clean flags and parallax uncertainties better than 50\%. A 20\% limit would remove all BHB stars.
\label{fig:gspphot-astrophysical-types-in-CMD}
}
\end{center}
\end{figure}

In Fig.~\ref{fig:gspphot-astrophysical-types-in-CMD} we identify stars in our CMDs using classifications from the literature.  This demonstrates that different classes of stars appear where we would expect them to.\footnote{Variable stars of course move around in the CMD according to their phase. The photometry is averaged over many observation epochs, so the positions here correspond to some kind of time average which may not be very representative.}  However, Fig.~\ref{fig:gspphot-astrophysical-types-in-CMD}b makes it clear that our temperature estimates are unreliable for highly extincted stars, which is the case for these red clump stars. This is consistent with the fact that we assume low extinction when estimating \teff\ in \priam.
This restriction was necessary given the strong \teff--\ag\ degeneracy in the colours used (see section \ref{sec:overview}). Note also that even most of the white dwarfs appear in the right location in Fig.~\ref{fig:gspphot-astrophysical-types-in-CMD}b, even though we excluded all white dwarfs from the \priam training sample. There are, however, quite a few white dwarfs that fall between the main, obvious white dwarf sequence and the lower envelope of the main sequence. We see this in both panels of Fig.~\ref{fig:gspphot-astrophysical-types-in-CMD}, so it is not an artefact of the Apsis results but rather some problem with the photometry (e.g.\ blending with a low-mass companion in an unresolved binary system).

\subsection{Temperature estimates for very hot and very cool stars}
\label{ssect:teff-for-OB-and-cool-stars}

\begin{figure}
\begin{center}
\includegraphics[width=0.45\textwidth]{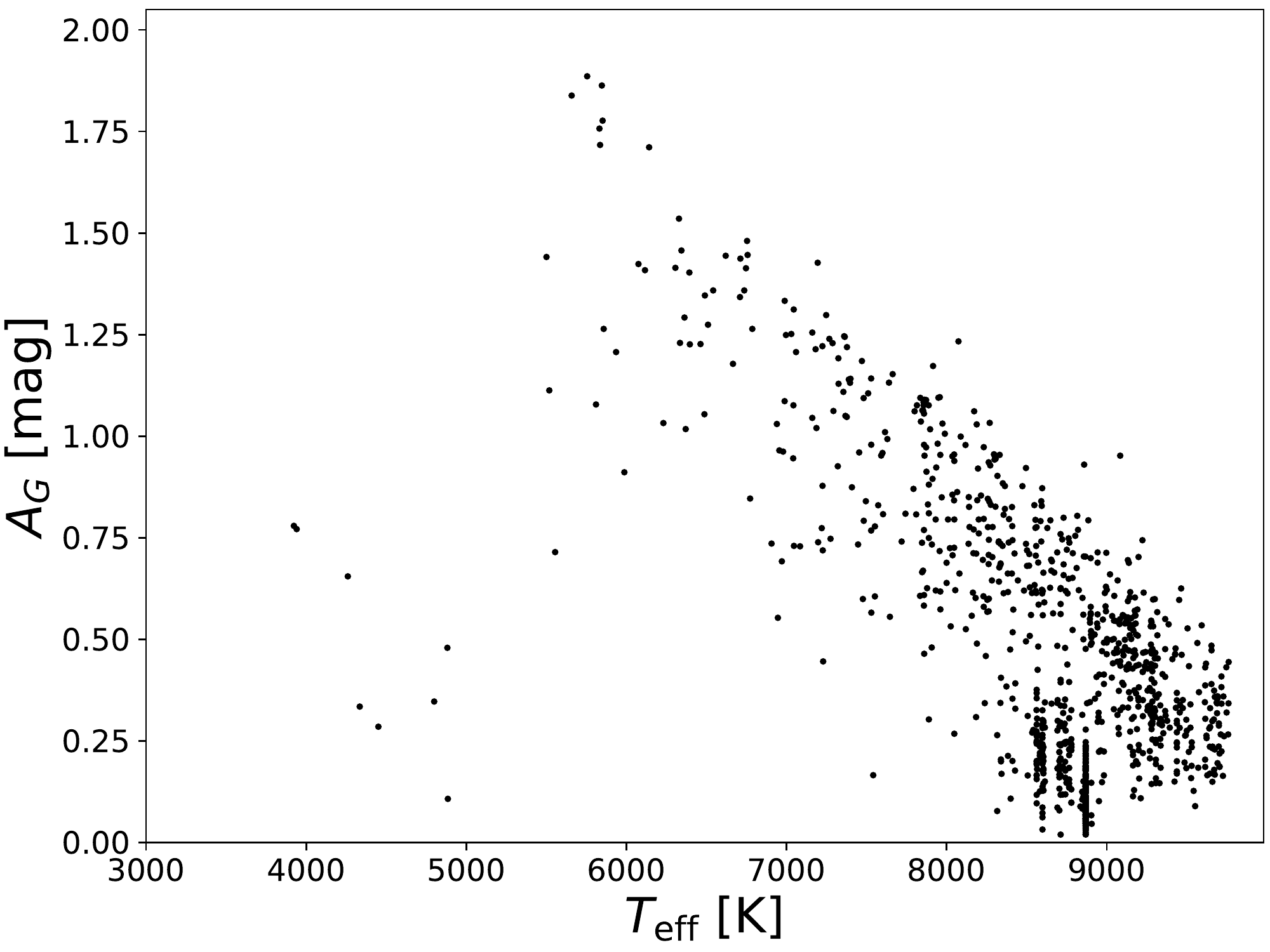}
\caption{\priam estimates of \teff~and \ag~for 1174 stars classified as OB stars by \citet{SimonDiaz2011,SimonDiaz2017} and \citet{RamirezAgudelo2017}. Only stars with clean flags are shown.
}
\label{fig:gspphot-OB-stars-Teff-vs-AG}
\end{center}
\end{figure}

What happens to stars which have true \teff\ outside the training interval 3000K--10\,000K? Intuitively, we would expect that hotter stars are assigned temperatures just below 10\,000K, whereas cooler stars are assigned temperatures just above 3000K. This can be seen in Fig.~\ref{fig:gspphot-astrophysical-types-in-CMD}b, where most of the OB stars indeed have \teff~estimates above 8000K up to the limit of the grid. However, there are also a few cases where OB stars are assigned significantly lower temperatures. 
\priam\ infers high extinctions for these (see Fig.~\ref{fig:gspphot-OB-stars-Teff-vs-AG}).
Since OB stars often reside in regions with substantial dust, the resulting reddening makes them appear cooler, leading \priam\ to assign a lower \teff\ (as explained earlier).
The univariate nature of \extratrees is a limitation here, since it prevents a high extinction estimate
from being a signal that observed colours may be substantially reddened. (We will use multivariate models in \gdr{3}; see section~\ref{sec:outlook}.) Furthermore, the degeneracy between \teff\ and \ag\ 
cannot lead to overestimating OB star extinctions by much, since there are no other models that are intrinsically even bluer that these.

At the lower \teff\ end, we have eight stars in our test set with literature estimates below 3000K.
Of these, seven are assigned \teff\ below 4000K and the eighth is assigned 4120K. 
The lower \teff\ limit is obviously better behaved than the upper temperature limit, where extinction is leading to additional confusion.

The inability of \extratrees to extrapolate from its training label range of 3000K -- 10\,000K does not only apply to the median but also to the 16th and 84th percentiles serving as uncertainty estimates. Consequently, the uncertainties cannot appropriately reflect the underestimation of \teff\ for OB stars, for example.

\subsection{Solar analogues and Gaia benchmark stars}
\label{ssec:Solar-analogs}

\begin{figure}
\begin{center}
\includegraphics[width=0.48\textwidth]{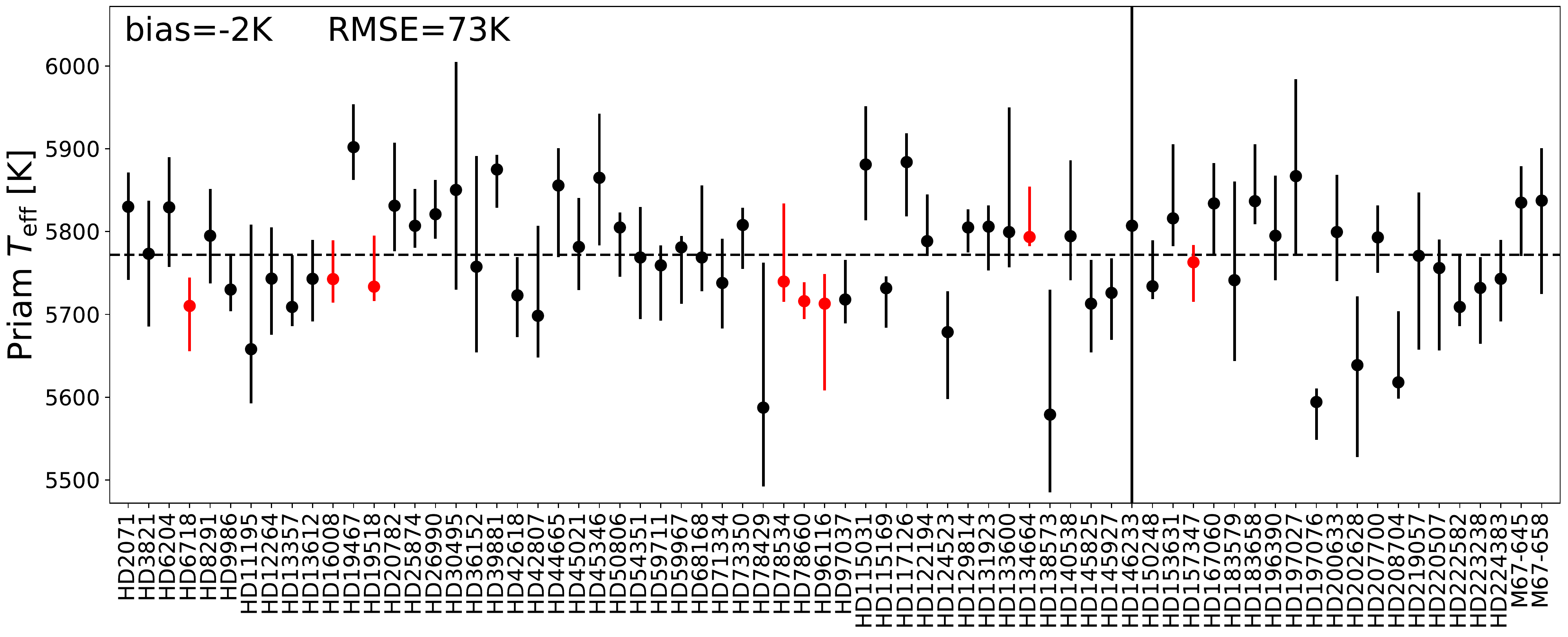}
\caption{\priam estimates of \teff\ for 70 solar analogues with clean flags from \citet{TucciMaia2016}. The horizontal dashed line indicates the accepted solar value of 5772K (see~Table~\ref{tab:refparameters}). Red points were part of the \extratrees training sample so are excluded from the bias and RMS error computations shown in the top left corner.}
\label{fig:gspphot-Teff-Solar-analogs}
\end{center}
\end{figure}

As mentioned in section~\ref{ssec:results-teff}, the error in our temperature estimates depends strongly on the temperature distribution of the sample under consideration. Here we look at a sample of 88 solar analogues from \citet{TucciMaia2016}, of which 70 were in our processing and have clean flags. 
As shown in Fig.~\ref{fig:gspphot-Teff-Solar-analogs}, most of our \teff\ estimates are close to the solar value of 5772K (see~Table~\ref{tab:refparameters}). Excluding the eight solar analogues that were part of the training sample for \extratrees, the RMS test error for the remaining 62 stars is just 73K (1.3\%), which is much smaller than the mean RMS error of 324\,K reported in section~\ref{ssec:results-teff}. The likely explanation for this excellent performance is that the solar temperature is very close to the mean temperature in our training sample (see Fig.~\ref{fig:gspphot-Priam-training-Teff-catalogs}), which is where machine learning algorithms usually perform best.

Regarding our uncertainty estimates, we find that the solar temperature is below our lower uncertainty level in  23\% of cases, and above the upper uncertainty level for 29\% of cases. This agrees with our findings from section~\ref{ssec:results-teff}.

\begin{figure}
\center{\includegraphics[width=0.8\columnwidth]{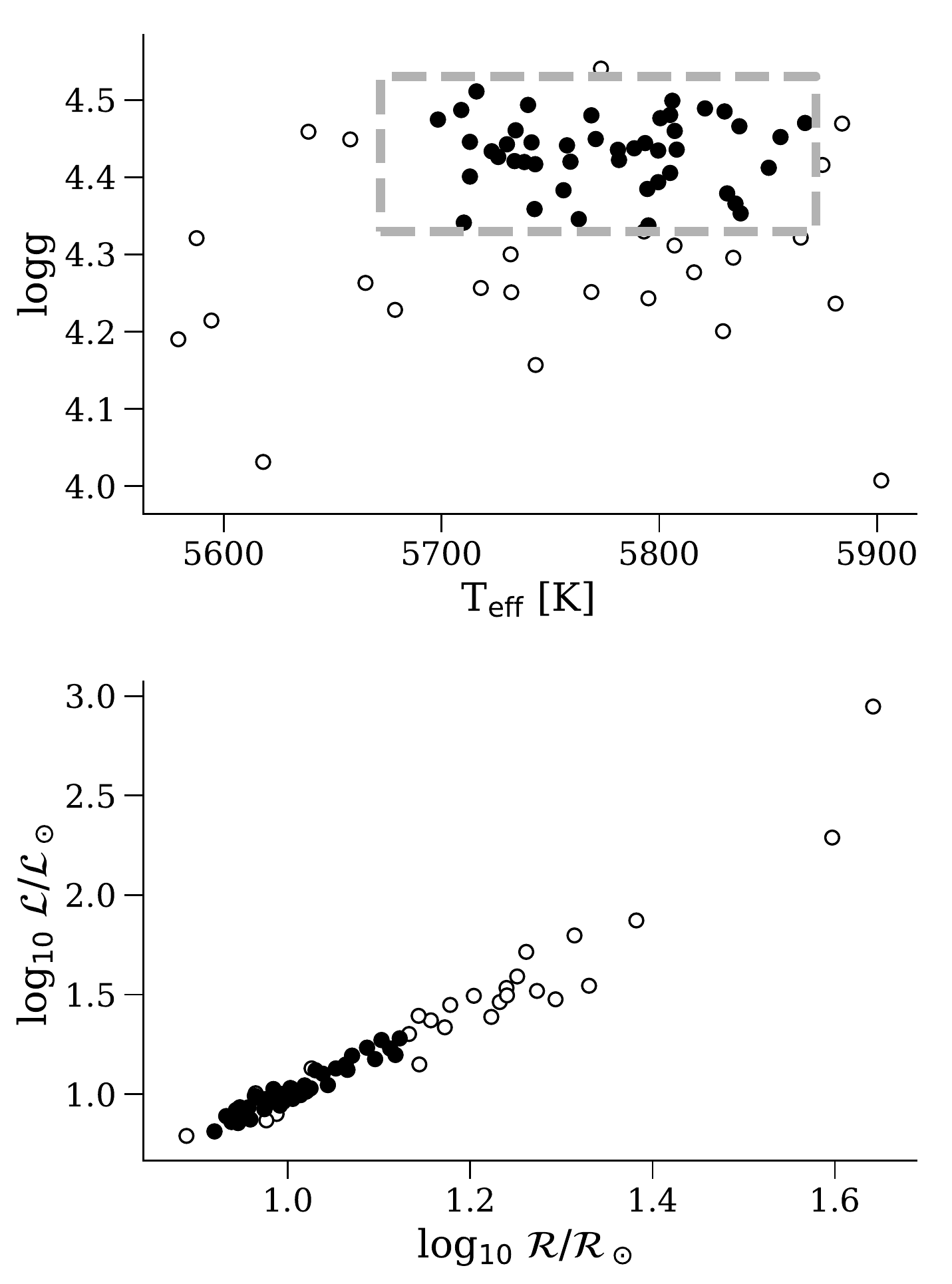}}
\caption{\teff, \logg, \radius, and \lum\ (computed from our results) for the 70 solar analogues. The solid points have \teff\ within 100\,K of the Sun and \logg\ within 0.1\,dex of the Sun (highlighted by the box in the upper panel). \label{fig:flame-solaranalogs}}
\end{figure}

All of the 70 solar analogues found in our catalogue have valid parallaxes and so \lum\ and \radius\ were derived.
We can estimate their surface gravities ($\log g$) using \radius\ and assuming solar mass.
These are shown in Fig.~\ref{fig:flame-solaranalogs}.
We refine our selection of solar analogues
by requiring \teff\ to be within 100 K of the Sun and $\log g$ within 0.1 dex.
This leaves 46 better analogues, shown as
filled circles in Fig.~\ref{fig:flame-solaranalogs}.
We can see this more clearly in the lower panel of Fig.~\ref{fig:flame-solaranalogs}.
The importance of adding the parallax, magnitude, and \radius\ to distinguish
between what we consider solar analogues is evident.
Using this set of 46 better solar analogues
we derive a mean 
$\mg  = 4.65 \pm 0.13$ from their \gmag\ and \parallax\ values.
By assuming their \mbol\ = \mbolsun, we obtain a mean value of $\bcg = +0.09$\,mag for these
solar analogues, which is in good agreement with our adopted value for the Sun of $\bcgsun = +0.06 \pm 0.10$\,mag from section \ref{sec:bc} (derived in 
appendix~\ref{sec:flame-app}).

16 of the 34 Gaia benchmark stars \citep{Heiter2015} are in \gdr{2} (the missing 18 are too bright for the current processing). All have \teff\ estimates from \priam: the RMS difference with respect to \citet{Heiter2015} is 230\,K, which includes a mean difference of $-$24\,K. However, only 6 of the 16 benchmark stars have parallaxes and thus \flame results, and the RMS difference of \lum\ to \citet{Heiter2015} is 13.1\%, which includes a mean difference of $-$3.7\%.

\subsection{Ensemble extinction at high Galactic latitudes}
\label{ssec:extinction-in-halo}

We can validate our extinction estimates using stars at high Galactic latitudes, $|b|>50^\circ$, where we expect the true extinction to be close to zero, especially for nearer stars \citep[e.g.][]{1998ApJ...500..525S,2015ApJ...799..116S}.
However, as our method cannot produce negative extinctions, and given the presence of noise, it is obvious that the average extinction for any ensemble of stars must be larger than zero. Here we study this effect and compare it to our expectations from simulations. We use extinction and reddening estimates for the full sample which is not available to the reader due to the filtering of outliers described in this section. We also derive global uncertainty estimates for \ag\ and \ebpminrp.

\begin{figure*}
\begin{center}
\includegraphics[width=0.95\textwidth]{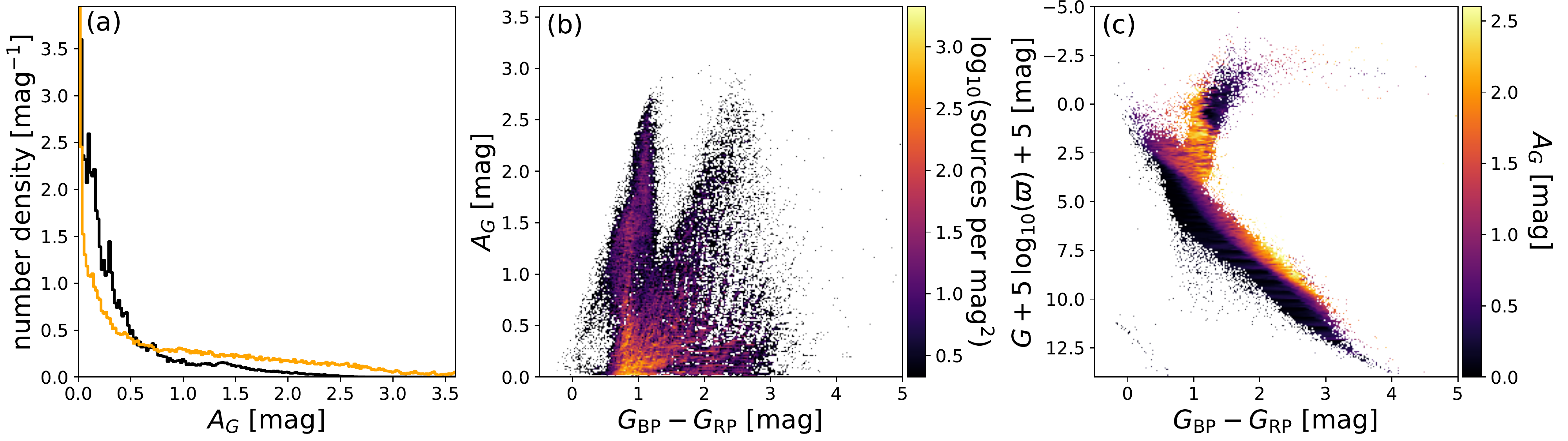}
\caption{Extinction estimates of high Galactic latitude stars with $|b|>50^\circ$ and parallax uncertainty better than 20\%  before applying the filters of Eqs.~(\ref{eq:quality-cut-1-AG-and-reddening})--(\ref{eq:quality-cut-4-AG-and-reddening}). Panel~a: Distribution of \ag\ values (black), and distribution of simulation results with PARSEC photometry for $\a0=0$ and $\feh=0$ (orange). Panel~b: Density of sources in colour--extinction space showing two sets of outliers with unexpectedly large \ag\ estimates at $\gbp-\grp\simeq$1\,mag and $\simeq$2.3\,mag. Panel~c: Colour-magnitude diagram highlighting the location of these two sets of outliers.}
\label{fig:gspphot-AG-halo-investigation}
\end{center}
\end{figure*}

Fig.~\ref{fig:gspphot-AG-halo-investigation}a shows the distribution of \ag\ for high Galactic latitude stars (black histogram). The distribution is clearly peaked at zero extinction and roughly follows an exponential distribution (see below). In particular, 74\% of these high Galactic latitude stars have extinctions below 0.5\,mag (in the \gdr{2} sample 82\% of them have $\ag<0.5$\,mag due to the removal of outliers). Nonetheless, there is a prominent tail extending to extinction values as large as $\ag\simeq 3$\,mag. We also show simulation results (orange histogram) where we applied \priam to synthetic sources from PARSEC which had zero extinction. This produces a similarly heavy tail (note that the distribution is hardly changed if we add typical Gaia noise to the simulated photometry).

The origin of this heavy tail is revealed by Fig.~\ref{fig:gspphot-AG-halo-investigation}b, where two pronounced groups of stars with large extinctions (``outliers'') emerge, one around $\gbp-\grp\simeq 1$\,mag and the other at $\gbp-\grp\simeq 2.3$\,mag. These outliers, while relatively few in number (note the colour scale), are assigned unreasonably large extinctions. We see these in Fig.~\ref{fig:gspphot-AG-halo-investigation}c as the two yellow patches on the upper envelope of the main sequence, one around $\gbp-\grp\simeq 1$\,mag 
and the other around $\gbp-\grp\simeq 2.3$\,mag. Both cases are a consequence of the strong degeneracy of extinction with temperature: The observed red colour can be incorrectly interpreted as an intrinsically bluer and brighter star being made redder and fainter through extinction. We see high extinction in those regions of the CMD where such hotter but extincted stars in the \priam\ training set can appear.
The same behaviour occurs in the simulation shown for comparison in Fig.~\ref{fig:gspphot-AG-halo-investigation}a, where we see a long tail to high extinctions. We conclude that these outliers are the result of an unfortunate alignment of the extinction vector with the astrophysically allowed states in the CMD as defined by the PARSEC training sample. This is a fundamental limitation imposed by having only three broad optical bands (\gmag, \gbp, \grp). The BP/RP spectra should enable us to largely overcome this problem in \gdr{3} (see section~\ref{sec:outlook}).

\begin{figure}
\begin{center}
\includegraphics[width=0.49\textwidth]{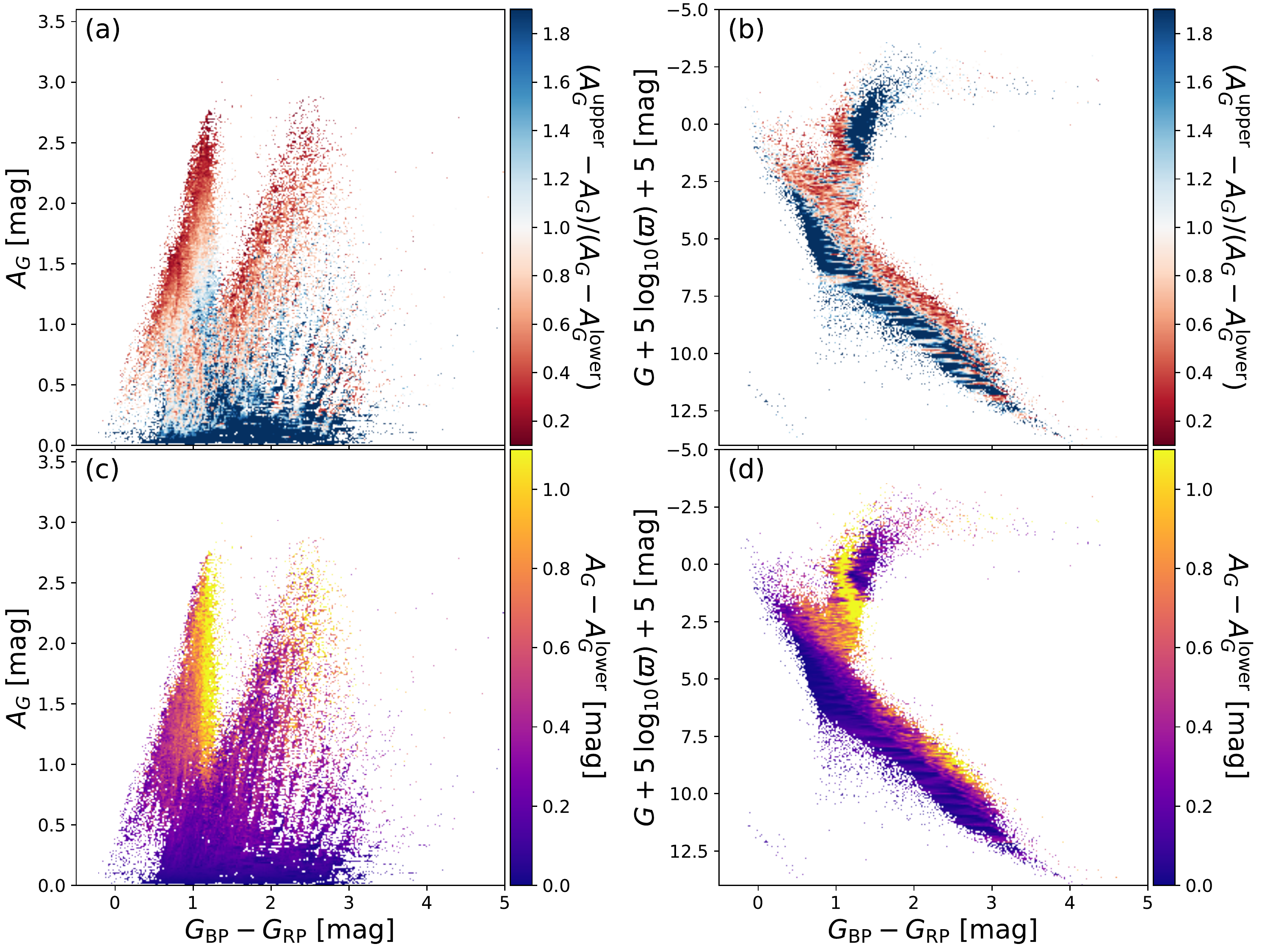}
\caption{Identification of the most degenerate extinction estimates for high Galactic latitude stars with $|b|>50^\circ$ before applying the filters of Eqs.~(\ref{eq:quality-cut-1-AG-and-reddening})--(\ref{eq:quality-cut-4-AG-and-reddening}). Panels~a and~b show the identification via the asymmetry of confidence intervals. Panels~c and~d show the identification via the lower confidence interval. A corresponding plot for \ebpminrp\ is provided in the online documentation.}
\label{fig:gspphot-AG-halo-outlier-selection}
\end{center}
\end{figure}

\begin{figure}
\begin{center}
\includegraphics[width=0.49\textwidth]{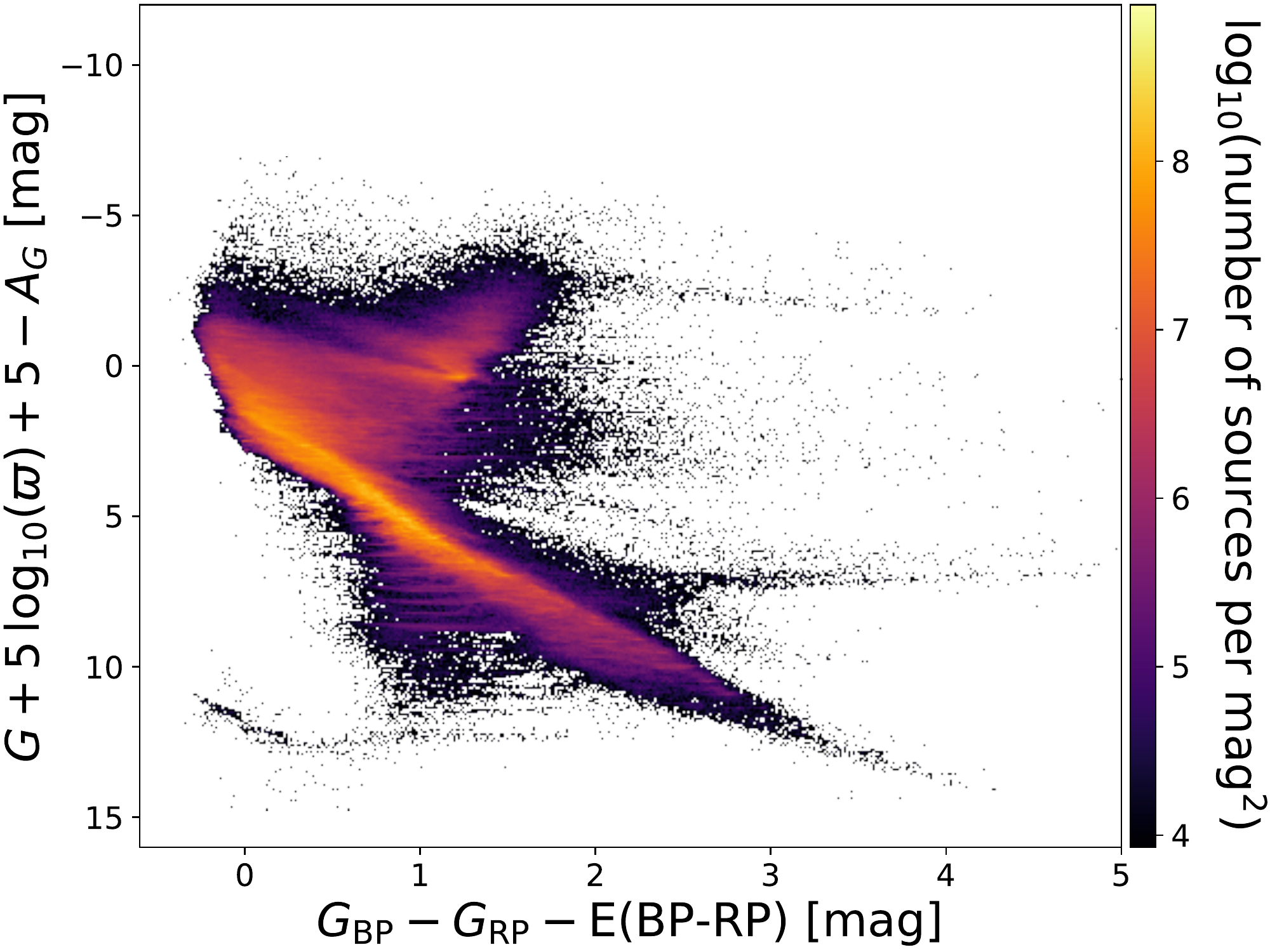}
\caption{Dust-corrected colour-magnitude diagram from complete results (not available in \gdr{2}). Same as Fig.~\ref{fig:gspphot-de-reddened-CMD}d but before applying the filters from Eqs.~(\ref{eq:quality-cut-1-AG-and-reddening})--(\ref{eq:quality-cut-4-AG-and-reddening}).}
\label{fig:gspphot-de-reddened-CMD-apsis-25}
\end{center}
\end{figure}

These outliers were removed from the \gdr{2} catalogue according to the following criteria. The degeneracies are caused by model stars with different extinctions and reddenings (and different stellar parameters) having the same apparent magnitudes. In such a case, the \extratrees ensemble provides a wide range of estimates over all these degenerate states, since it cannot distinguish among them. As our uncertainty estimates are a measure of this spread, such outliers will have large uncertainty intervals. Thus, \gdr{2} only retains estimates of \ag\ and \ebpminrp\ for sources which satisfy the following conditions
\begin{alignat}{2}
\label{eq:quality-cut-1-AG-and-reddening}
\frac{\ag^\textrm{upper}-\ag}{\ag-\ag^\textrm{lower}} \ &>  0.4 \ ,  \\
\ag-\ag^\textrm{lower} \ &< \ 0.5 \ , \\
\frac{\ebpminrp^\textrm{upper}-\ebpminrp}{\ebpminrp-\ebpminrp^\textrm{lower}} \ &>  0.4 \hspace{1em}\text{and}\\
\label{eq:quality-cut-4-AG-and-reddening}
\ebpminrp-\ebpminrp^\textrm{lower} \ &< \ 0.3 \ .
\end{alignat}
$\ag^\textrm{lower}$ and $\ag^\textrm{upper}$ refer to the 16th and 84th percentiles respectively and likewise for 
$\ebpminrp^\textrm{lower}$ and $\ebpminrp^\textrm{upper}$.
The first two filters are justified by Fig.~\ref{fig:gspphot-AG-halo-outlier-selection}. A similar figure justifying the latter two can be found in the online documentation.
Of the 161 million sources with extinctions from \apsis, 88 million pass these criteria and so have extinctions in \gdr{2}.
While this removes most of the outliers, it is of course not perfect.  Unfortunately, the filters also remove many stars on the main sequence turn-off and the lower giant branch. The complete CMD before applying these filters is shown in Fig.~\ref{fig:gspphot-de-reddened-CMD-apsis-25}, which should be compared directly to Fig.~\ref{fig:gspphot-de-reddened-CMD}d to see the impact of the filtering.

\begin{figure}
\begin{center}
\includegraphics[width=0.49\textwidth]{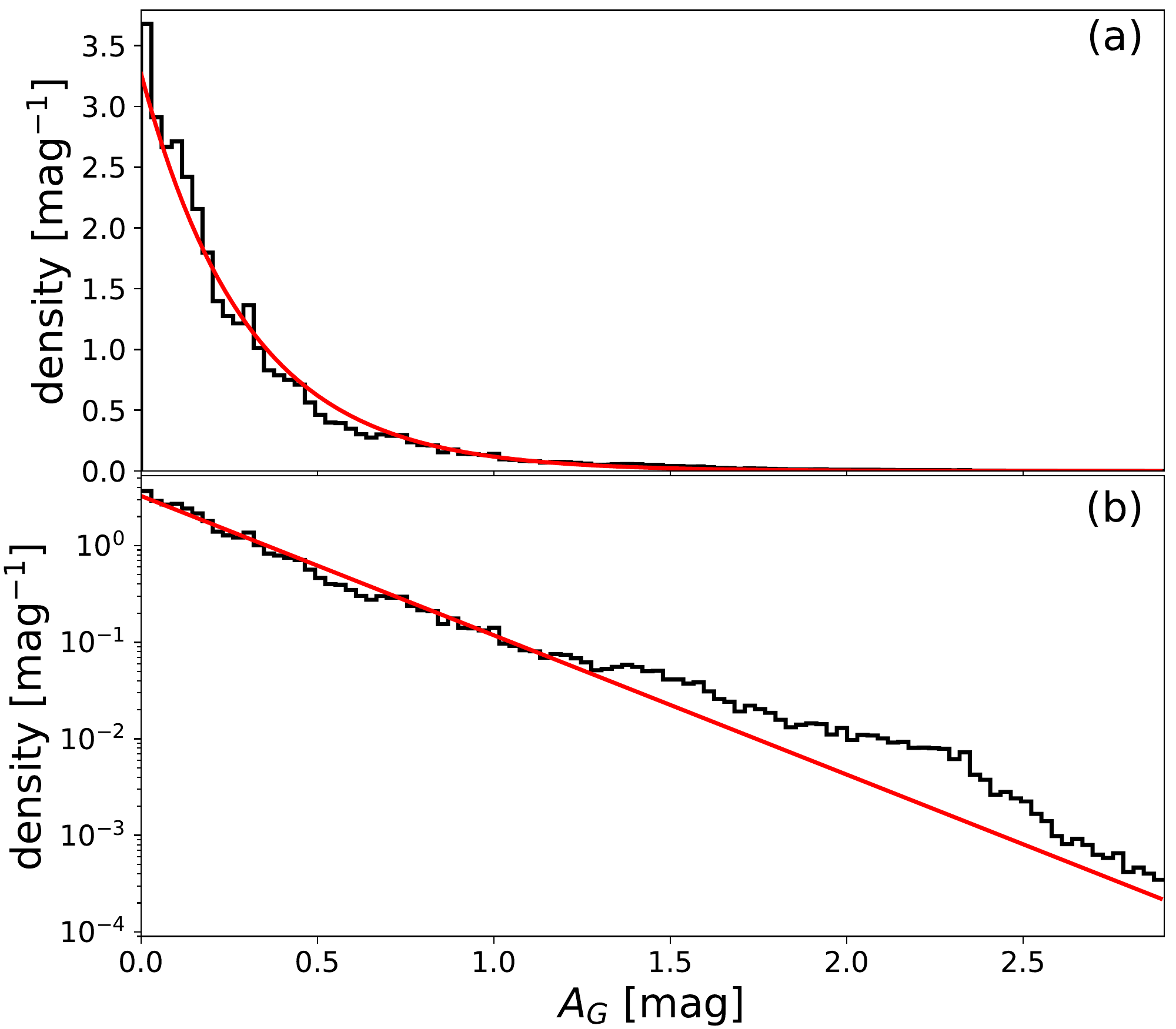}
\caption{Distribution of \ag\ for high Galactic latitude stars with $|b|>50^\circ$ after applying the filters from Eqs.~(\ref{eq:quality-cut-1-AG-and-reddening})--(\ref{eq:quality-cut-4-AG-and-reddening}) (black histogram). The red line shows an exponential whose mean value is set to the mean extinction of this sample, which is 0.30\,mag. Panels a and b show both distributions in linear and logarithmic scale, respectively.}
\label{fig:gspphot-AG-halo-distribution-with-quality-cuts}
\end{center}
\end{figure}

We now return to the sample of high Galactic latitude stars ($|b|>50^\circ$) with this filtering applied. This results in the \ag\ distribution shown in Fig.~\ref{fig:gspphot-AG-halo-distribution-with-quality-cuts}. It agrees very well with an exponential distribution of mean 0.30\,mag and its standard deviation from zero extinction, $\sqrt{\langle\ag^2\rangle}$, is 0.46\,mag. 
The fact that Fig.~\ref{fig:gspphot-AG-halo-distribution-with-quality-cuts} is
largely consistent with an exponential distribution is important, because this
suggests these high Galactic extinction values are consistent with truly zero
with random noise according to information theory: if AG values are pure noise
then its distribution follows its maximum entropy distribution, and with a
positivity constraint this distribution is an exponential \citep[e.g.][]{Dowson1973}.
If the true extinction of high Galactic latitude stars is zero, then our method
will infer a positive value which is just noise; the distribution over these
extinctions must therefore be the maximum entropy distribution for a
non-negative real-valued quantity, which is an exponential. Since
Fig.~\ref{fig:gspphot-AG-halo-distribution-with-quality-cuts} is very consistent
with an exponential distribution, this is evidence that $\ag$ is pure noise for high Galactic latitude stars.\footnote{The agreement with an exponential is not perfect, as is obvious from Fig.~\ref{fig:gspphot-AG-halo-distribution-with-quality-cuts}b. Possible explanations are an imperfect removal of outliers and genuine extinction features at high Galactic latitudes such as young clusters.}

The reddening is also largely consistent with an exponential, of mean 0.15\,mag and a standard deviation about zero, $\sqrt{\langle\ebpminrp^2\rangle}$, of 0.23\,mag. 

From these results we obtain ``global'' uncertainty estimates for \ag\ and \ebpminrp\ of 0.46\,mag and 0.23\,mag, respectively. These are slightly larger than our comparisons to literature values of \av\ and $E(B-V)$ from Table~\ref{table:compare-extinctions-to-literature}, but we take these as estimates of our overall uncertainties.

\subsection{Red clump stars}
\label{ssec:red-clump-stars}

\begin{figure}
\begin{center}
\includegraphics[width=0.5\textwidth]{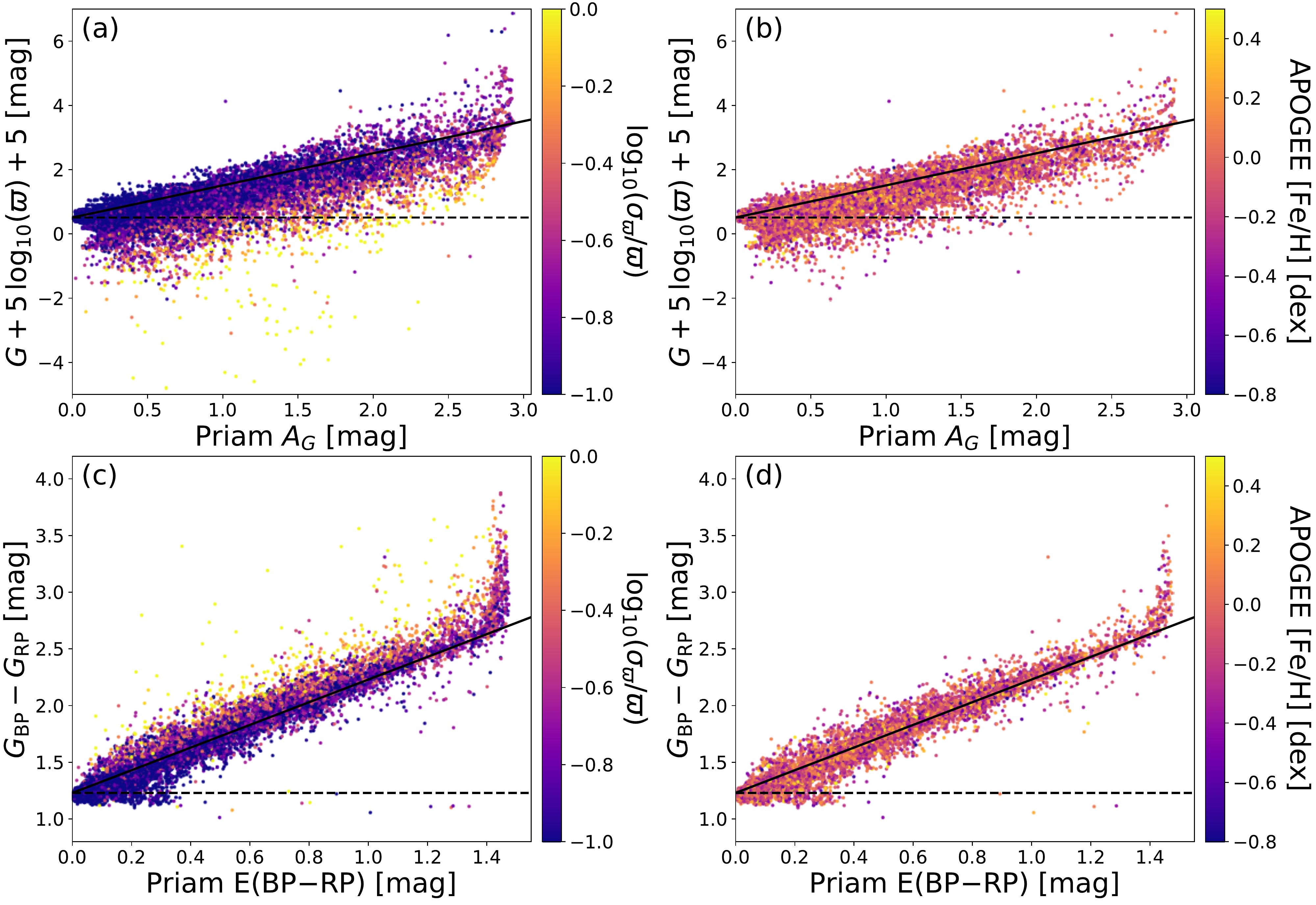}
\caption{\priam estimates of \ag~and \ebpminrp~for red clump stars from \citet{2014ApJ...790..127B}. 
Note that $\gmag+5\log_{10}\parallax+5$ ($=\mg+\ag$) is an observable. 
The horizontal dashed lines show $\mg=0.51$\,mag for $\ag=0$ and $\gbp-\grp=1.23$\,mag for $\ebpminrp=0$, which are the approximate absolute magnitude and intrinsic colour of the red clump. 
The diagonal line is not a fit but the locus of constant absolute magnitude where we expect extincted red clump stars to lie. 
Panels a and c: All red clump stars, colour-coded by relative parallax uncertainty. Panels b and d: Red clump stars with parallaxes uncertainties better than 20\%, colour-coded by metallicity estimates from APOGEE \citep{Alam2015}.
\label{fig:gspphot-red-clump-AG-validation}
}
\end{center}
\end{figure}

As discussed earlier (section~\ref{ssec:results-extinction}), our estimates of \ag~and \ebpminrp\ show a large RMS error on a test sample,
so are generally only useful when combined in an ensemble. We examine here our extinction estimates for red clump stars from \citet{2014ApJ...790..127B} in order to derive a global uncertainty estimate for \ag\ and \ebpminrp. We found 18\,957 of their red clump stars in the \gdr{2} sample with $\gmag<17$, whereby 15\,876 still had extinction and reddening estimates after applying the filters from equations~(\ref{eq:quality-cut-1-AG-and-reddening})--(\ref{eq:quality-cut-4-AG-and-reddening}). Using PARSEC models \citep{Bressan12} with $\a0=0$, $\Zabun=0.0152$,
and evolutionary stage ``4'', 
we estimate that red clump stars reside in a box with a central absolute magnitude of $\mg=0.51\pm 0.25$\,mag \citep[which agrees with][]{2017arXiv171005803R} and a central colour of $\gbp-\grp = 1.23\pm 0.05$\,mag.
In Fig.~\ref{fig:gspphot-red-clump-AG-validation} panels a and b we plot the observable $\gmag+5\log_{10}\parallax+5$
computed from the Gaia data (which is equal to $\mg+\ag$),
against our estimate of \ag. 
For 12\,127 red clump stars with parallax uncertainties below 20\%, we indeed find a relation that is consistent with the absolute magnitude inferred from the PARSEC models.
Likewise, if we plot the observed $\gbp-\grp$ colour against our estimated \ebpminrp, we also find results that are consistent
with the intrinsic colour inferred from PARSEC  (Fig.~\ref{fig:gspphot-red-clump-AG-validation} panels c and d). 
Even though many red clump stars have APOGEE metallicity and abundance estimates \citep{Alam2015} that differ from solar metallicity, thus violating our model assumptions from section~\ref{ssect:gspphot-extinction-description}, there are no obvious biases. Using the method described in Appendix~\ref{appendix:Deming-formalism-error-estimation}, we find that our estimates of \ag~and \ebpminrp~differ from the observables $\mg+\ag$ and $\gbp-\grp$ by about 0.21\,mag and  0.09\,mag, respectively.
These uncertainty estimates are substantially smaller than the global uncertainty estimates of 0.46\,mag for \ag\ and 0.23\,mag for \ebpminrp\ that we obtained from high Galactic latitude stars in section~\ref{ssec:extinction-in-halo}. This suggests that our extinction and reddening estimates work better for red clump stars. Furthermore, we find that our estimates of \ag~and \ebpminrp~are unbiased, since for panels b and d in Fig.~\ref{fig:gspphot-red-clump-AG-validation}, the resulting intercept
estimates (obtained from appendix~\ref{appendix:Deming-formalism-error-estimation}) of 0.27\,mag for $\gmag+5\log_{10}\varpi+5$ and 1.22\,mag for $\gbp-\grp$ agree very well with the absolute magnitude and intrinsic colour inferred from PARSEC models. Note that the red clump selection by \citet{2014ApJ...790..127B} may also contain a few RGB stars, contaminating our analysis. Furthermore, there is an intrinsic spread in the absolute magnitude and colour of the red clump. We neglect this, so our  uncertainty estimates of \ag\ and \ebpminrp\ for red clump stars are conservatively large. Finally, note that the sharp upturn around $\ebpminrp\simeq 1.5$ in Fig.~\ref{fig:gspphot-red-clump-AG-validation}d suggests that some of the red clump stars may be subject to such strong extinction that they are not covered by our model grid (which extends to $\a0=4$mag). 

\subsection{Dust towards Galactic clusters and nebulae}
\label{sec:dust_clusters}

\begin{figure}
\begin{center}
\includegraphics[width=0.5\textwidth]{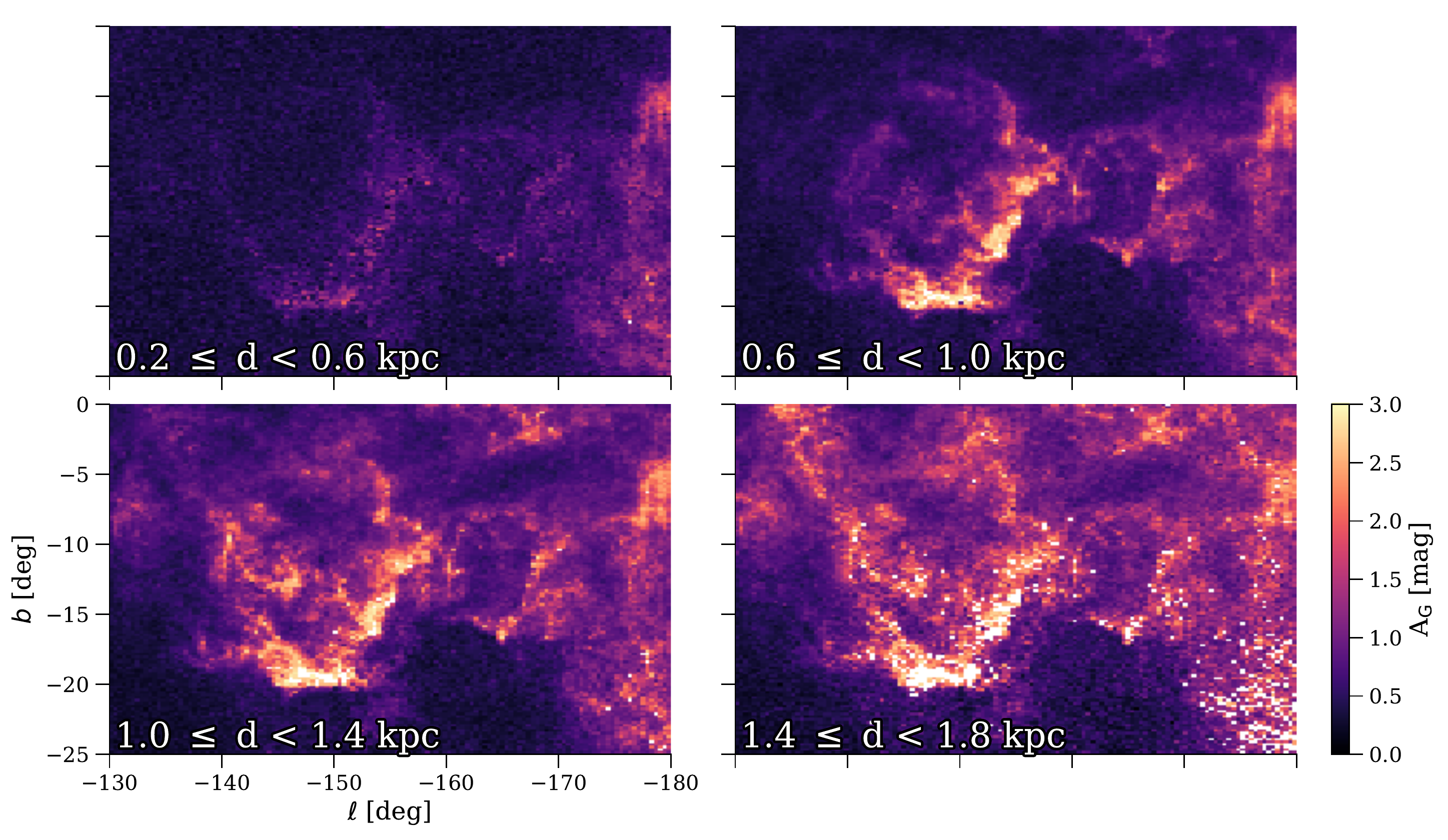}
\caption{Extinction estimates toward Orion for different distance slices, from sources with parallax precisions better than 20\%. White points lack extinction estimates. Compare to Fig.~2 in \citet{2015ApJ...799..116S}.}
\label{fig:gspphot-AG-Orion}
\end{center}
\end{figure}

\citet{2015ApJ...799..116S} have investigated the three-dimensional structure of the line-of-sight extinction \av~for the Orion nebula.  We show in Fig.~\ref{fig:gspphot-AG-Orion} our \ag~estimates in the same way as Fig.~2 of \citet{2015ApJ...799..116S}. Each cell shows the mean extinction over the distance range given.
As noted in appendix~\ref{appendix:mean-cluster-extinction}, this is likely to be an overestimate when the true extinction is low.
The agreement is nonetheless striking, although we are missing the fainter stars due to the $G\leq 17$ limit in our processing.

\begin{figure*}
\begin{center}
\includegraphics[width=0.99\textwidth]{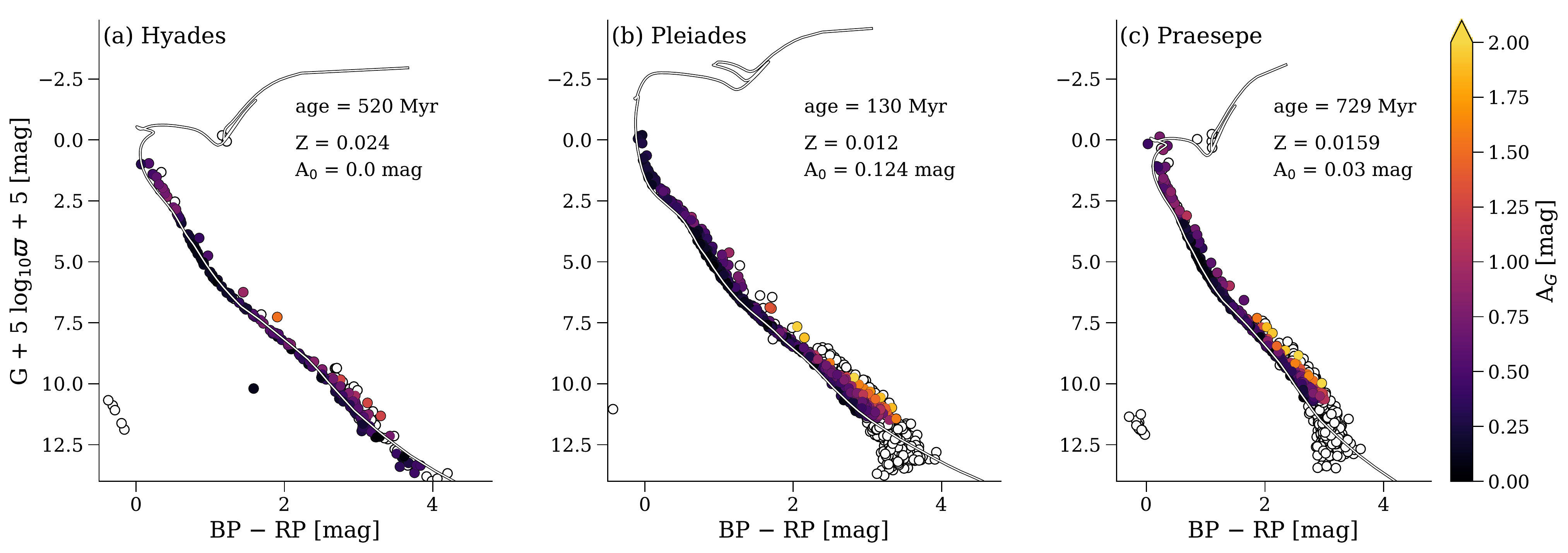}
\caption{
Colour magnitude diagrams for the Hyades, Pleiades, and Praesepe star clusters.
In each panel we indicate cluster members 
(defined by \citealt{DR2-DPACP-31})
using filled circles coloured according to their extinction estimates
(white symbols are members without published extinction estimates in \gdr{2}).
The solid lines indicate the PARSEC isochrones with parameter values, marked in the figure,
taken from the estimates for the clusters from \citet{2002MNRAS.334..193C} for the Hyades and Pleiades and
from \citet{2006AJ....132.2453T} for Praesepe.
\label{fig:pleiades}
}
\end{center}
\end{figure*}

To further assess the quality of our estimates of dust attenuation, we studied the colour--absolute magnitude diagrams of some Galactic star clusters.  Fig.~\ref{fig:pleiades} shows three examples, with members taken from \citealt{DR2-DPACP-31} who selected them based on their spatial positions and proper motions. The plot shows good agreement with isochrone models of matching age, metallicity, and extinction from \citet{2002MNRAS.334..193C}.
The colour scale shows the extinction estimates from \priam.  The
sequences above and parallel to the main sequence in both clusters are
presumably binaries. These tend to have larger (overestimated)
extinctions. As binaries have slightly redder colours than single stars of the
same brightness due to the generally lower mass companion, yet have the same
parallax, \priam\ -- assuming all stars to be single -- interprets their redder
colours to be a result of extinction. (Binary stars will receive explicit
treatment in future Gaia data releases; see section~\ref{sec:outlook}.)

We show in appendix \ref{appendix:mean-cluster-extinction} that,
because our extinction estimates are non-negative, the mean of a sample of stars
is often a biased estimator of the true (assumed common) extinction.
We can instead use a likelihood function for each star which respects this non-negativity, namely
a truncated Gaussian.
When we do this, the maximum of the likelihood for the
set of cluster members is
$\ag = 0.21 ^ {+0.05}_{-0.08}$\,mag for the Hyades,
$\ag = 0.11 ^ {+0.06}_{-0.07}$\,mag for the Pleiades, and
$\ag = 0.00 ^ {+0.08}_{-0.00}$\,mag for Praesepe 
(the uncertainties refer to the 16th and 84th percentiles on the likelihood).
These cluster extinction estimates are in very good agreement with the values from
\cite{2002MNRAS.334..193C} and \cite{2006AJ....132.2453T}, listed in Fig.~\ref{fig:pleiades}, for the Pleiades
and Praesepe clusters. Our estimate for the Hyades, in contrast, is much larger than 
the literature value. 
As discussed in appendix \ref{appendix:mean-cluster-extinction}, we suspect this is just a combination of bad luck and a relatively small number of members for precisely locating the maximum of the likelihood.
Finally, we note that even within our clean sample, our results in clusters may suffer from photometric errors
induced by crowding.

\subsection{Validation of radii and luminosities with seismology, interferometry, and surveys
of nearby stars}
\label{ssec:flame-validation}

\begin{figure*}
\begin{center}
\includegraphics[width=0.85\textwidth]{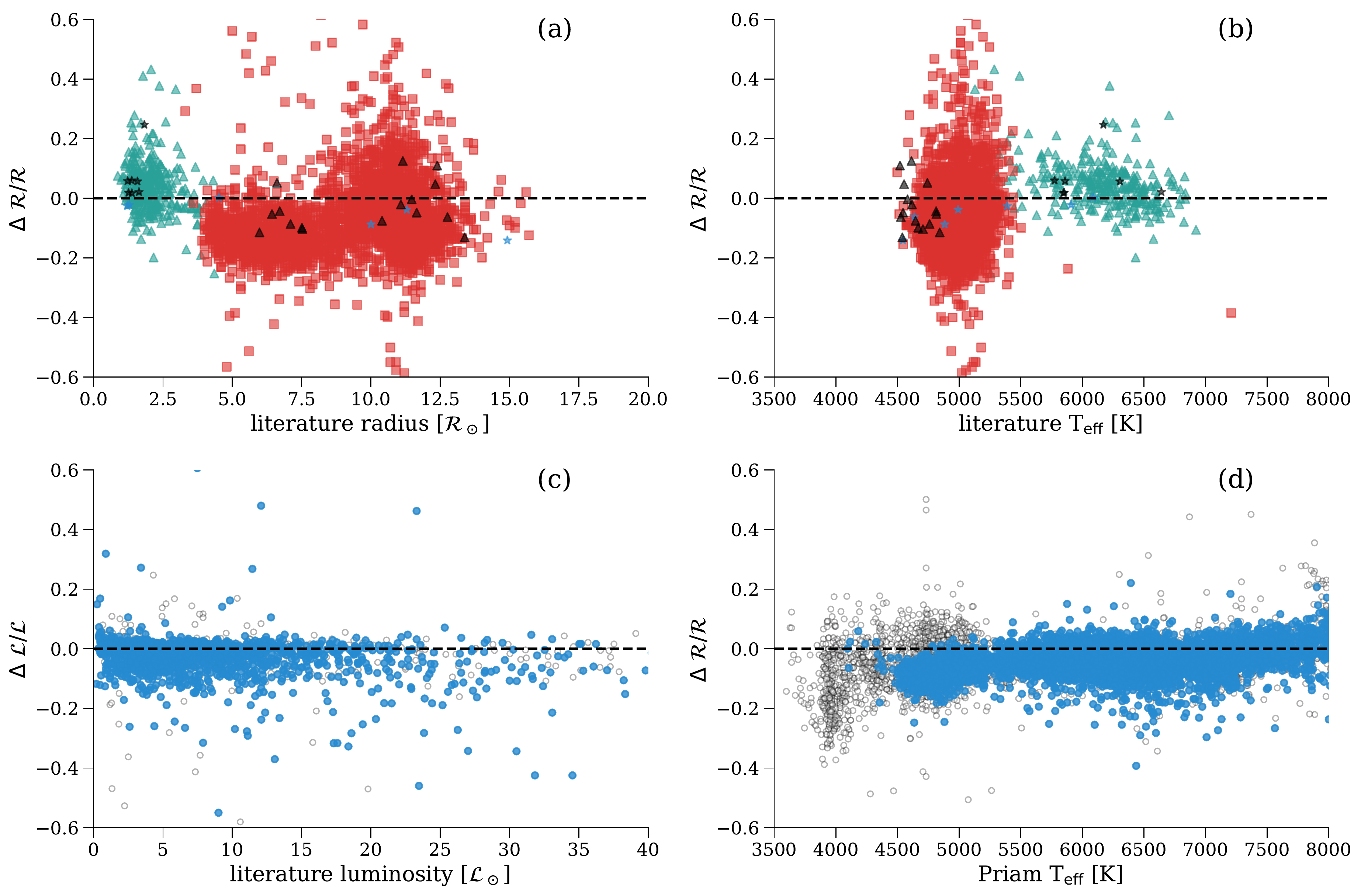}
\caption{
Comparison of \radius\ and \lum\ with external data. $\Delta$ is defined as 
(\flame$-$literature)/literature.
(Top) Comparison of radii as a function
of (a) radius and (b) \teff, for asteroseismic and interferometric targets.
The symbols indicate different literature sources:
red squares are \citet{2016A&A...588A..87V}; 
green triangles are \citet{cha14}; 
blue stars are \citet{Boyajian2013} and \citet{ligi16};
black stars are \citet{cre17};
black triangles are members of \object{NGC\,6819} from \citet{basu11}.
(Bottom) Comparison of (c) \flame luminosities with those from \citet{2011yCat..35300138C} 
and (d) \flame radii with those from \citet{2017yCat.2346....0B}. 
The blue circles are the subsample of stars used to estimate \bcgsun\
(see appendix~\ref{sec:flame-app}).
The open black circles are other stars not fulfilling the stricter criteria for this estimate.
\label{fig:flameval-rad1}
}
\end{center}
\end{figure*}
 
To validate the results for \flame, we compare the derived radii and luminosities with those from a selection of external catalogues.  These are shown in Fig.~\ref{fig:flameval-rad1}, whereby the targets are mostly bright ($G<12$) and nearby ($<1500$ pc).  The top panels compare the radii from a compilation of asteroseismic and interferometric references as a function of literature radii (panel a) and \teff\ (panel b).  For the less evolved stars (\radius\ $<$ 3.0 \Rsun) the \flame\ 
radii are slightly overestimated, but the differences are consistent with zero considering the radius uncertainties for this range (Fig.~\ref{fig:flameradiuserror}), due to the differences in the adopted temperature scales.
The green triangles represent radii from automatic asteroseismic analysis using scaling relations, where typical uncertainties in the radius can be 5\% and the actual values of the radii are rather sensitive to the input \teff\ (see \citealt{cha14}).  For this sample we also find a systematic trend in the radii which increases with decreasing \teff\ (panel b).  This suggests that the differences in the radii arises from the different temperature scales used.  The other stars in this less evolved sample range have been studied in much finer detail using interferometry or detailed asteroseismic analysis (black and blue stars).
In these cases, the literature radii are much less sensitive to the adopted \teff. For these better-studied stars, we do find much better agreement with the \flame\ radii ($<5\%$), with no significant differences as a function of \teff.

The largest sample in this figure is that from \citet{2016A&A...588A..87V}, who studied red giants using asteroseismic scaling relations in an automatic manner.  For the giants, typical uncertainties in the asteroseismic radii are of the order 7--10\% and the radii are sensitive to the adopted \teff.  As explained earlier, we ignore interstellar extinction, and thus the luminosities will be systematically underestimated.  This is particularly problematic for giants which are more distant, as extinction could be non-negligible.  However, for a fixed \teff\ a smaller luminosity implies a smaller radius.  
In most cases we see that the radii are underestimated (negative differences). However, in the range $8 < \radius /\Rsun < 12$ there is a large scatter around zero (there are many overestimated radii too). This zero offset is a direct result of the cancellation effects of the adopted temperature scales and of ignoring extinction.
For this sample of stars, we find systematically lower \teff\ with \priam, where the typical differences are between --5\% and --1\%.  As a consequence, for fixed luminosity, the radius will be overestimated by \flame.  As with the main sequence stars, the interferometric sample (blue stars) and the giants in NGC~6819 (black triangles), which were studied in much finer detail, agree with the \flame\ radii to within 10\%, and show no trend in their differences as a function of radius or \teff.

In the lower panels of Fig.~\ref{fig:flameval-rad1} we compare our estimates of \lum\ and \radius\ with those derived from external analyses based on photometric measurements.  The left panel (c) compares our luminosities with those derived from the bolometric flux estimates from \cite{2011yCat..35300138C}.  We also compared our radii using their bolometric flux and temperature measurements (not shown).  We obtain a negligible mean difference in the luminosities and radii which can be rectified by changing the \bcgsun\ by a few mmag (see appendix~\ref{sec:flame-app}).  We find a dispersion below $5$\%, suggesting no unknown systematic effects in our results.  In the lower right panel (d) we compare \flame\ radii with those derived from predicted angular diameters from \cite{2017yCat.2346....0B}.  Again we find a mean offset in the differences close to zero and a dispersion about this mean of less than 10\%.  In both figures, the blue filled circles represent a subset of these data which fulfil the criteria \priam\ $\sigma(\teff)/\teff < 3\%$, $7 < G < 10$, Gaia $\sigparallax/\parallax < 5\%$, and $1/\parallax < 300$pc.  These were used to test the \bcgsun\ discussed in appendix~\ref{sec:flame-app}.  
The open black circles are stars retrieved in our validation data which don't fulfil these criteria.

\section{Using the data}\label{sec:usage}

In this work we have, independently for each source, inferred five parameters from three partially degenerate flux measurements and (for all but \teff) a parallax. This has necessarily demanded a number of extreme simplifications and assumptions. The data should therefore be used with great care. We recommend always using the flags, defined in appendix \ref{sec:flags}, to filter out poorer data. In particular, for astrophysical analyses, we recommend only using parameters for stars in the clean \teff\ sample (defined in that appendix).  When using extinctions, luminosity, and radii users may also want to only select stars with small fractional parallax uncertainties 
(Fig.~\ref{fig:gspphot-de-reddened-CMD} gives an example of the impact of this filtering on the CMD.)

\subsection{Limitations and caveats}
\label{sec:limitations}

The following caveats should be kept in mind.
\begin{enumerate}
\item All sources are treated as single stars. We do not perform source classification and our processing does not make use of external classifications. We do not filter out results for sources we know from other data to be galaxies, unresolved binaries, etc.
\item The training sample for \teff~of \extratrees contains stars with a certain range of (low) extinctions (see section~\ref{ssect:gspphot-teff-explanation}). The \teff\ estimates will be systematically too low when the true extinction is significantly below this range (e.g.\ in the halo). Likewise, \teff\ estimates will be systematically too high when the extinction is above this range. One manifestation of this is a systematic trend in temperature errors (from comparison to literature estimates) with Galactic latitude (see Fig.~\ref{fig:gspphot-Teff-test-errors-skymap}). Hot stars with high extinction get \teff~estimates that are systematically too low (see Fig.~\ref{fig:gspphot-Teff-vs-AG-bias}). See also Fig.~\ref{fig:gspphot-Teff-test-error-vs-other-data}e. 
\item For \teff~estimation, \extratrees was only trained on the range
  3000K--10\,000K. Stars which are truly hotter or cooler will therefore be
  systematically under- or overestimated. However, for \ag~and
  \ebpminrp~estimation, \extratrees was trained on PARSEC models with $0 \leq
  \a0 \leq 4$\,mag and 2500\,K $\leq \teff \leq$ 20\,000\,K.  While we cannot get \teff~estimates from those models, we still get reliable extinction and reddening estimates for intrinsically blue sources such as OB stars.
\item Due to the distribution of parameters in the training sample, the resulting temperature distribution exhibits artificial stripes (see Fig.~\ref{fig:gspphot-Teff-vs-AG-bias}). This also affects the Hertzsprung-Russell diagram (see Fig.~\ref{fig:flame-hrdiag} and Fig.~\ref{fig:gspphot-astrophysical-types-in-CMD}b).
\item The (empirical) \teff\ training sample lacks low metallicity stars, so systematic errors in \teff\ can be expected for truly low metallicity stars. The same can be said for the extinction estimates, as our (synthetic) training sample is solar metallicity.
\item As we use three broad optical bands, our estimates of  \ag~and \ebpminrp\ are highly degenerate with our \teff\ estimates. This leads to unreliable extinction and reddening estimates in parts of the CMD (see section~\ref{ssec:extinction-in-halo}).
\item Our extinction estimates are strictly non-negative, with uncertainties of similar size to the estimate itself. Hence neither \ag\ nor \ebpminrp\ can be considered as a Gaussian random variable, not even approximately.
The likelihood (probability density) is intrinsically skewed, which is why we report 16th and 84th percentiles to reflect the uncertainty.
As explained in appendix \ref{appendix:mean-cluster-extinction}, a truncated Gaussian is a more appropriate likelihood model for the extinctions. The non-negativity can feign a systematic overestimation of extinction in regions where very low extinction is expected, such as at high Galactic latitudes (section~\ref{ssec:extinction-in-halo}) or for some stellar clusters (section~\ref{sec:dust_clusters}). As we show in appendix \ref{appendix:mean-cluster-extinction}, the mean extinction in such regions is a poor estimator of the true extinction.
\item The estimates of \ag~and \ebpminrp\ generally have such large uncertainties that their usefulness for individual stars is limited. These estimates should generally only be used statistically, by applying them to ensembles of stars.
The central limit theorem applies despite the non-negativity, ensuring that the variance in the mean
of a sample will drop as $1/N$, even though the mean may be a biased estimator.
The Galactic extinction map (Fig. \ref{fig:gspphot-AG-skymap-Galactic}) and the Orion map (Fig.~\ref{fig:gspphot-AG-Orion}) suggest that mean extinctions are reliable. The \teff--extinction degeneracy in the photometric data nonetheless mean that some extinctions are quite erroneous, although many of these were removed by the cuts applied by equations
\ref{eq:quality-cut-1-AG-and-reddening} to \ref{eq:quality-cut-4-AG-and-reddening}.
\item The extinction estimates satisfy $\ag\sim 2\cdot\ebpminrp$. This is a consequence of the models which we are using (see~Fig.~\ref{fig:gspphot-PARSEC-models-AG-vs-EBPminRP}).
\item Our extinction estimates are based on PARSEC 1.2S models with extinction
assuming fixed $\relext=3.1$. Using a different \relext, a different extinction
  law, or a different set of stellar models may lead to systematic differences
  compared to our current estimates.
\item We infer parameters assuming solar or near-solar metallicity (see sections \ref{sec:gspphot} and \ref{sec:flame}). Our results are therefore likely to be wrong for low metallicity stars, such as in most globular clusters. The impact of this on extinction estimates for lower metallicity red clump stars can be seen in 
Fig.~\ref{fig:gspphot-red-clump-AG-validation}.
\item Stellar clusters, in particular globular clusters, are crowded regions where the integrated photometry is sometimes compromised. The BP and RP photometry are obtained by integrating over an area of $3.5 \times 2.1$\,arcsec$^2$. As the two Gaia fields-of-view are projected onto a common focal plane, sources can overlap even in low density regions. For crowding and other reasons, users may want to additionally filter out sources according to the ``BP/RP flux excess'' (see \citealt{DR2-DPACP-40}).
\item Parallaxes are used to estimate \ag, \ebpminrp, \lum, and \radius, by giving a distance somewhat naively as
$1/\parallax$ \citep[see][]{2015PASP..127..994B,DR2-DPACP-38}.
Sources without positive parallaxes therefore have no estimates of these parameters, and those
with large fractional parallax uncertainties ($\sigparallax/\parallax$) will have particularly noisy estimates of these parameters. This applies in particular to distant and/or faint stars. We recommend only using estimates for these parameters for stars with fractional parallax uncertainties of 20\% or less.
\item The uncertainties in the \lum\ and \radius\ are most likely underestimated.
The luminosity may suffer from a systematic error based on the adoption of the value of \bcgsun\ (see appendix~\ref{sec:flame-app}), although we estimate this to be within 0.1 mag. 
Luminosity will also be systematically underestimated for extincted stars, since we assumed extinction to be zero when inferring the absolute magnitude.
As \radius\ is derived directly from \lum\ and \teff, 
both of these effects also impact the radius estimates, albeit to a smaller degree 
(see section~\ref{sec:results-radius}).
\end{enumerate}
The application and implication of the above guidelines is illustrated in the following sections.

\subsection{Selection of targets for interferometry}

\begin{figure*}
\begin{center}
\includegraphics[width=0.98\textwidth]{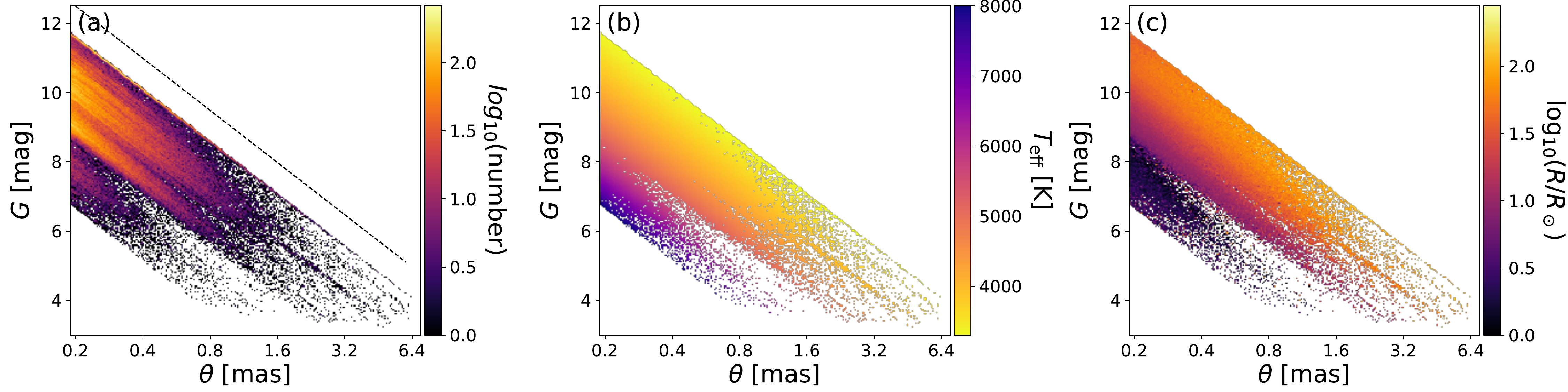}
\caption[]{Selection of target candidates for interferometry colour-coded by (a) number of stars, (b) \priam estimate of \teff, and (c) \flame estimate of radius. The stellar angular diameter $\theta$ is plotted on a log scale. We find 213\,139 candidates with angular diameter $\theta>0.2$\mas, 95\,794 candidates with $\theta>0.3$\mas, and 52\,196 candidates with $\theta>0.4$\mas. The black dashed line in panel (a) corresponds to $G=\textrm{const}-5\log_{10}\theta$.}
\label{fig:flame-angular-diameter-vs-Gmag}
\end{center}
\end{figure*}
 
Measurements of apparent diameters of stars with ground-based interferometry play a crucial role for direct estimation of stellar temperatures \citep[e.g.][]{Heiter2015} or the validation of asteroseismic radii \citep[e.g.][]{2013MNRAS.433.1262W}. Long-baseline optical interferometry can reliably measure angular diameters as small as 0.2\mas\ \citep[e.g.][]{Boyajian2015}. Our results may be used to select potential targets for such interferometric observations. We suggest the following selection criteria:
\begin{equation}
\begin{aligned}
\label{eq:cuts-for-interferometry-targets}
\parallax/\sigparallax>5 \\
\theta=2\,\radius\cdot\parallax>0.2\mas \\
\textrm{\priam flag 0100001 or 0100002}
\end{aligned}
\end{equation}
The cut on relative parallax uncertainty is obviously necessary in order to obtain a reliable distance estimate to infer a reasonably accurate diameter from the \flame\ radius. We find 213\,139 targets satisfying these criteria. As shown in Fig.~\ref{fig:flame-angular-diameter-vs-Gmag}, all of these are brighter than $G=12$. Most of these targets are cool giants but some are main sequence stars.  Since $\theta=2R/\dist$ and the flux scales as $(R/\dist)^2\propto\theta^2$, we expect $G=\textrm{const}-5\log_{10}\theta$, where the constant depends on \teff, among other things. This scaling is confirmed by Fig.~\ref{fig:flame-angular-diameter-vs-Gmag}a. The diagonal stripes are due to the \flame\ radius estimation relying on \priam\ temperatures which are based on a training sample with a inhomogeneous temperature distribution (see Fig.~\ref{fig:gspphot-Teff-vs-AG-bias}).
\gdr{2} is incomplete at the bright end {\citep{DR2-DPACP-40}}, which unfortunately reduces the overlap with potential interferometric targets.

\subsection{Selection of red clump candidates}

The red clump is prominent in the de-reddened CMD in Fig.~\ref{fig:gspphot-de-reddened-CMD}d. We suggest the following selection criteria to isolate candidates for red clump stars:
\begin{equation}
\begin{aligned}
\label{eq:cuts-for-red-clump}
\parallax/\sigparallax>5 \\
1.2<\bpminrp-\ebpminrp<1.3 \\
0.25<M_G<0.75
\end{aligned}
\end{equation}
This selection produces 1\,877\,297 candidates. Of the 19\,937~red clump candidates from \citet{2014ApJ...790..127B}, 19\,240 are in \gdr{2} and have $\gmag<17$, and 14\,510 of these pass the astrometry criterion above. The cuts in intrinsic colour and absolute magnitude are then passed by 4415 known red clump stars.

\subsection{Selection of solar analogue candidates}

Based on the results from Fig.~\ref{fig:gspphot-Teff-Solar-analogs}, we adopt the following selection criteria for solar analogue candidates:
\begin{equation}
\begin{aligned}
\parallax/\sigparallax>5 \\
|\teff - 5772|<70 \\
0.75<\bpminrp<0.9 \\
3.9<\gmag+5\,\log_{10}\parallax+5<5.1 \\
0.33<\gbp-\gmag<0.40 \\
0.45<\gmag-\grp<0.49 \\
\textrm{\priam flag 0100001 or 0100002}
\end{aligned}
\end{equation}
This selection results in 124\,384 candidates for $\gmag\leq 17$. However, we caution that this selection probably has significant contamination from non-solar-like stars with poorly estimated temperatures.
\section{Looking ahead to the third Gaia data release}
\label{sec:outlook}

The broad band fluxes used in this paper were derived from Gaia's low resolution spectrograph BP/RP.  The full spectra are not made available in \gdr{2} due to insufficient calibration at this stage in the data processing. For the next data release -- \gdr{3} -- we plan to infer more detailed and more precise astrophysical parameters using these spectra.\footnote{Initial results with preliminary BP/RP spectra in an 
empirically-trained algorithm show very promising results when comparing to the literature, not only for \teff, but also for \feh\ and \logg.} This will be done using further algorithms within the astrophysical parameter analysis system (\apsis), of which the algorithms \priam\ and \flame\ discussed in the present paper are just a small part.
For details of \apsis\ see \cite{2013A&A...559A..74B}. 

The full BP/RP spectrophotometry permit estimates of both the effective temperature and the extinction without the need to use parallaxes. Simulations show that these spectra should also allow good estimates of \feh\ and \logg\ \citep{2010MNRAS.403...96B, 2011MNRAS.411..435B, 2012MNRAS.426.2463L}. A separate algorithm will use the spectra together with the parallaxes to estimate \teff, \ag, \feh, and \logg\ together with luminosity and distance, self-consistently. This should provide a more reliable distance estimate than using only the parallax \citep[e.g.][]{2016ApJ...833..119A}, especially for sources beyond a few kpc. The radius estimates will then be more precise, and using the \logg\ estimate we will also be able to infer mass. Using evolutionary tracks we plan also to release estimates of stellar ages for some types of stars.

Independent estimates of stellar parameters for bright stars will be possible using the high resolution radial velocity spectrograph \citep{2016A&A...595A...1G}, which gives information sensitive to \teff, \mh, \aabun, and \logg\ in the region 845--872\,nm \citep{2016A&A...585A..93R}. Further algorithms in \apsis\ will be dedicated to exploring ultra cool dwarfs \citep{2013A&A...550A..44S}, emission line stars, and unresolved binary stars.
 
In the future we will no longer have to treat all sources as single stars. The Discrete Source Classifier in \apsis\ will classify sources using not only BP/RP \citep{2008MNRAS.391.1838B}, but also the astrometry and apparent magnitude. Quasars and unresolved galaxies identified in this way will be treated by source-specific algorithms \citep{2017arXiv170909378D, 2012BlgAJ..18b...3B} to derive redshift, morphology type, etc. In addition to classifying source-by-source, we also plan some global analyses to identify sources which don't fit into our supervised classification schemes \citep[e.g.][]{2013A&A...559A...7F}, and to do large scale statistical cluster analyses to find empirical relations between sources. How much of this can be released in \gdr{3} depends in part on the quality of the data and calibrations at the next stage of the data processing.

\section{Summary}\label{sec:summary}

We have presented the methods used for estimating stellar parameters and line-of-sight extinctions in \gdr{2}, as well as an analysis and validation of the results. The parameter estimates are based on very limited input data, namely the parallax and the integrated photometry in three bands, \gmag, \gbp, and \grp.  More detailed and accurate parameters should be released in \gdr{3} once the BP/RP spectrophotometry and RVS spectra can be used. 

Broadly speaking, we can estimate effective temperature \teff\ with an accuracy of about 324\,K in the range 3000K to 10\,000K. For solar analogues and Gaia benchmark stars \citep{Heiter2015}, we achieve RMS errors as low as 75K and 230K, respectively. Our test and training samples were drawn from the same parent distribution, which are not necessarily representative of all stellar populations. For the same reason, there are some systematic misestimations of \teff, e.g.\ for sources with  very low or very high extinctions (see Fig.~\ref{fig:gspphot-Teff-test-error-vs-other-data}e and Fig.~\ref{fig:gspphot-Teff-test-errors-skymap}). Likewise, there appear to be systematic trends with the actual temperature (Fig.~\ref{fig:gspphot-Teff-test-error-vs-other-data}c), the colour (Fig.~\ref{fig:gspphot-Teff-test-error-vs-other-data}f) and with metallicity (Fig.~\ref{fig:gspphot-Teff-test-error-vs-logg-and-FeH}b). These figures attempt to quantify these systematics, but they are not included in the 324K uncertainty estimates which, by definition, quantify only \textit{random} errors. The reported 16th and 84th percentiles appear to be consistent with differences to literature estimates when also accounting for literature uncertainty estimates.

The line-of-sight extinction \ag~and the reddening \ebpminrp~are estimated with global uncertainties of 0.46\,mag and 0.23\,mag, respectively (section~\ref{ssec:extinction-in-halo}). Better estimates can be achieved for red clump stars (section~\ref{ssec:red-clump-stars}). 
Note that \ag\ is the extinction in the G-band rather than an extinction parameter, so 
also depends on the spectral energy distribution (and thus \teff) of the star. Thus a given amount of dust can correspond to different values of $\ag$, depending on the type of star.
Given only three broad optical bands as input data, there are obviously strong degeneracies; we have explored the systematics these can introduce into our extinction estimates. After applying quality cuts to remove outliers, there is no evidence for further biases. Nevertheless, with such large random errors \ag~and \ebpminrp~are of limited use for individual stars. However, \ag~and \ebpminrp~can be used statistically on a set of stars, for example to apply a dust correction in the colour-magnitude diagram. 
Given the degeneracy,
for our \teff\ estimations and our subsequent estimations of bolometric luminosity \lum~and radius \radius, we have assumed the sources have zero (or at least low) extinction. 

Adopting a bolometric correction offset determined
empirically, we have estimated \lum\ to an accuracy of about 15\% and \radius\ to about 10\% (RMS errors), again without significant systematics. 

Given the limited input data, we necessarily rely on a number of assumptions and simplifications. We list some known limitations of our results in section \ref{sec:limitations}. These must be considered when using our data products. 
We recommend that only our clean \teff\ sample be used (appendix \ref{sec:flags}). For better estimates of extinction, users may want to only use stars with some maximum fractional parallax uncertainty (see Fig.~\ref{fig:gspphot-de-reddened-CMD} for an example of this).

Finally, we emphasise again that the objective of this work was to provide
stellar parameter estimates using Gaia data only. Substantially better results
are likely achievable when combining with other data, such as GALEX
\citep{Morrissey2007}, PanSTARRS \citep{2016arXiv161205560C} or WISE
\citep{AllWiseCat}. 
\citet{2017MNRAS.471..770M} and
\citet{2017arXiv170805025S} have attempted that already with \gdr{1} data,
although they faced severe problems with cross-matching the different
catalogues due to the high spatial resolution of the Gaia catalogue. Great care must be exercised when
using multi-catalogue spectral energy distributions resulting from such
cross-matches, since the passbands are not always well defined for some
photometric systems (e.g., APASS; \citealt{APASSDR9Cat}).

Although limited in precision and accuracy, our results should nonetheless be useful as the largest, all-sky catalogue of homogeneously-inferred stellar parameters published to date, and the first to use parallaxes on a large scale.

\begin{acknowledgements}
For their constructive comments on the manuscript we thank 
Carine Babusiaux,
Anthony Brown,
Ron Drimmel,
Yves Fr\'emat,
Carme Jordi,
Jan Rybizki,
Antonella Vallenari, 
and the anonymous referee.
We thank Floor van Leeuwen for providing membership lists of stellar clusters in \gdr{2}.
This work is based on data from the European Space Agency (ESA) mission {\it Gaia} (\url{https://www.cosmos.esa.int/gaia}), processed by the {\it Gaia} Data Processing and Analysis Consortium (DPAC, \url{https://www.cosmos.esa.int/web/gaia/dpac/consortium}). Funding for the DPAC has been provided by national institutions, in particular the institutions participating in the {\it Gaia} Multilateral Agreement. Funding specifically for the DPAC work in this paper has been provided by the DLR (German space agency) via grants 50\,QG\,0602, 50\,QG\,1001, and 50\,QG\,1403. AJK acknowledges support by the Swedish National Space Board (SNSB).  This research has made extensive use of IPython \citep{Perez2007}, matplotlib \citep{Hunter2007}, and Vaex \citep{Breddels2017}.
\end{acknowledgements}
\bibliographystyle{aa}
\bibliography{apsis,apsis-GDR2}

\appendix
\section{Filtering applied to produce the \gdr{2} catalogue}
\label{sec:results-filtering}

Some of the sources processed by \apsis\ have poor data and cannot yield useful astrophysical parameters. Likewise, our assumptions may sometimes yield invalid results. We therefore apply various filters in post-processing in order to remove some or all parameter estimates. (These filters are applied during the creation of the final catalogue.) Table \ref{tab:results-filtering} lists these filters.
In addition we always require complete photometry -- fluxes available for all of G, BP, and RP -- to produce any results, and
we only processed sources with $\gmag\leq 17.068766$.
Filters applied by other parts of the processing could remove some sources entirely, regardless of the \apsis\ results. 
For more details on filtering, see the online documentation with
\gdr{2}. The number of sources in the catalogue with an extant estimate for each parameter is as follows:
\teff: 161\,497\,595; \ag\ and \ebpminrp: 87\,733\,672; \lum\ and \radius: 76\,956\,778.

\begin{table}[h]
\begin{center}
\caption{Filters applied to the \apsis\
  results to determine whether certain parameters are excluded from
  \gdr{2} for individual stars. 
$\sigma(\teff)$ 
is defined as half the confidence interval, i.e.\ half the difference
between the 84th and 16th percentiles. 
Excluding the estimate of a parameter also excludes its uncertainty
estimates.
The \priam\ flags are defined in Table~\ref{table:Priam-flags}.
\label{tab:results-filtering}
}
\begin{tabular}{llllll}
\hline
condition & \multicolumn{5}{c}{parameters excluded} \\
\hline
$\parallax \leq 0$ & & \ag & \ebpminrp & \lum & \radius \\
$\sigparallax/\parallax > 0.2$ & & & & \lum &\radius \\
$\teff <3300$\,K or $\teff >8000$\,K & & & & \lum &\radius \\
$\sigma(\teff)/\teff > 0.2$  & & & & \lum &\radius \\
$\sigma(\lum)/\lum > 0.3$  & & & & \lum &\radius \\
$\radius < 0.5\Rsun$   & & & & \lum &\radius \\
violation of equations~(\ref{eq:quality-cut-1-AG-and-reddening})-(\ref{eq:quality-cut-4-AG-and-reddening}) & & \ag & \ebpminrp &  &  \\
 \priam\ flag $>0100002$ & & \ag & \ebpminrp &  &  \\
\hline
\end{tabular}
\end{center}
\end{table}

\section{The \apsis flags}
\label{sec:flags}

\begin{table*}
\caption{Definition of \priam processing flags, which have format {\tt 01ABCDE}. 
\label{table:Priam-flags}}
\begin{center}
\begin{tabular}{ccl}
\hline
position & value & meaning \\
\hline
\multirow{2}{*}{\tt A}  & 0 & parallax value is strictly positive ($\varpi>0$) \\
 & 1 & parallax value is non-positive ($\varpi\leq 0$) such that extinction estimate
does not work \\
 & 2 & while $\varpi>0$ the parallax error is $\sigma_\varpi>1$mas\\
\hline
\multirow{4}{*}{\tt B}  & 0 & both colours are close to the standard locus \\
 & 1 & below standard locus, i.e., $G_\textrm{BP}-G>0.1$ and $(G-G_\textrm{RP})<(G_\textrm{BP}-G-0.1)^{0.4} - 0.3$ \\
 & 2 & above standard locus, i.e., $(G-G_\textrm{RP})>(2.5\cdot((G_\textrm{BP}-G)+0.02))$ or \\
 & & $(G-G_\textrm{RP})>(0.5\cdot((G_\textrm{BP}-G)-1.0)+1.1)$ \\
\hline
\multirow{3}{*}{\tt C}  & 0 & $G-G_\textrm{RP}$ colour is inside union of \teff\ and extinction training sets \\
 & 1 & $G-G_\textrm{RP}$ colour is smaller than union of \teff\ and extinction training sets \\
 & 2 & $G-G_\textrm{RP}$ colour is larger than union of \teff\ and extinction training sets \\
\hline
\multirow{3}{*}{\tt D}  & 0 & $G_\textrm{BP}-G$ colour is inside union of \teff\ and extinction training sets \\
 & 1 & $G_\textrm{BP}-G$ colour is smaller than union of \teff\ and extinction training sets \\
 & 2 & $G_\textrm{BP}-G$ colour is larger than union of \teff\ and extinction training sets \\
\hline
\multirow{2}{*}{\tt E} & 1 & input data was gold photometry \\
 & 2 & input data was silver photometry \\
\hline
\end{tabular}
\end{center}
\end{table*}
 
Various flags are written by \apsis\ during processing to indicate the
quality of the input data and/or the results. They have the format {\tt XYABCDE}, where each letter represents a decimal digit.
{\tt XY}=01 indicates a \priam flag, the values of which are shown in Table \ref{table:Priam-flags}.

Digits {\tt C} and {\tt D} refer to colours in the union of the \teff\ and extinction training sets. These can be used to remove sources which are unlikely to get good results from \priam. The colour ranges for the separate models are as follows (so a colour selection could be used for the parameter of interest, instead of using these digits in the flags):
For \teff\: $G_\textrm{BP}-G$: $-0.06$ to $4.38$\,mag, $G-G_\textrm{RP}$: $-0.15$ to $2.08$\,mag.
For extinction: $G_\textrm{BP}-G$: $-0.12$ to $4.66$\,mag, $G-G_\textrm{RP}$: $-0.20$ to $1.69$\,mag.

To get a clean sample of \teff\ estimates, use only sources with \priam\ flag values equal to
0100001, 0100002, 0110001, 0110002, 0120001, or 0120002,
i.e.\ {\tt A} and {\tt E} can have any value, but {\tt B}, {\tt C}, and {\tt D} are zero.
When using \teff\ for astrophysical analyses, we recommend that this clean sample be used.

For \ag\ and \ebpminrp\ estimates, the filtering on best
  \priam\ flag values has already been applied during the catalog
  production for \gdr{2}. The user may want to make further cuts to
  only retain sources with low fractional parallax uncertainties.

After the filtering described in Table~\ref{tab:results-filtering}, 
only one flag corresponding to \flame\ is left in the catalogue (the one beginning with {\tt
  02}), so can be ignored.
Since \flame results depend on \teff, we recommend to only use \flame\
results for the clean \teff\ sample.

\section{Uncertainty estimates for \ag\ and \ebpminrp}
\label{appendix:Deming-formalism-error-estimation}

In section \ref{ssec:red-clump-stars}, we estimate the uncertainties in \ag\ for red clump stars from \citet{2014ApJ...790..127B} by comparing our estimate of \ag~to the observable $\gmag+5\log_{10}\parallax+5$ (which is equal to $\mg+\ag$). 
Likewise, we also compare \ebpminrp\ to the observed colour $\gbp-\grp$.
In these cases, both variables, call them $x$ and $y$, have uncertainties,
so a standard least-squares regression would lead to systematically
wrong results for the slopes and intercepts in
Fig.~\ref{fig:gspphot-red-clump-AG-validation} \citep[e.g.][section 1.1.1 therein]{Fuller2009}. 
For each $y_n$ (the observable), the corresponding uncertainty $\sigma_n$ can be obtained by the usual propagation of  uncertainties in the G-band flux and the parallax. 
The uncertainty $\sigma_x$ in $x$ and the intercept are our desired estimates. These tell us the uncertainties in \ag\ and \ebpminrp, respectively, and whether we are consistent with the expected absolute magnitude and intrinsic colour of the red clump.

When both $x$ and $y$ have uncertainties, then in order to obtain an
unbiased estimate of $\sigma_x$, we use a modification of the
Deming formalism \citep[e.g.][]{Deming1943}. We introduce the true $\hat x_n$ and true $\hat y_n$, which satisfy the linear relation
\begin{equation}
\hat y_n = c_0 + \hat x_n
\,\textrm{.}
\end{equation}
The intercept $c_0$ is the absolute magnitude of the red clump stars for \ag\ or the intrinsic colour for \ebpminrp, which remains a free fit parameter. We set the slope to $c_1=1$, since Fig.~\ref{fig:gspphot-red-clump-AG-validation}b and d have already established an approximate one-to-one relation between $x$ and $y$ and we now seek the uncertainty of $x$ under this relation. We then estimate the true $\hat x_n$, the intercept $c_0$ and the uncertainty $\sigma_x$ by minimising
\begin{equation}
\chi^2 = 2N\log\sigma_x + \sum_{n=1}^N\left[
\left(\frac{y_n - c_0 - \hat x_n}{\sigma_n}\right)^2
+\left(\frac{x_n - \hat x_n}{\sigma_x}\right)^2
\right]
\,\textrm{.}
\end{equation}
Note the first term which ensures that the likelihood function is normalised, while we fit for the unknown uncertainty $\sigma_x$. This minimisation has analytic results 
\begin{equation}
\hat x_n = x_n + \frac{y_n - c_0 - x_n}{1 + \sigma_n^2/\sigma_x^2}
\qquad\forall n=1,\ldots,N
\end{equation}
and
\begin{equation}
c_0 = \frac{\sum_{n=1}^N\frac{y_n-x_n}{\sigma_x^2 + \sigma_n^2}}{\sum_{n=1}^N\frac{1}{\sigma_x^2 + \sigma_n^2}}
\,\textrm{.}
\end{equation}
Unfortunately, there is no analytic solution for $\sigma_x$. However, a numerical solution can be found easily.
\section{Bolometric correction scale}
\label{sec:flame-app}

The bolometric correction used to compute the luminosity \lum\ is calculated on a grid of synthetic stellar spectra for
varying values of \teff, \logg, \feh, and \aabun.
While the variation of the bolometric correction with stellar parameters can be easily evaluated 
considering any filter band pass, the absolute value can only be derived
if we know the absolute magnitude \mg\ for one source with known bolometric flux and distance.
The Sun is the obvious choice.  
From hereon, we refer to this as the offset of the bolometric correction, and this 
is the value of $a_0$ in equation \ref{eqn:bcgfgkstars} for the \teff\ range 4\,000 -- 8\,000 K.

Adopting the previously-mentioned IAU resolution 2015 B2,  
the Sun's bolometric magnitude is \mbolsun\ = 4.74\,mag.  
Given its distance and its measured \vmag-band magnitude, $V_{\odot} = -26.76 \pm 0.03$,
the absolute magnitude $M_{\rm V\odot} = 4.81$  
is calculated and the $BC_{\rm V}$ is derived: $BC_{\rm V} = 4.74 - 4.81 = -0.07$ mag \citep{2010AJ....140.1158T}.  
However, we do not have a measure of $\gmag_{\odot}$, and so we cannot apply these equations directly to determine \bcg.  
Thus to derive the offset to the bolometric correction scale there are four options: 
(1) estimate $G_{\odot}$ from stellar models; 
(2) use a $V$ to $G$ conversion; 
(3) externally calibrate using stars with accurately measured luminosities and radii; 
(4) use solar twins to measure $M_{\rm G}$ (see section~\ref{sec:validation}).  
We chose to adopt solution (3) and this is described in detail here. 
It's also appropriate given that our \teff\ estimates are based on
empirically trained models.

We derive our bolometric correction offset by comparing our overall estimates of luminosity and radius with those from other studies. This works provided the stars in question have accurately determined extinctions (often taken to be zero). Net offsets can indicate a problem with the offset of the bolometric correction (in our study or in the other studies, or both).
We perform this comparison on three samples of stars:
(1) the \citet{2011yCat..35300138C} analysis of the Geneva Copenhagen Survey data;
(2) the JMMC Catalogue of Stellar Diameters \citep{2017yCat.2346....0B}; 
(3) an asteroseismic sample of giant stars from \cite{2016A&A...588A..87V}.
For these samples we also selected those stars where \priam\ $\sigma_{T_{\rm eff}}/T_{\rm eff} < 3\%$ and
Gaia $\sigma_\varpi/\varpi < 5\%$.

\citet{2011yCat..35300138C} provide bolometric flux and \teff\ in their catalogue.  Using these along with 
the Gaia parallaxes we calculated stellar radii and luminosities.  
From this sample we selected the stars within 100 pc ($N = 307$ 
stars in the sample), 
200 pc ($N = 809$) and 300 pc ($N=895$) of 
the Sun, and we imposed  $7 < G < 10$.
We first compared the \priam\ \teff\ with theirs, and by adding an offset to compensate for 
the differences in our \teff\ we rederived our radii. 
This was done to isolate the effect of the \bcg. 
Then we adjusted \bcgsun\ until we minimized the 
mean difference between our results (luminosities and rederived radii) and theirs.   
This resulted in \bcgsun\ = +0.10, +0.09, +0.09 mag for the 
three distance cuts respectively. 

The JMMC catalogue \citep{2017yCat.2346....0B} predicts angular diameters from 
magnitudes and colours; this is the so-called surface-brightness relation.  It is
calibrated using interferometric measurements of stellar diameters.
We performed the same analysis on this catalogue 
for stars within 100 pc ($N = 1\,182$), 200 pc ($N = 5\,427$), and 300 pc ($N = 6\,332$).
By minimizing the mean differences between the radii
we derived a \bcgsun\ of +0.00, +0.01, and +0.01 mag for the three distance cuts respectively.

The third catalogue we used consists of thousands of giants in the {\it Kepler} field analysed
using asteroseismology \citep{2016A&A...588A..87V}. Our validation sample comprises 3355 stars. 
In this case we can not assume that extinction is zero or negligible.  
The effect of the bolometric correction offset and extinction for deriving \lum\ 
is degenerate.  Thus we are required to assume a mean extinction for these stars if we wish to  
estimate \bcgsun.
We adopted \av $= 0.25 \pm 0.10$ mag \cite[see for example][]{2014MNRAS.445.2758R,2015ASSP...39...83Z}.
We then used an \ag\ -- \av\ conversion 
to fix \ag = +0.21 mag in our subsequent analysis. 
By repeating the analysis described above, we obtained $\bcgsun = +0.04$\,mag 
as the best overall agreement between our results and those from \citet{2016A&A...588A..87V}.

Following these analyses, along with subsequent validation
(section~\ref{ssec:flame-validation}), we conclude that the
uncertainty in the offset of the bolometric correction is about 0.10
mag, and we somewhat arbitrarily set the offset to fall within the
extremes that we found here, and define it as \bcgsun\ = +0.060 mag.
This is the value of the $a_0$ coefficient for the temperature range
4\,000 -- 8\,000\,K in Table~\ref{tab:bcgcoeff}.
This result implies $M_{\rm G\odot} = 4.68$ mag and consequently
$(G-V)_{\odot} = -0.13$ mag.  This value is corroborated by the results on 
solar analogues (section~\ref{ssec:Solar-analogs}).

\section{Estimation of cluster extinction}
\label{appendix:mean-cluster-extinction}

As our extinction and reddening estimates are very noisy and also constrained to be non-negative, their combination (for parameter estimation) becomes non-trivial. Here we first outline how to use our results, using the example of estimating the extinction in a star cluster. Second, we show that the sample mean is generally a poor estimator that suffers from strong biases. Third, we use the clusters to obtain another global uncertainty estimate for our extinctions (independent of section~\ref{ssec:extinction-in-halo}).

Suppose we have estimated \ag\ for $N$ cluster members, whereby each estimate has a common uncertainty $\sigma$. (We will not use out inferred confidence interval -- 16th and 84th percentiles -- for each extinction.) 
Suppose further that the intrinsic scatter in the true extinctions is negligible compared to this uncertainty.\footnote{As $\ag$ also depends on the spectral energy distribution of a star, this is true only for low-extinction clusters.  Taking all PARSEC models with $\Zabun=\Zabun_\odot=0.0152$, $\a0=2$\,mag and $\log_{10}$($age$/yr)=9.7, then \ag\ ranges from 1.10 to 1.68\,mag and \ebpminrp\ ranges from 0.73 to 0.84\,mag.}  
Then the true (but unknown) extinctions are all equal to the cluster extinction, $\mu$.  We want to infer $\mu$ and $\sigma$ from the $N$ measurements. 
The likelihood $P(\ag|\mu,\sigma)$ which makes the fewest assumptions in this case is the Gaussian. If $\ag$ is additionally restricted to a finite range (as is the case for our \extratrees outputs), then the least-informative likelihood distribution is a Gaussian truncated over this range \citep{Dowson1973}.\footnote{The least-informative distribution is derived from a maximum entropy argument, as used in section~\ref{ssec:extinction-in-halo} to arrive at the exponential distribution for high Galactic latitudes.
The maximum entropy distribution is different in the present case because we now impose a mode $\mu$ and a
variance $\sigma^2$, which are the parameters of our model we want to find.}
This we can write (properly normalized) as
\begin{equation}\label{eq:single-star-likelihood-cluster-ag}
p(\ag|\mu,\sigma)=
\left\{
\begin{array}{ll}
\frac{\frac{1}{\sigma\sqrt{2\pi}}\exp\left[-\frac{1}{2}\left(\frac{\ag-\mu}{\sigma}\right)^2\right]}{\frac{1}{2}\left[
{\rm erf}\left(\frac{\ag^{\rm max}-\mu}{\sqrt{2}\sigma}\right)
+{\rm erf}\left(\frac{\mu-\ag^{\rm min}}{\sqrt{2}\sigma}\right)
\right]}
& \textrm{for} \ \ag\in[\ag^{\rm min}, \ag^{\rm max}] \\
\\
0 & \textrm{otherwise}
\end{array}
\right.
\,\textrm{,}
\end{equation}
where
%
\begin{equation}
{\rm erf}(z)=\frac{2}{\sqrt{\pi}}\int_0^z e^{-t^2}dt
\,\textrm{.}
\end{equation}
Throughout this appendix we fix $\ag^{\rm min}=0$ and $\ag^{\rm
max}=3.609$\,mag, which is the \ag\ range in the \priam training sample
($\a0\in[0,4]$\,mag).

\begin{figure}
\begin{center}
\includegraphics[width=0.45\textwidth]{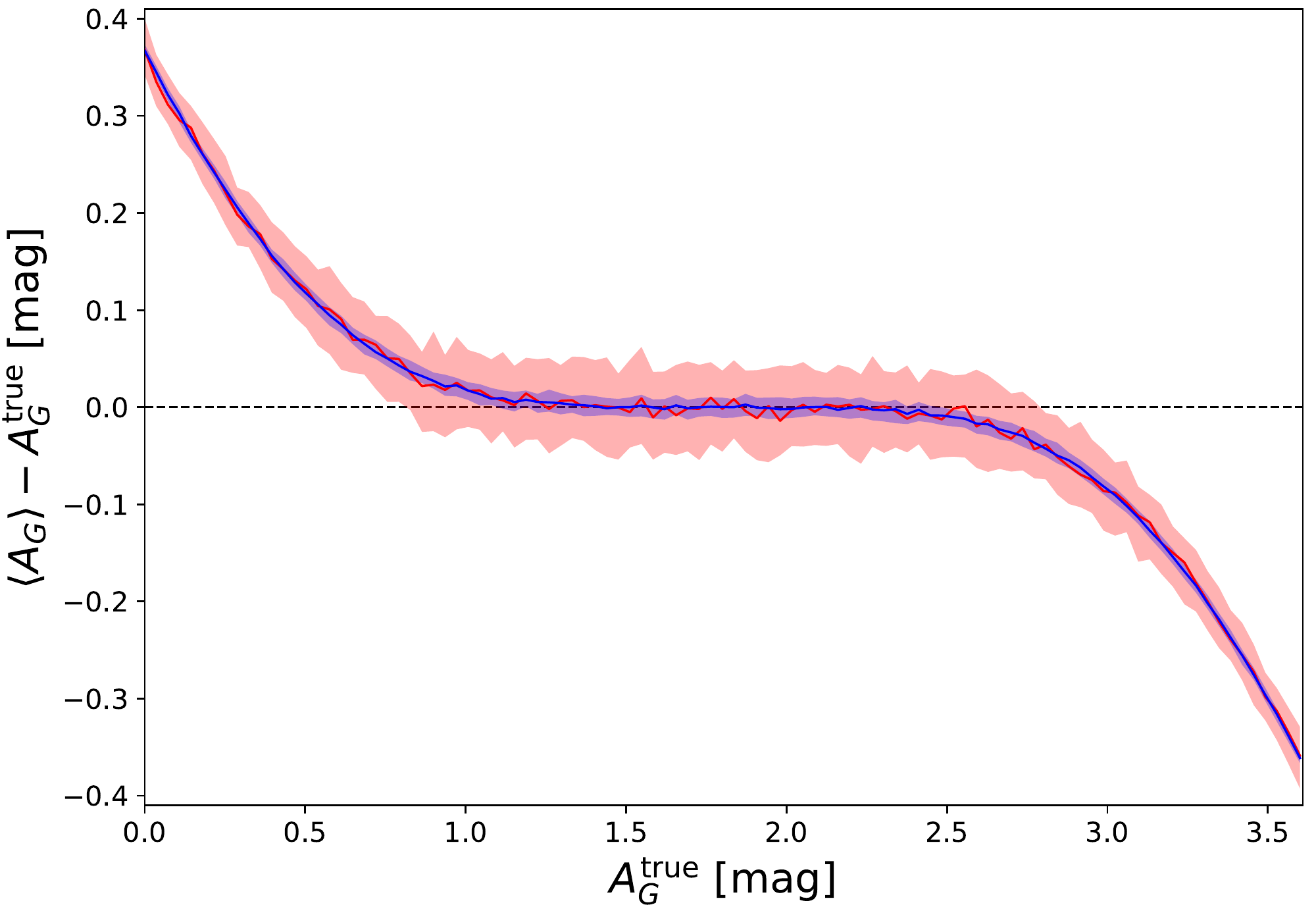}
\caption{Simulation showing how the sample mean is a biased estimator of the cluster extinction when using 100 cluster member stars (red) and 2000 members (blue) with $\sigma=0.46$\,mag. Shaded areas show the central 68\% confidence interval estimated from 101 simulations.}
\label{fig:gspphot-cluster-bias-sample-mean-demo}
\end{center}
\end{figure}

For the set of $N$ cluster members, the total likelihood is just the product of the single-star likelihoods. Note that due to the limitation $\ag\in[0,3.609]$ the normalisation constant of the likelihood now depends on the cluster extinction $\mu$ (and not only on $\sigma$). For this reason, there is no analytic solution for either of the parameters.
In particular, the sample mean is no longer a useful estimator for the cluster extinction $\mu$, which is 
not the mean of the truncated Gaussian (it is the mode, provided it lies within $[\ag^{\rm min}, \ag^{\rm max}]$). We illustrate this with a simulation in which we draw a number of samples from a truncated Gaussian with specified true extinction $\mu$ and fixed $\sigma=0.46$\,mag and compare the sample mean, $\langle\ag\rangle$, to the true extinction of the simulation. As is obvious from Fig.~\ref{fig:gspphot-cluster-bias-sample-mean-demo}, if the true extinction approaches the lower or upper limit, the sample mean becomes biased. Only for clusters with intrinsic extinctions between about 1.3\,mag and 2.4\,mag can we expect the sample mean to be a reliable estimator, i.e.\ when the intrinsic extinction is about $3\sigma$ or more away from the lower and upper limits. For low-extinction clusters, the bias of the sample mean will be largest and of the same order as $\sigma$ for our \ag\ estimates, i.e.\ we may obtain a 0.4\,mag or larger sample mean $\langle\ag\rangle$ for clusters whose expected intrinsic extinction is zero. We emphasise that this bias does \textit{not} diminish if we have more cluster members, as is also clear from Fig.~\ref{fig:gspphot-cluster-bias-sample-mean-demo}.

\begin{figure*}
\begin{center}
\includegraphics[width=0.95\textwidth]{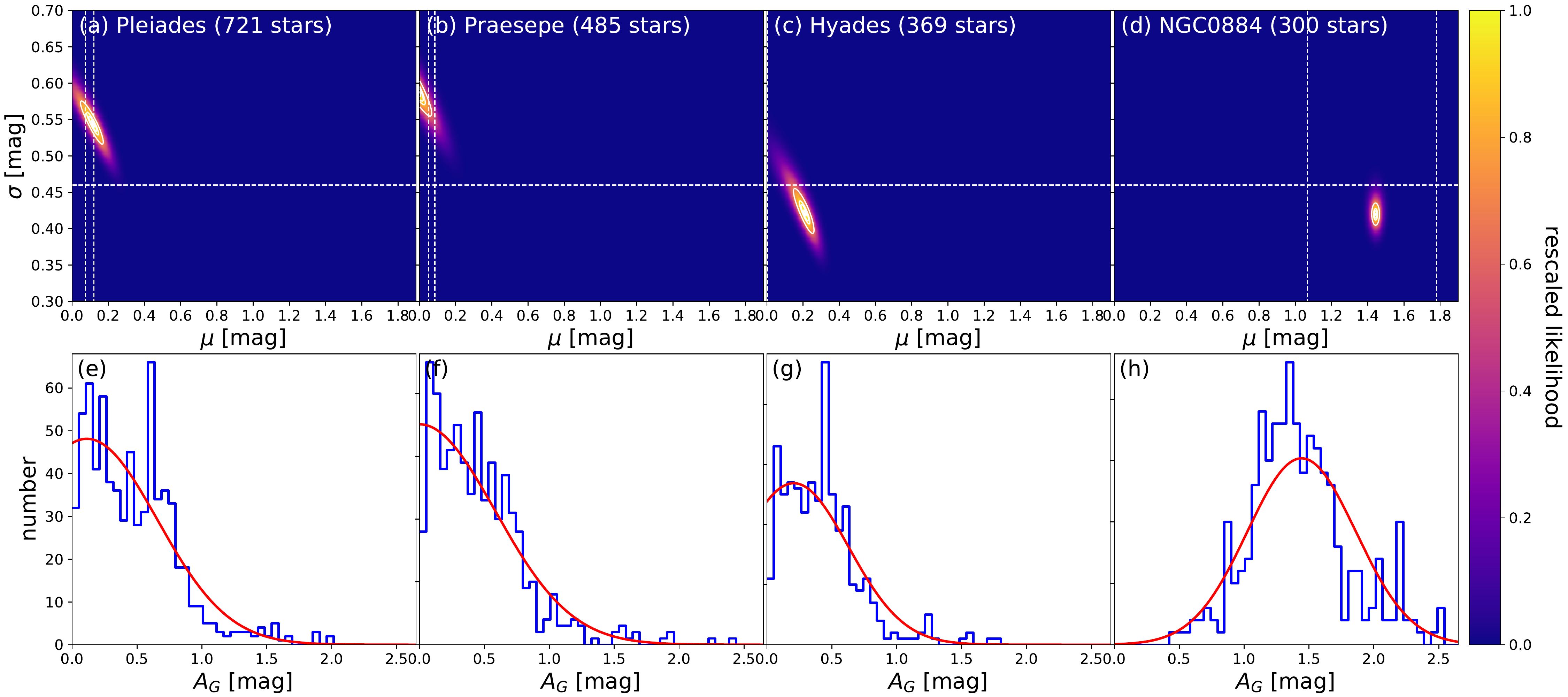}
\caption{Estimating cluster extinctions for Pleiades (a,e), Praesepe (b,f), Hyades (c,g) and NGC884 (d,h).
The top row shows likelihood maps of cluster extinction $\mu$ and global uncertainty estimate $\sigma$.
Contours show 68\%, 90\%, 95\% and 99\% confidence regions. The horizontal dashed line indicates our global uncertainty estimate of 0.46\,mag from section~\ref{ssec:extinction-in-halo}. Vertical dashed lines are $\ag\sim(0.6-1.0)\cdot\a0$
\citep[][Fig.~17 therein]{2010A&A...523A..48J} for literature estimates of  
$\a0\sim 3.1\cdot E(B-V)$ from \citet{2003A&A...397..191P}.
The bottom row shows, for each cluster, the histogram of estimated \ag\ (black) and the maximum-likelihood estimate of the truncated Gaussian (red).
}
\label{fig:gspphot-cluster-extinction-estimation}
\end{center}
\end{figure*}

Since the sample mean is no longer a useful estimator in general, we must use the likelihood from equation~\ref{eq:single-star-likelihood-cluster-ag} in order to estimate parameters.
We find the maximum of the likelihood (Bayesians may add priors).\footnote{The maximum likelihood estimate $\hat{\mu}$ can be negative, but we truncate this to zero on the grounds that true cluster extinctions cannot be negative (this is just a prior). Highly extinct clusters could have true $\mu\geq\ag^{\rm max}=3.609$\,mag, in which case the mode may lie outside our parameter range and $\hat{\mu}$ would be an underestimate.} In Table~\ref{table:cluster-extinction-estimates} we provide our estimates for the 46 clusters from  \citet{DR2-DPACP-31}. Fig.~\ref{fig:gspphot-cluster-extinction-estimation} shows the likelihood function over $\mu$ and $\sigma$ for the Pleiades, Praesepe, Hyades and NGC884. For Praesepe and the Pleiades, the estimated cluster extinctions agree very well with the literature values. Emphasizing the point from Fig.~\ref{fig:gspphot-cluster-bias-sample-mean-demo}, the sample mean extinction for Praesepe is $\langle\ag\rangle=0.4$\,mag, whereas the maximum likelihood estimate is actually $\hat{\mu}=0$\,mag (see Table~\ref{table:cluster-extinction-estimates}), which compares 
favourably with the literature estimate of 0.03\,mag used in section \ref{sec:dust_clusters}. For the Pleiades, the sample mean is 0.48\,mag while the maximum likelihood estimate is 0.11\,mag (cf.\ literature estimate of 0.12\,mag). We obtain $\hat\sigma= 0.59$\,mag and 0.54\,mag for Praesepe and Pleiades respectively, which are larger than our global uncertainty estimate of 0.46\,mag from section~\ref{ssec:extinction-in-halo}. In the case of the Hyades we clearly overestimate the cluster extinction with $\hat{\mu}=0.21$\,mag while $\hat\sigma=0.42$\,mag is slightly below our global uncertainty estimate. (Note from Fig.~\ref{fig:gspphot-cluster-extinction-estimation} that these two quantities are negatively correlated when approaching the lower boundary.) We suspect this to be shot noise
from the finite number of cluster members with \ag\ measurements.
Indeed, simulating 100 Hyades-like samples, each with true $\mu=0.01$\,mag, true $\sigma=0.46$\,mag, and 369 stars, we observe poor estimates in 8\% of the cases. It may come as a surprise that 369 stars is not necessarily sufficient to reliably estimate the cluster extinction and the scatter. The explanation is that for a low-extinction cluster such as the Hyades, the truncation of the Gaussian is dominant. We are trying to infer a pathological distribution, so
the estimate of the mode of the distribution, $\mu$, is more sensitive to the set of samples than is the sample mean.
As an example of a cluster with higher extinction, the maximum-likelihood estimate of NGC884
(panels d and h of Fig.~\ref{fig:gspphot-cluster-extinction-estimation})
also agrees reasonably well with the literature ($E(B-V)=0.58$\,mag from \citealt{2003A&A...397..191P}, which corresponds to $\a0\simeq1.8$\,mag). 
Nevertheless, although our maximum-likelihood estimates of the cluster extinctions are largely consistent with the expected values, the random errors are still far too large for detailed studies of Hertzsprung-Russell diagrams in clusters. 

\begin{table}
\caption{Estimates of the cluster extinction using the
sample mean $\langle\ag\rangle$ (which is biased) and the maximum likelihood (ML) estimates of $\mu$, as well as the global uncertainty $\sigma$.
Clusters are sorted with descending number of member stars. Nearby clusters within 250pc are marked in bold face.
Cluster memberships are taken from \citet{DR2-DPACP-31}.}
\label{table:cluster-extinction-estimates}
\begin{center}
\begin{tabular}{lrrrr}
\hline
      & no.\ of & & \multicolumn{2}{c}{ML estimates} \\
\cline{3-5}
name & stars & $\langle\ag\rangle$  & $\hat{\mu}$ & $\hat{\sigma}$ \\
 &  & mag & mag & mag \\
\hline
NGC3532 & 1217 & 0.293 & $0.000^{+0.005}_{-0.000}$ & $0.406^{+0.007}_{-0.009}$ \\
NGC2437 & 962 & 0.538 & $0.503^{+0.012}_{-0.013}$ & $0.309^{+0.011}_{-0.008}$ \\
NGC2516 & 894 & 0.373 & $0.083^{+0.042}_{-0.052}$ & $0.427^{+0.025}_{-0.020}$ \\
NGC2682 & 724 & 0.371 & $0.000^{+0.025}_{-0.000}$ & $0.479^{+0.008}_{-0.020}$ \\
\textbf{Pleiades} & 721 & 0.478 & $0.113^{+0.056}_{-0.073}$ & $0.544^{+0.034}_{-0.027}$ \\
NGC2168 & 709 & 0.629 & $0.619^{+0.010}_{-0.012}$ & $0.283^{+0.009}_{-0.008}$ \\
NGC6475 & 612 & 0.373 & $0.165^{+0.042}_{-0.055}$ & $0.381^{+0.031}_{-0.020}$ \\
\textbf{Praesepe} & 485 & 0.464 & $0.000^{+0.078}_{-0.000}$ & $0.585^{+0.005}_{-0.047}$ \\
IC4651 & 474 & 0.506 & $0.360^{+0.037}_{-0.047}$ & $0.423^{+0.031}_{-0.022}$ \\
NGC0188 & 450 & 0.585 & $0.504^{+0.026}_{-0.032}$ & $0.402^{+0.025}_{-0.018}$ \\
NGC2360 & 421 & 0.444 & $0.085^{+0.055}_{-0.074}$ & $0.516^{+0.030}_{-0.036}$ \\
Stock2 & 410 & 0.802 & $0.779^{+0.021}_{-0.022}$ & $0.394^{+0.019}_{-0.016}$ \\
NGC2447 & 377 & 0.267 & $0.000^{+0.012}_{-0.000}$ & $0.375^{+0.011}_{-0.017}$ \\
\textbf{Hyades} & 369 & 0.424 & $0.209^{+0.054}_{-0.077}$ & $0.421^{+0.044}_{-0.028}$ \\
NGC2422 & 343 & 0.358 & $0.000^{+0.054}_{-0.000}$ & $0.457^{+0.008}_{-0.036}$ \\
NGC6281 & 309 & 0.546 & $0.510^{+0.020}_{-0.025}$ & $0.316^{+0.020}_{-0.016}$ \\
NGC0884 & 300 & 1.445 & $1.444^{+0.024}_{-0.024}$ & $0.419^{+0.019}_{-0.016}$ \\
IC4756 & 299 & 0.597 & $0.453^{+0.047}_{-0.062}$ & $0.473^{+0.043}_{-0.029}$ \\
NGC1039 & 296 & 0.400 & $0.000^{+0.050}_{-0.000}$ & $0.518^{+0.011}_{-0.037}$ \\
\textbf{alphaPer} & 284 & 0.736 & $0.284^{+0.110}_{-0.163}$ & $0.778^{+0.087}_{-0.052}$ \\
NGC2548 & 284 & 0.331 & $0.000^{+0.060}_{-0.000}$ & $0.421^{+0.006}_{-0.039}$ \\
NGC6405 & 275 & 0.688 & $0.578^{+0.043}_{-0.054}$ & $0.489^{+0.042}_{-0.029}$ \\
NGC0869 & 264 & 1.423 & $1.423^{+0.026}_{-0.027}$ & $0.424^{+0.020}_{-0.017}$ \\
IC4725 & 258 & 1.088 & $1.086^{+0.022}_{-0.022}$ & $0.356^{+0.018}_{-0.015}$ \\
NGC2423 & 255 & 0.474 & $0.118^{+0.067}_{-0.099}$ & $0.537^{+0.042}_{-0.043}$ \\
NGC6025 & 243 & 0.558 & $0.534^{+0.022}_{-0.023}$ & $0.296^{+0.020}_{-0.015}$ \\
\textbf{Blanco1} & 191 & 0.265 & $0.000^{+0.014}_{-0.000}$ & $0.422^{+0.020}_{-0.023}$ \\
NGC6633 & 190 & 0.562 & $0.478^{+0.041}_{-0.053}$ & $0.392^{+0.041}_{-0.027}$ \\
Trumpler10 & 175 & 0.428 & $0.000^{+0.029}_{-0.000}$ & $0.636^{+0.029}_{-0.040}$ \\
NGC7092 & 174 & 0.338 & $0.000^{+0.025}_{-0.000}$ & $0.497^{+0.023}_{-0.032}$ \\
\textbf{IC2602} & 173 & 0.805 & $0.000^{+0.203}_{-0.000}$ & $1.020^{+0.018}_{-0.127}$ \\
\textbf{NGC2451} & 145 & 0.747 & $0.000^{+0.146}_{-0.000}$ & $0.970^{+0.030}_{-0.104}$ \\
NGC2323 & 143 & 0.637 & $0.601^{+0.034}_{-0.042}$ & $0.357^{+0.036}_{-0.024}$ \\
NGC0752 & 141 & 0.408 & $0.000^{+0.039}_{-0.000}$ & $0.586^{+0.028}_{-0.044}$ \\
\textbf{IC2391} & 134 & 0.732 & $0.000^{+0.176}_{-0.000}$ & $0.938^{+0.024}_{-0.116}$ \\
Trumpler02 & 134 & 0.848 & $0.842^{+0.030}_{-0.032}$ & $0.339^{+0.026}_{-0.021}$ \\
NGC2158 & 132 & 1.107 & $1.078^{+0.050}_{-0.057}$ & $0.537^{+0.049}_{-0.035}$ \\
NGC6774 & 127 & 0.495 & $0.339^{+0.071}_{-0.115}$ & $0.424^{+0.076}_{-0.039}$ \\
NGC2547 & 122 & 0.729 & $0.000^{+0.088}_{-0.000}$ & $1.028^{+0.052}_{-0.086}$ \\
\textbf{ComaBer} & 109 & 0.257 & $0.000^{+0.025}_{-0.000}$ & $0.391^{+0.024}_{-0.030}$ \\
NGC6793 & 103 & 0.962 & $0.961^{+0.031}_{-0.031}$ & $0.310^{+0.026}_{-0.019}$ \\
NGC1901 & 71 & 0.407 & $0.000^{+0.086}_{-0.000}$ & $0.554^{+0.032}_{-0.068}$ \\
NGC2232 & 70 & 0.681 & $0.000^{+0.121}_{-0.000}$ & $0.967^{+0.064}_{-0.108}$ \\
Coll140 & 63 & 0.647 & $0.000^{+0.109}_{-0.000}$ & $0.956^{+0.072}_{-0.105}$ \\
NGC3228 & 60 & 0.469 & $0.000^{+0.137}_{-0.000}$ & $0.616^{+0.032}_{-0.095}$ \\
IC4665 & 56 & 0.691 & $0.666^{+0.052}_{-0.068}$ & $0.354^{+0.062}_{-0.034}$ \\
\hline
\end{tabular}
\end{center}
\end{table}

Finally, we can see from Table~\ref{table:cluster-extinction-estimates} that the maximum-likelihood estimates of $\sigma$ vary quite considerably, sometimes being below and sometimes above our global uncertainty estimate of 0.46\,mag from section~\ref{ssec:extinction-in-halo}. 
Unsurprisingly, the estimation of cluster extinctions works better for some samples than for others.


\end{document}